
%
\input amstex
\input amsppt.sty
\input epsf
%
%
%

\CenteredTagsOnSplits
\NoBlackBoxes
\def\today{\ifcase\month\or
 January\or February\or March\or April\or May\or June\or
 July\or August\or September\or October\or November\or December\fi
 \space\number\day, \number\year}
\define\({\left(}
\define\){\right)}

\define\Aut{\operatorname{Aut}}
\define\CC{{\Bbb C}}

\define\Diff{\operatorname{Diff}}

\define\End{\operatorname{End}}

\define\Hom{\operatorname{Hom}}

\define\RR{{\Bbb R}}

\define\Tr{\operatorname{Tr}}
\define\ZZ{{\Bbb Z}}
\define\[{\left[}
\define\]{\right]}

\define\chiup{\raise.5ex\hbox{$\chi$}}
\define\cir{S^1}

\define\exertag #1#2{\removelastskip\bigskip\medskip\eightpoint\noindent%
\hbox{\rm\ignorespaces#2\unskip} #1.\ }

\define\inv{^{-1}}
\define\mstrut{^{\vphantom{1*\prime y}}}
\define\protag#1 #2{#2\ #1}

\define\res#1{\negmedspace\bigm|_{#1}}
\define\temsquare{\raise3.5pt\hbox{\boxed{ }}}

\define\theprotag#1 #2{#2~#1}

\define\zmod#1{\ZZ/#1\ZZ}

\define\Atil{\tilde{A}}
\define\CG{\Cal{G}}
\define\CW{\Cal{W}}
\define\Ctil{\tilde{C}}
\define\ER{E_\RR}
\define\Eap{E\ua(\pt)}
\define\Ep{E(\pt)}
\define\Gb#1{\CG_{x_#1}}
\define\Mor{\operatorname{Mor}}
\define\Obj{\operatorname{Obj}}
\define\Pb{[P]}
\define\Qtriv{Q_{\text{triv}}}
\define\Qx{Q_{[x]}}
\define\RZ{\RR/\ZZ}
\define\Rtriv{R_{\text{triv}}}
\define\TT{\Bbb T}
\define\T{$\TT$}
\define\Vb{\overline{V}}
\define\Vect{\operatorname{Vect}}
\define\WG{\CW_\CG}
\define\Wb{\bold{W}}
\define\bM{\partial M}
\define\bP{\partial P}
\define\bX{\partial X}
\define\bY{\partial Y}
\define\bal{\boldsymbol\lambda }
\define\bfld#1{{\Cal C}^\prime_{#1}}
\define\bfldb#1{\overline{\bfld{#1}}}
\define\bo{\bold{1}}
\define\ca#1#2{T_{#1}(#2)}
\define\cao{T\ua_{\pt}(\Qtriv)}
\define\cat#1{\Cal{C}_{#1}}
\define\cut{^{\text{cut}}}
\define\eac#1#2{e^{2\pi iS_{#1}({#2})}}
\define\eint#1#2{\exp\( 2\pi i\int_{#1}{#2}\)}
\define\etpi#1{e^{2 \pi i #1}}
\define\fld#1{\Cal{C}_{#1}}
\define\fldb#1{\overline{\Cal{C}_{#1}}}
\define\form{(\cdot,\cdot)}
\define\fugo{\Cal{F}(\goid)}
\define\fug{\Cal{F}(\Gamma)}
\define\fun#1{\Cal{F}_{#1;\alpha}}
\define\goidtilde{\tilde{\goid}}
\define\goid{\Cal G}
\define\grot{\operatorname{Groth}\bigl(E(\cir)\bigr)}
\define\hl#1#2{L(#1,#2)}
\define\id{\operatorname{id}}
\define\iline#1{I_{#1,\alpha}}
\define\ilines#1{I_{#1;\alpha}}
\define\intline#1#2{I_{#1,#2}}
\define\ivr#1#2{(#1)^{#2}}
\define\lc#1#2{L_C(#1,#2)}
\define\lce#1#2{\ell_C(#1,#2)}
\define\lin#1#2{L_{#1}(#2)}
\define\lt#1#2{L_P(#1\,|\,#2)}
\define\lte#1#2{\ell_P(#1\,|\,#2)}
\define\mytimes{\odot}
\define\nG{\#\Gamma}
\define\nte#1#2{\ell_{#1,#2}}
\define\ooGd{\frac{1}{\#\Gamma }\cdot}
\define\phibar{\bar{\varphi}}
\define\pt{pt}
\define\spann{\operatorname{span}}
\define\tcat#1{\Cal{T}_{#1}}
\define\te#1#2{\ell (#1,#2)}
\define\tors#1#2{T_{#1}(#2)}
\define\tpi{2 \pi i}
\define\ttor#1#2{T(#1,#2)}
\define\ua{^{(\alpha)}}
\define\vect#1{\Cal{V}_{#1}}
\define\voll{\operatorname{vol}}
\define\zot{[0,1]\times }
\define\zo{[0,1]}
\CenteredTagsOnSplits
\NoRunningHeads
\loadbold
\refstyle{A}
\widestnumber\key{DVVV}

	\topmatter
 \pretitle{$$\boxed{\boxed{\text{REVISED VERSION}}}$$\par\vskip 3pc}
 \title\nofrills Higher Algebraic Structures and Quantization \endtitle
 \author Daniel S. Freed  \endauthor
 \thanks The author is supported by NSF grant DMS-8805684, a Presidential
Young Investigators award DMS-9057144, and by the O'Donnell Foundation.  He
warmly thanks the Geometry Center at the University of Minnesota for their
hospitality while this work was undertaken.\endthanks
 \affil Department of Mathematics \\ University of Texas at Austin\endaffil
 \address Department of Mathematics, University of Texas, Austin, TX
78712\endaddress
 \email dafr\@math.utexas.edu \endemail
 \date April 22, 1993\enddate
	\abstract\nofrills{\smc Very Abstract.}
 We derive (quasi-)quantum groups in $2+1$~dimensional topological field
theory directly from the classical action and the path integral.  Detailed
computations are carried out for the Chern-Simons theory with finite gauge
group.  The principles behind our computations are presumably more general.
We extend the classical action in a $d+1$~dimensional topological theory to
manifolds of dimension less than~$d+1$.  We then ``construct'' a generalized
path integral which in $d+1$~dimensions reduces to the standard one and in
$d$~dimensions reproduces the quantum Hilbert space.  In a $2+1$~dimensional
topological theory the path integral over the circle is the category of
representations of a quasi-quantum group.  In this paper we only consider
finite theories, in which the generalized path integral reduces to a finite
sum.  New ideas are needed to extend beyond the finite theories treated here.
	\endabstract
	\endtopmatter

\document

Recent work on invariants of low dimensional manifolds utilizes complicated
algebraic structures, for both theory and computation.  New invariants of
3-manifolds, and of knots and links in 3-manifolds, are constructed from
certain types of Hopf algebras~\cite{RT} or more generally from special sorts
of categories~\cite{KR}.  These invariants are known to arise from a
$2+1$~dimensional quantum field theory~\cite{W}.  In this paper we derive the
algebraic structure from the field theory, starting with the classical
lagrangian, and so express the relationship between the algebra and the
geometry directly.  With this understanding the algebra can be put to work to
calculate invariants.  The guiding principle for us is the {\it locality\/}
of field theory, as expressed in {\it gluing laws\/}.  The gluing laws
resonate well with cut and paste techniques in topology.  They are important
tools field theory offers for both theoretical work and computations.  We
generalize the standard constructs in a $d+1$~dimensional field theory---{\it
classical action\/} and {\it path integral\/}---to spaces of dimension less
than~$d+1$, retaining the essential property of locality.  Whereas the
classical action is always a finite dimensional integral, the path integral
over the space of fields usually involves infinitely many variables.  Our
focus here is not on the analytical difficulties of path integrals over
infinite dimensional spaces; we only treat path integrals in a ``toy model''
where they reduce to finite sums.  Nevertheless, our generalizations of the
classical action and path integral most likely pertain to other topological
field theories.

In a $d+1$ dimensional field theory the classical action of a field~$\Theta $
on a $(d+1)$-manifold~$X$ is usually a real\footnote{The theories we consider
in this paper are unitary.} number~$S_X(\Theta )$.  Often in topological
theories only the exponential~$\eac X\Theta $ is well-defined.  The simplest
example is the holonomy of a connection: $X=\cir$~is the circle and the field
$\Theta $~is a connection on a principal circle bundle $P\to\cir$.  Notice
that the action is not as straightforward if $X=[0,1]$ has
boundary---interpreted as a number the parallel transport of a connection
over the interval depends on boundary conditions.  Rather, the dependence on
boundary conditions is best expressed by regarding the parallel transport as
a map~$P_0\to P_1$ from the fiber of the circle bundle over~0 to the fiber
over~1.  This is the classical action over the interval.  Our generalization
of the classical action asserts that the classical action of a connection
over a point, which is just a principal circle bundle $Q\to pt$, is the
fiber~$Q$.  The value of that action is a space on which the circle
group~$\TT$ acts simply transitively, a so-called {\it \T-torsor\/}.  Notice
that the action of a field (connection) on the interval takes values in the
action of the restriction of the field to the boundary.  The Chern-Simons
invariant in 3~dimensions is similar---the action in 2~dimensions is a
\T-torsor---and the story continues to lower dimensions~\cite{F1},
{}~\cite{F2}.

At the crudest level of structure the classical action in $d$~dimensions is a
{\it set\/}.  (The classical action in $d+1$~dimensions is a {\it number\/}.)

The usual path integral in a $d+1$~dimensional theory may be written
schematically as
  $$ \int_{\fld X}d\mu _X(\Theta )\;\eac X\Theta , $$
where $X$~is a $(d+1)$-manifold without boundary, $\fld X$~is the space of
fields on~$X$, and $d\mu _X$ is a measure on~$\fld X$.  Of course, in many
examples of interest this is only a formal expression since the measure does
not exist, or has not been constructed.  This integral is a sum of positive
numbers (the measure) times complex numbers (the exponentiated action), so is
a complex number.  Our generalization to $d$~dimensions is as follows.  The
action is now a \T-torsor, which we extend to a {\it hermitian line\/}, i.e.,
a one dimensional complex inner product space.  The original \T-torsor is the
set of elements of unit norm in the associated hermitian line.  The integral
is then a sum of positive numbers times hermitian lines.  If $L$~is a
hermitian line and $\mu $~a positive number, let $\mu \cdot L$~be the same
underlying one dimensional vector space with inner product multiplied by~$\mu
$.  We sum hermitian lines via direct sum; the sum is a hermitian vector
space, or Hilbert space.  Formally, then, this generalized path integral is
the space of $L^2$~sections of a line bundle over the space of fields.  When
the space of fields has continuous parameters we can formally reinterpret
canonical quantization, or geometric quantization, as the regularization
needed to make sense of the integral.

In higher codimensions the classical action and path integral take values in
certain generalizations of \T-torsors and vector spaces.  The next step after
a \T-torsor is a {\it \T-gerbe\/}~\cite{Gi}, ~\cite{Br}, ~\cite{BMc} and the
next step after a vector space is a {\it 2-vector space\/}~\cite{KV},
{}~\cite{L}.  The underlying structure in both cases is not a set, but rather a
{\it category\/}.  The continuation to higher codimensions leads to {\it
multicategories\/}, and the foundations become rather murky, at least to this
author.  We attempt an exposition of these ``higher algebraic structures''
in~\S{1} and~\S{3}.  Our treatment has no pretensions of rigor.  For this
reason throughout this paper we use the term `Assertion' as opposed to
`Theorem' or `Proposition', except when dealing with ordinary sets and
categories.  Since we deal with {\it unitary\/} theories our quantum spaces
have an inner product, so are Hilbert spaces.  In codimension two we
therefore obtain {\it 2-inner product spaces\/} or {\it 2-Hilbert spaces\/}.
The terminology may be confusing: A 2-inner product space is an ordinary
category, not a 2-category.

The particular model we treat is gauge theory with finite gauge group.  It
exists in any dimension.  This theory was introduced by
Dijkgraaf/Witten~\cite{DW} and further developed by many authors~\cite{S2},
{}~\cite{Ko}, ~\cite{Q1}, ~\cite{Q2}, ~\cite{Fg}, ~\cite{Y3}, ~\cite{FQ}.  In
some ways this paper is a continuation of ~\cite{FQ}, though it may be read
independently.  The space of fields (up to equivalence) on a compact manifold
is a finite set in this model, hence all path integrals reduce to finite
sums.  The lagrangian in the $d+1$~dimensional theory is a singular
$(d+1)$-cocycle, and the generalized classical action is defined as its
integral over compact oriented manifolds of dimension less than or equal
to~$d+1$.  Only the cohomology pairing with the fundamental class of a closed
oriented $(d+1)$-manifold is standardly defined.  In the appendix we briefly
describe an integration theory which extends this pairing.  It is the origin
of the torsors, gerbes, etc. that we encounter.  We define the generalized
classical action in~\S{2} and the generalized path integral in~\S{4}.  Our
assertions in these sections are formulated for all codimensions
simultaneously, and we suggest that the reader decipher them starting in the
top dimension, where they reduce to the corresponding theorems in~\cite{FQ}.

In~\S{5} we explore the structure of the generalized path integral~$E$ over a
circle in $1+1$~dimensional theories and in $2+1$~dimensional theories.  The
treatment here is based on the generalized axioms of topological field
theory\footnote{These axioms are not meant to be complete, and in any case
they must be modified in other examples to allow for central extensions of
diffeomorphism groups.  See~\cite{A}, ~\cite{Q2} for a discussion of the
general axioms in topological field theory.  See~\cite{F3} for a discussion
of central extensions.} set out in \theprotag{2.5} {Assertion} and
\theprotag{4.12} {Assertion}, not on any particular features of finite gauge
theory.  In a $1+1$~dimensional theory $E$~is an inner product space and we
construct a compatible algebra structure and a compatible real structure.
The argument here is standard.  In a $2+1$~dimensional theory $E$~is a
2-inner product space, which in particular is a category.  The analogue of
the real algebra structure, here derived from the generalized path integral,
makes this a {\it braided monoidal category\/} with compatible ``balancing''
and duality.  Such categories arise in rational conformal field
theory~\cite{MS}, and have been much discussed in connection with topological
invariants and topological field theory.  Reconstruction theorems in category
theory~\cite{DM}, ~\cite{Ma1} assert that such a category is the category of
representations of a {\it quasitriangular quasi-Hopf algebra\/}, or {\it
quasi-quantum group\/}~\cite{Dr}.\footnote{I believe that the quasitriangular
quasi-Hopf algebras we obtain will always have a ``ribbon element''~\cite{RT}
as well.  This certainly holds in the finite gauge theory.} In fact, the
reconstruction also requires a special functor from the category~$E$ to the
category of vector spaces.  We remark that a different quasi-Hopf algebra
related to field theories was proposed in~\cite{Ma3}.

We put the abstract theory of \S\S{1--4} to work in~\S\S{6--9}, where we
carry out the computations for the finite gauge theory.  We warmup in~\S{6}
by discussing some features of the $1+1$~dimensional theory.  The remainder
of the paper treats the $2+1$~dimensional Chern-Simons theory (with finite
gauge group).  The quasi-Hopf algebras we compute via the generalized path
integral are the quasi-Hopf algebras introduced by
Dijkgraaf/Pasquier/Roche~\cite{DPR}.  They were further studied by
Altschuler/Coste~\cite{AC}.  The computations are not difficult, but they are
nerve-racking!  When dealing with categories (and, even worse,
multicategories) one must be very careful about equality versus isomorphism,
at the next level about equality of isomorphisms versus isomorphisms of
isomorphisms, and so on.  This sort of algebra seems well-adapted to the
geometry of cutting and pasting, but as I said it {\it is\/} nerve-racking.
We keep close track of the trivializations we need to introduce at various
stages of the computation.  Some of these trivializations are used to define
the functor to the category of vector spaces which we need to reconstruct the
quasi-Hopf algebra.  Our reconstructions do not follow the procedures in the
abstract category theory proofs.  Rather, in our examples the algebras are
apparent from appropriate descriptions of the braided monoidal category.
In~\S{9} we use more sophisticated gluing arguments to choose special bases
of the algebras, and so derive the exact formulas in~\cite{DPR}.  This
involves cutting and pasting manifolds with the simplest kind of corners.  We
formulate a generalized gluing law for the classical action in
\theprotag{9.2} {Assertion}.  Clearly it generalizes to higher codimensional
gluing and to the quantum theory.  Segal~\cite{S1} gives a proof of the
``Verlinde diagonalization''~\cite{V} using a quantum version of this gluing
law.  This sort of generalized gluing should be useful in other problems as
well.  We also briefly describe at the end of~\S{7} how Segal's {\it modular
functor\/}~\cite{S1} fits in with our approach.

In gauge theory one usually makes special arguments to account for {\it
reducible connections\/}.  In these finite gauge theories every
``connection'' is reducible, that is, every bundle has nontrivial
automorphisms, and all of the constructions must account for the automorphism
groups.

We formulate everything in terms of manifolds, whereas others prefer to work
more directly with knots and links.  The relationship is the following
(cf.~\cite{W}).  Suppose $K$~is a knot in a closed oriented 3-manifold~$X$.
Let $X'=X-\nu (K)$ denote the manifold~$X$ with an open tubular
neighborhood~$\nu (K)$ of the knot removed.  Then a {\it framing\/} of the
normal bundle of~$K$ in~$X$ determines an isotopy class of diffeomorphisms
from the standard torus~$\cir\times \cir$ to~$\partial X'=-\partial \bigl(\nu
(X) \bigr)$.  In a $2+1$~dimensional topological field theory this induces an
isometry between the quantum Hilbert space of~$\partial X'$ and the quantum
Hilbert space of the standard torus.  So the path integral over~$X'$ takes
values in the Hilbert space of the standard torus.  As we explain at the end
of~\S{9} this Hilbert space is the ``Grothendieck ring'' of the monoidal
category discussed above, and it has a distinguished basis consisting of
equivalence classes of irreducible representations.  These are the ``labels''
in the theory, and the coefficients of the path integral over~$X'$ are the
knot invariants for labeled, framed knots.  The generalization to links is
immediate.

An expository version of some of this material appears in~\cite{F3}.

I warmly thank Larry Breen, Misha Kapranov, Ruth Lawrence, Nicolai
Reshetikhin, Jim Stasheff, and David Yetter for informative discussions.

\newpage
\head
\S{1} Higher Algebra I
\endhead
\comment
lasteqno 1@  4
\endcomment

Whereas the classical action in a $d+1$~dimensional field theory typically
takes values in the real numbers, often in topological theories only its
exponential with values in the circle group
  $$ \TT = \{\lambda \in \CC: |\lambda |=1\} $$
is defined.  We remark that nonunitary versions of these theories would
replace~$\TT$ by the group~$\CC^\times $ of nonzero complex numbers.  For the
algebra in this section we could replace~$\TT$ by any commutative group.  The
usual action is defined for fields on closed\footnote{Here `closed' means
`compact without boundary'.  There is also a (relative) action on compact
manifolds with boundary, which we describe below.} spacetimes of
dimension~$d+1$.  In~\S{2} we describe ``higher actions'' which are defined
for fields on manifolds of dimension less than~$d+1$ and take their values in
``higher groups''.  For example, over closed $d$-manifolds the action takes
its values in the {\it abelian group-like category\/} of $\TT$-{\it
torsors\/}.  On a closed $(d-1)$-manifold the action takes its values in the
{\it abelian group-like 2-category\/} of $\TT$-{\it gerbes\/}.  And so on.
In this section we briefly describe these ``higher groups''.  We also use the
term ``higher torsors''.  As stated in the introduction we only attempt a
heuristic treatment, not a rigorous one.  Our goal in this section, then, is
to explain a hierarchy:
  $$ \alignedat 2
      &\tcat 0=\TT&\qquad&\text{circle group}\\
      &\tcat 1&\qquad&\text{``group'' of $\TT$-torsors (1-torsors)}\\
      &\tcat 2&\qquad&\text{``group'' of $\TT$-gerbes (2-torsors)}\\
      &\text{etc.}&\qquad& \endalignedat \tag{1.1} $$
Each of these is an abelian group in the sense that there is a commutative
associative composition law, an identity element, and inverses.  However,
only $\tcat 0$~is an honest group; in fact, only $\tcat0$~is a set!  The
$\TT$-torsors~$\tcat1$ form a {\it category\/},\footnote{We refer
to~\cite{Mac} for the basics of category theory as well as plenty of
examples.  Roughly, a category~$\Cal{C}$ is a collection of
objects~$\Obj(\Cal{C})$ and for every $A,B\in \Obj(\Cal{C})$ there is a {\it
set\/} of morphisms $\Mor(A,B)$.  Morphisms $A @>f>> B$ and $B @>g>> C$
compose to give a morphism $A @>gf>> C$.  This composition is associative and
there are identity morphisms.  Notice that $\Obj(\Cal{C})$~is not necessarily
a set.  We often write `$A\in \Cal{C}$' for `$A\in \Obj(\Cal{C})$'.} the
$\TT$-gerbes a {\it 2-category\/},\footnote{A {\it 2-category\/}~$\Cal{C}$
has a collection of objects~$\Obj(\Cal{C})$ and for each $A,B\in \Obj(\Cal{C})$
a {\it category\/} of morphisms~$\Mor(A,B)$.  In other words, if
$f,g\in \Mor(A,B)$, then there is a set of {\it 2-morphisms\/} which map
from~$f$ to~$g$.  The composition $\Mor(A,B)\times \Mor(B,C)\to\Mor(A,C)$ is
now assumed to be a functor.  One obtains different notions depending on
whether one assumes that this composition is exactly (strictly) associative
or whether one postulates that it is associative up to a given 2-morphism.
The former notion generalizes to $n$-categories.  The latter notion was
introduced by Benabou~\cite{Be} for 2-categories (these are called
``bicategories''), and apparently a complete list of axioms for the higher
case has not been written down.  (See the lists of axioms in~\cite{KV} to see
the complications involved.)  Since for three \T-torsors~$A,B,C$ the torsors
$(A\otimes B)\otimes C$ and $A\otimes (B\otimes C)$ are strictly speaking
different, but isomorphic, the category~$\tcat1$ does not have a strictly
associative tensor product.  This propagates through to the higher~$\tcat n$.
Our use of the work `$n$-category' is in the latter, yet undefined, sense.}
etc.  So the group structure must be understood in that framework.  Although
this will not be relevant for us in this paper, we note that $\TT$~is a {\it
Lie\/} group and the higher~$\tcat n$ also have some smooth structure.

We begin with a definition.

     \proclaim{\protag{1.2} {Definition}}
 A {\it $\TT$-torsor\/}~$T$ is a manifold with a simply transitive (right)
$\TT$-action.
     \endproclaim

\flushpar
 So `\T-torsor' is a short equivalent to `principal homogeneous \T-space'.  Of
course, $\TT$~itself is a $\TT$-torsor, the {\it trivial\/} \T-torsor.  A
nontrivial example, which is of no particular relevance to us, is the
nonidentity component of the orthogonal group~$O(2)$.  An example of more
relevance: Let $L$~be any one dimensional complex inner product space.  Then
the set of elements of unit norm is a \T-torsor.  Any \T-torsor takes this
form for some hermitian line~$L$ (cf.~\thetag{3.2}).  Now if $T_1,T_2$ are
\T-torsors, then a {\it morphism\/} $h\:T_1\to T_2$ is a map which commutes
with the \T~action: $h(t\cdot \lambda ) = h(t)\cdot \lambda $ for all $t\in
T_1$,\; $\lambda \in \TT$.  The collection of all \T-torsors and morphisms
forms a category~$\tcat1$.  The group of automorphisms~$\Aut(T)$ of any $T\in
\tcat1$ is naturally isomorphic to~$\TT$: any $\mu \in \TT$ acts as the
automorphism $t\mapsto t\cdot \mu $.  Also, the set of morphisms
$\Mor(T_1,T_2)$ is naturally a $\TT$-torsor.  Finally, every morphism
in~$\tcat1$ has an inverse.\footnote{So $\tcat1$~is called a {\it
groupoid\/}, which is not to be confused with the abelian group-like
structure we introduce below.}

So far we have only described the category structure on~$\tcat1$, which is
analogous to the set structure on~$\TT$.  The important point is this:
Elements of~$\tcat1$ have automorphisms.  We do not identify isomorphic
elements which are not equal; the choice of isomorphism matters.  In fact,
any two elements of~$\tcat1$ are isomorphic, so all of the information is in
the isomorphism.  It does make sense to say that two isomorphisms are equal,
since $\Mor(T_1,T_2)$ is a set for any $T_1,T_2\in \tcat 1$.

To describe the abelian group structure we need to introduce new operations
which serve as the group multiplication and group inverse.  These are the
product of two torsors and the inverse torsor.  So if $T_1,T_2\in \tcat1$ are
\T-torsors, define the product $T_1\cdot T_2$ as
  $$ T_1\cdot T_2 = \{\langle t_1,t_2 \rangle\in T_1\times T_2\} \bigm/
     \langle t_1\cdot \lambda \,,\,t_2 \rangle\sim \langle t_1\,,\,t_2\cdot
     \lambda \rangle  $$
for all $\lambda \in \TT$.  The \T~action on~$T_1\cdot T_2$ is
  $$ \langle t_1,t_2 \rangle\cdot \lambda =\langle t_1\cdot \lambda\,,\,t_2
     \rangle = \langle t_1\,,\,t_2\cdot \lambda \rangle.  $$
The inverse~$T\inv $ of a torsor~$T$ with \T~action~$\cdot $ has the same
underlying set but a new \T~action~$*$ given by
  $$ t*\lambda =t\cdot \lambda \inv . $$
We denote the element in~$T\inv $ corresponding to~$t\in T$ as~$t\inv \in
T\inv $.  The trivial torsor~\T\ acts as the identity element under the
multiplication.  One must remember the maxim that elements in~$\tcat1$ cannot
be declared equal, only isomorphic.  So we do not have $T\cdot T\inv =\TT$,
but rather an isomorphism
  $$ \aligned
     T\cdot T\inv &\longrightarrow \TT\\
     \langle t\cdot \lambda \,,\,t  \rangle&\longmapsto \lambda .\endaligned
     $$
This isomorphism is part of the data describing~$\tcat1$.  All other axioms
for an abelian group, such as commutativity and associativity, must be
similarly modified.  For example, now the associative law is not an axiom but
a piece of the structure---a system of isomorphisms---and these isomorphisms
satisfy a higher-order axiom called the {\it pentagon diagram\/}.

We remark that there is a natural identification
  $$ T\mstrut _2 \cdot T_1\inv \cong \Hom(T_1,T_2) \tag{1.3}$$
for any~$T_1,T_2\in \tcat1$.

Starting with the group~\T\ we have outlined the construction of an abelian
group-like category~$\tcat1$.  Now we want to repeat the construction
replacing~\T\ with~$\tcat1$.  In other words, we consider
``$\tcat1$-torsors'' and then introduce a product law and inverse so as to
obtain what is now an abelian group-like {\it 2-category\/}~$\tcat2$ of the
collection of all ``$\tcat1$-torsors''.  The terminology is that a
``$\tcat1$-torsor'' is a {\it \T-gerbe\/}.

The definitions are analogous to those for \T-torsors, so we will be brief
and incomplete.  A \T-gerbe is a category~$\Cal{G}$ equipped with a simply
transitive action of~$\tcat1$.  The action is a functor $\Cal{G}\times
\tcat1\to\Cal{G}$ whose action is denoted $\langle G,T  \rangle\mapsto G\cdot
T$.  The simple transitivity means that the functor
  $$ \aligned
     \Cal{G}\times \tcat1&\longrightarrow \Cal{G}\times \Cal{G}\\
     \langle G,T  \rangle&\longmapsto \langle G,G\cdot T  \rangle\endaligned
     $$
is an equivalence, and we are given an ``inverse'' function and equivalences
of the composites to the identity.  This amounts to the specification of a
torsor~$T(G_1,G_2)$ for $G_1,G_2\in \Cal{G}$ together with natural
equivalences $G_2\cong G_1\cdot T(G_1,G_2)$ and $T\cong T(G,G\cdot T)$.  This
definition may be more rigid than the standard definition, but it fits our
examples.

Now if $\Cal{G}_1$~and $\Cal{G}_2$~are \T-gerbes, then a morphism
$\Cal{G}_1\to\Cal{G}_2$ is a functor which commutes with the $\tcat1$~action.
This means that part of the data of the morphism is a natural transformation
between the two functors obtained by traveling from northwest to southeast
around the square
  $$ \CD
     \Cal{G}_1\times \tcat1 @>>> \Cal{G}_1\\
     @VVV @VVV\\
     \Cal{G}_2\times\tcat1 @>>> \Cal{G}_2 \endCD$$
It is easy to see that the collection of morphisms $\Cal{G}_1\to\Cal{G}_2$
forms a category and that the morphisms $\Mor(\Cal{G}_1,\Cal{G}_2)$ form a
\T-gerbe.  The collection of \T-gerbes forms a 2-category~$\tcat2$.  One can
introduce an abelian group-like structure on this 2-category by defining the
product of two \T-gerbes and the inverse of a \T-gerbe, which we leave to the
reader.

I hope that at this stage it is in principle clear how I mean to define the
series of abelian group-like structures listed in~\thetag{1.1}, and that it
is clear what their basic properties are, though the detailed definition
promises to be a combinatorial mess.  We need one more notion, which is a
symmetry of such abelian group-like structures.  Suppose $A$~is a finite
group.  To say that $A$~acts on~\T\ by symmetries means that we have a
homomorphism $A\to\TT$, i.e., a character of~$A$, and then $A$~acts on~\T\ as
multiplication by this character.  If $T$~is a \T-torsor, then since
$\Aut(T)\cong \TT$, an action of~$A$ on~$T$ is again given by a character
of~$A$.  Note that the characters form the cohomology group $H^1(A;\TT)$.
Next, an action of~$A$ on~$\tcat1$ means that we have a ``homomorphism''
$A\to\tcat1$.  More precisely, for each $a\in A$ we have a \T-torsor~$T_a$
and for $a_1,a_2\in A$ an isomorphism $T_{a_1}\cdot T_{a_2}\cong T_{a_1a_2}$.
These isomorphisms must satisfy an associativity constraint.  Such a system
of torsors describes a {\it central extension\/} $\tilde{A}=\cup_{a\in A}T_a$
of~$A$:
  $$ 1 @>>> \TT @>>>\tilde{A} @>\pi >> A @>>> 1. \tag{1.4} $$
The fiber of~$\pi $ over~$a$ is~$T_a$. {\it Up to isomorphism\/} the central
extension is classified by an element of the cohomology group $H^2(A;\TT)$.
An action of~$A$ on a \T-gerbe~$\Cal{G}$ also leads to a cohomology class,
since different trivializations of~$\CG$ lead to equivalent extensions
of~$A$.  The continuation of this discussion to higher~$\tcat n$ leads to
representatives of higher group cohomology (with abelian coefficients).

\newpage
\head
\S{2} Classical Theory
\endhead
\comment
lasteqno 2@ 12
\endcomment

In this section we describe a classical (gauge) field theory in
$d+1$~dimensions with {\it finite\/} gauge group~$\Gamma $.  We generalize
the classical theory to higher codimensions, that is, to lower dimensional
manifolds.  The (exponentiated) action on fields on a $(d+1)$-manifold takes
values in~$\TT$.  For fields on a d-manifold the action takes values
in~$\tcat1$, i.e., the value of the action is a \T-torsor.  More generally,
over a $(d+1-n)$-manifold the action takes values in~$\tcat n$.  We construct
the action using the integration theory of the Appendix.  Since this is a
straightforward generalization of~\cite{FQ,\S1}, given the algebra in~\S{1}
and the integration theory in the Appendix, we defer to that reference for
more details and exposition.

Throughout this paper we use a procedure to eliminate the dependence of
quantities on extra variables or choices.  In~\cite{FQ,\S1} we call this the
{\it invariant section construction\/} after the special case mentioned in
the footnote below.  Here, following MacLane~\cite{Mac} (cf.~Quinn~\cite{Q1})
we call it an {\it inverse limit\/} of a functor.  Let $\Cal{C}$~be a
groupoid and $\Cal{F}\:\Cal{C}\to\Cal{D}$ a functor to a category (or
multicategory)~$\Cal{D}$.  We define\footnote{Think of the following example.
Let $\Cal{C}$~be the category whose objects are the points of a manifold~$M$
and whose morphisms are paths on~$M$.  Let $\Cal{D}$~be the category of
vector spaces and linear isomorphisms.  A vector bundle with connection
over~$M$ determines a functor $\Cal{F}\:\Cal{C}\to\Cal{D}$ (the morphisms act
by parallel transport), and the inverse limit is the space of {\it flat\/}
sections.} an element of the inverse limit to be a collection $\{v(C)\in
\Cal{F}(C)\}$ such that $\Cal{F}(C\to C')v(C)=v(C')$ for all morphisms $C\to
C'$.  The inverse limit is an object in~$\Cal{D}$.  In our applications
$\Cal{D}$~is $\tcat n$ for some~$n$ or is the multicategory~$\vect n$ of
higher inner product spaces which we introduce in~\S{3}.  Also, in our
applications the groupoid~$\Cal{C}$ has only a finite number of components.
For $\Cal{D}=\vect n$ the inverse limit always exists.  If $\Cal{D}=\tcat n$
we must also assume that $\Cal{F}(C\to C)$ is trivial for all automorphisms
$C\to C$, i.e., that ``$\Cal{F}$~has no holonomy''.

Fix a finite group~$\Gamma $.  For any manifold~$M$ we let $\cat M$ denote
the category of principal $\Gamma $~bundles over~$M$.  This is the collection
of {\it fields\/} in the theory.  There are symmetries as well: A morphism
$f\:P'\to P$ is a smooth map which commutes with the $\Gamma $~action and
induces the identity map on~$M$.  Notice that every morphism is invertible.
Define an equivalence relation by setting $P'\cong P$ if there exists a
morphism~$P'\to P$.  Let $\fldb M$~denote the space of equivalence classes of
fields; it is a finite set if $M$~is compact.  If $M$~is connected there is a
natural identification
  $$ \fldb M\cong \Hom\bigl( \pi _1(M,m),\Gamma \bigr)\bigm/\Gamma
     $$
for any basepoint $m\in M$.  Here $\Gamma $~acts on a homomorphism by
conjugation.

Let $B\Gamma $ be a classifying space for~$\Gamma $, which we fix together
with a universal bundle $E\Gamma \to B\Gamma $.  If $P\to M$ is a principal
$\Gamma $~bundle, then there exists a $\Gamma $~map $P\to E\Gamma $ and any
two such {\it classifying maps\/} are homotopic through $\Gamma $~maps.

Fix a singular $(d+1)$-cocycle $\alpha \in C^{d+1}(B\Gamma ;\RZ)$.  This is
the {\it lagrangian\/} of our theory.  The action is constructed as follows.
Suppose $M$~is a compact oriented manifold of dimension at most~$d+1$.
Let~$P\in \cat M$.  Then if $F\:P\to E\Gamma $ is a classifying map for~$P$,
with quotient $\overline{F}\:M\to B\Gamma $, consider the integral
  $$ \eint M{\overline{F}^*\alpha },  $$
which is defined via the integration theory of the Appendix.  We need then to
determine the dependence on~$F$ and obtain something independent of~$F$.  We
treat closed manifolds and arbitrary compact manifolds (possibly with
boundary) separately, though the second case clearly includes the first.

Suppose first that $Y$~is a closed oriented $(d+1-n)$-manifold, $n>0$, and
$Q\in \cat Y$ is a $\Gamma $~bundle over~$Y$.  Define a category~$\cat Q$
whose objects are classifying maps $f\:Q\to E\Gamma $ and whose morphisms are
homotopies $f@>h>> f'$.  Define a functor $\fun Q\:\cat Q\to\tcat n$ by
  $$ \fun Q(f)=\eint Y{\bar{f}^*\alpha } = I_{Y,\bar{f}^*\alpha },
     \tag{2.1}$$
where $\bar{f}\:Y\to B\Gamma $ is the quotient map determined by $f\:Q\to
E\Gamma $.  For a homotopy $f@>h>> f'$, let $\fun Q(f @>h>> f')$ be the
morphism
  $$ \eint{\zo\times Y}{\bar{h}^*\alpha }\:I_{Y,\bar{f}^*\alpha
     }\longrightarrow I_{Y,\bar{f}^{\prime*}\alpha }. \tag{2.2}$$
Here the homotopy $h\:\zo\times Q\to E\Gamma $ has quotient map
$\bar{h}\:\zo\times Y\to B\Gamma $.  Since $\partial (\zo\times Y)=
\{1\}\times Y\,\sqcup \,-\{0\}\times Y$, the isomorphisms \thetag{A.6},
\thetag{A.8}, and~\thetag{1.3} identify the integral~\thetag{2.2} as a map
between the spaces shown.  The gluing law~\thetag{A.10} applied to gluings of
cylinders shows that $\fun Q$~is indeed a functor.  An automorphism $f@>h>>
f$ determines a classifying map $h\:\cir\times Q\to E\Gamma $, by gluing, and
so extends to a classifying map $H\:D^2\times Q\to E\Gamma $.  Then
$\bar{h}\:\cir\times Y\to E\Gamma $ extends to $\overline{H}\:D^2\times Y\to
E\Gamma $, and by Stokes' theorem~\thetag{A.11} the morphism $\fun Q(f @>h>>
f)$ acts trivially.  So there is an inverse limit of~$\fun Q$ in~$\tcat n$,
which we denote~$T_Y^\alpha (Q)=T_Y(Q)$.  (We omit the~`$\alpha $' if it is
understood from the context.)  It should be thought of as the value of the
classical action on~$Q$.

Now suppose $X$~is a compact oriented $(d+2-n)$-manifold, possibly with
boundary, and $P\in \cat X$ is a $\Gamma $~bundle over~$X$.  Let $\cat P$~be
the category of classifying maps $F\:P\to E\Gamma $ and homotopies $F @>H>>
F'$.  Restriction to the boundary defines a functor $\cat P @>\partial >>
\cat{\partial P}$.  If $F\in \cat P$ then by integration we obtain
  $$ \eint X{\overline{F}^*\alpha }\in I_{\bX,\partial \overline{F}^*\alpha }
     = \fun{\partial P}(\partial F). \tag{2.3}$$
Furthermore, one can check that if $F @>H>> F'$ is a homotopy, then
\thetag{A.11}~implies that
  $$ \fun{\partial P}(\partial F @>\partial H>> \partial F')\, \eint
     X{\overline{F}^* \alpha } = \eint X{\overline{F}^{\prime*}\alpha }.
     $$
These equations imply that \thetag{2.3}~determines an element
  $$ e^{\tpi S_X(P)}\in \tors{\bX}{\bP}. \tag{2.4}$$

We state the properties of this action without proof.

     \proclaim{\protag{2.5} {Assertion}}
 Let $\Gamma $~be a finite group and $\alpha\in C^{d+1}(B\Gamma ;\RZ)$ a
cocycle.  Then the assignments\footnote{It is possibly better notation to
write $\eac XP\in \eac{\bX}{\bP}$ for {\it any\/} compact oriented~$X$, or
perhaps instead $\tors XP\in \tors{\bX}{\bP}$.  We will sometimes use the
latter notation, especially in~\S{9}.}
  $$ \alignedat 2
      Q &\longmapsto \tors YQ\in \tcat n ,&&\qquad Q \in \cat Y,\\
      P &\longmapsto \eac XP\in \tors{\bX}{\bP},&&\qquad P \in \cat
     X\endalignedat \tag{2.6} $$
defined above for closed oriented $(d+1-n)$-manifolds~$Y$ and compact
oriented $(d+2-n)$-manifolds~$X$ satisfy:\newline
 \rom(a\rom)\ \rom({\it Functoriality\/}\rom)\ If $\psi \:Q'\to Q$ is a bundle
map covering an orientation preserving diffeomorphism $\overline{\psi}\:Y'\to
Y$, then there is an induced isomorphism
  $$ \psi _*\:\tors Y{Q'} \longrightarrow \tors Y{Q} \tag{2.7} $$
and these compose properly.  If $\varphi \:P'\to P$ is a bundle map covering
an orientation preserving diffeomorphism $\phibar\:X'\to X$, then there is an
induced isomorphism\footnote{If $n=1$ then \thetag{2.8}~is an equality of
elements in a \T-torsor.  Similarly for~\thetag{2.10}, \thetag{2.11},
and~\thetag{2.12}.}
  $$ (\partial \varphi )_*\(\eac {X'}{P'}\) \longrightarrow \eac {X}{P},
      \tag{2.8} $$
where $\partial \varphi \:\partial P'\to\partial P$ is the induced map over
the boundary.\newline
 \rom(b\rom)\ \rom({\it Orientation\/}\rom)\ There are natural isomorphisms
  $$ \tors{-Y}Q \cong \bigl(\tors YQ\bigr)\inv ,  \tag{2.9} $$
and
  $$ \eac{-X}P \cong  \bigl(\eac XP\bigr)\inv .  \tag{2.10} $$
 \rom(c\rom)\ \rom({\it Additivity\/}\rom)\ If $Y=Y_1\sqcup Y_2$ is a
disjoint union, and $Q _i$~are bundles over~$Y_i$, then there is a natural
isomorphism
  $$ \tors Y{Q _1\sqcup Q _2} \cong \tors Y{Q _1}\cdot \tors Y{Q _2}.
       $$
If $X=X_1\sqcup X_2$ is a disjoint union, and $P _i$~are bundles
over~$X_i$, then there is a natural isomorphism
  $$ \eac{X_1\sqcup X_2}{P _1\sqcup P _2} \cong  \eac{X_1}{P _1}
     \cdot  \eac{X_2}{P _2}.  \tag{2.11} $$
 \rom(d\rom)\ \rom({\it Gluing\/}\rom)\ Suppose $Y\hookrightarrow X$ is a
closed oriented codimension one submanifold and $X\cut$~is the manifold
obtained by cutting~$X$ along~$Y$.  Then $\partial X\cut = \partial X\sqcup Y
\sqcup -Y$.  Suppose $P $~is a bundle over~$X$, $P \cut$~the induced bundle
over~$X\cut$, and $Q $~the restriction of~$P $ to~$Y$.  Then there is a
natural isomorphism
  $$ \Tr_Q \( \eac{X\cut}{P \cut}\)\longrightarrow \eac XP , \tag{2.12} $$
where $\Tr_Q $ is the contraction
  $$ \Tr_Q \:\tors{X\cut}{\partial P \cut} \cong \tors X{\partial P }\cdot
     \tors YQ \cdot \tors YQ\inv \longrightarrow \tors X{\partial P }.
      $$
     \endproclaim

The Functoriality Axiom~(a) means in particular that for any $Q\in \cat Y$
there is an action of the finite group~$\Aut Q$ on~$\tors YQ$.  As explained
in~\S{1} the isomorphism class of this action is an element of $H^n(\Aut
Q;\TT)$.  For $n=2$ this action determines a central extension of~$\Aut Q$
by~$\TT$.  We use an additional property of gluing in~\S{9}: Iterated gluings
commute.  As always, we must interpret `commute' appropriately in categories.

\newpage
\head
\S{3} Higher Algebra II
\endhead
\comment
lasteqno 3@  8
\endcomment

The quantum integration process is this: We integrate the classical action
over the space of equivalence classes of fields on some manifold.  As
explained in~\S{2} the classical action in codimension~$n$ takes values
in~$\tcat n$ (or in a $\tcat n$-torsor for manifolds with boundary).  For
example, in the top dimension it takes values in~$\TT$.  But we cannot add
elements of~$\TT$.  Rather, to form the quantum path integral we embed
$\TT\hookrightarrow \CC$ and add up the values of the classical action as
complex numbers.  In higher codimensions we introduce ``higher inner product
spaces'' where we can perform the sum.\footnote{Since our basic group
is~$\TT$ (as opposed to~$\CC^\times $) we obtain complex inner product spaces
(as opposed to simply complex vector spaces).  Presumably one can generalize
to other base fields or rings.} The collection~$\vect n$ of all {\it complex
$n$-inner product spaces\/}\footnote{It is probably better to consider the
category of {\it virtual\/} complex $n$-inner product spaces, that is, formal
differences of complex $n$-inner product spaces.  This provides additive
inverses and is more closely analogous to a ring.  However, we will only
encounter ``positive'' elements of this ``ring'' so do not insist on the
inclusion of virtual inner product spaces.} is an $n$-category, which is in
some sense the trivial complex $(n+1)$-inner product space, and there is an
embedding $\tcat n\hookrightarrow \vect n$ onto the set of elements of ``unit
norm''.  We view the action as taking values in~$\vect n$ and then take sums
there to perform the path integral.  Our goal in this section, then, is to
describe this hierarchy:
  $$ \alignedat 2
      &\vect 0=\CC&\qquad&\text{field of complex numbers}\\
      &\vect 1&\qquad&\text{``ring'' of (virtual) finite dimensional complex
     inner product spaces}\\
      &\vect 2&\qquad&\text{``ring'' of (virtual) finite dimensional complex
     2-inner product spaces}\\
      &\text{etc.}&\qquad& \endalignedat \tag{3.1} $$
The inner product space notions of dual space (or conjugate space), direct
sum, and tensor product generalize to~$\vect n$, and this gives it a
structure analogous to a commutative ring with involution.

The notion of a 2-vector space appears in work of Kapranov and
Voevodsky~\cite{KV}, and also in lectures of Kazhdan and in recent
work of Lawrence~\cite{L}.  We in no way claim to have worked out the
category theory in detail, and we feel that this sort of ``higher linear
algebra'' merits further development.

The terminology is confusing:  An $n$-inner product space is an
$(n-1)$-category.  Thus a 2-inner product space is an ordinary category.

Recall that an inner product space~$V$ is a set with an commutative vector
sum $V\times V\to V$, a scalar multiplication $\CC\times V\to V$, and an
inner product $\form\:V\times \Vb\to\CC$.  (The conjugate inner product
space~$\Vb$ is defined below.)  We will not review all of the axioms here.
There are two trivial examples: the zero inner product space~$O$ consisting
of one element, and $\CC$~with its usual inner product $(z,w)=z\cdot
\bar{w}$.  If $V_1,V_2$~are inner product spaces, then a {\it morphism\/} is
a linear map $V_1\to V_2$ which preserves the inner product.  The collection
of inner product spaces and linear maps forms a category~$\vect 1$.

Suppose $T\in \tcat1$ is a \T-torsor.  From~$T$ we form the one dimensional
complex inner product space (hermitian line)
  $$ \aligned
     L_T &= T\times _\TT \CC\\
     &= \{\langle t,z  \rangle\in T\times \CC\}\bigm / \langle t\cdot \lambda
     ,z  \rangle \sim \langle t,\lambda \cdot z  \rangle\endaligned
     \tag{3.2}$$
for all $\lambda \in \TT$.  Note that $L_{\TT}\cong \CC$.  The inner product
on~$L_T$ is
  $$ \bigl( \langle t,z  \rangle\,,\,\langle t,w  \rangle\bigr) = z\cdot
     \bar{w}. $$

If $V\in \vect1$ is an inner product space, we form the {\it dual space\/}
$V^*=\Hom(V,\CC)$ with its usual inner product.  The {\it conjugate\/} inner
product space~$\Vb$ has the same underlying abelian group as~$V$ but the
conjugate scalar multiplication and the transposed inner product.  There is a
natural isometry $\Vb\cong V^*$ given by the inner product.  If $V_1,V_2\in
\vect 1$ then one can form the {\it direct sum\/}~$V_1\oplus V_2$ and the
{\it tensor product\/}~$V_1\otimes V_2$ with the inner products
  $$ \aligned
     (v_1\oplus v_2\,,\,w_1\oplus w_2) &= (v_1,w_1) + (v_2,w_2)\\
     (v_1\otimes v_2\,,\,w_1\otimes w_2) &= (v_1,w_1)\;(v_2,w_2).\endaligned
     $$
Notice that there are natural isomorphisms $O\oplus V\cong V$ and $\CC\otimes
V\cong V$.  Also, if $T_1,T_2\in \tcat 1$ then $L_{T\inv }\cong L_T^*$ and
$L_{T_1\cdot T_2}\cong L_{T_1}\otimes L_{T_2}$.  The direct sum and tensor
product give~$\vect 1$ a {\it commutative ring-like\footnote{As we mentioned
above, we should include {\it virtual\/} inner product spaces to have
additive inverses.} structure with involution\/}, the involution being the
conjugation or duality.

It is useful to observe that for any inner product space~$V$, the induced
inner product on~$V^*\otimes V$ is
  $$ (T\mstrut _1,T\mstrut _2) = \Tr(T\mstrut _1T_2^*),\qquad T_i\in \Hom(V),
      $$
where we identify $V^*\otimes V\cong \Hom(V)$ via the canonical isomorphism,
and $T^*$~is the hermitian adjoint of~$T$.

Finally, we introduce an ``inner product''
  $$ \form\:\vect1\times \overline{\vect1}\longrightarrow \vect1 $$
by
  $$ (V_1,V_2) = V_1\otimes \overline{V_2}, $$
and the associated ``norm'' $|V|^2 = V\otimes \Vb$.  Notice that the elements
of ``unit norm'', that is of norm~$\CC$, are precisely the hermitian lines,
i.e., the image of the embedding $\tcat1\hookrightarrow \vect1$.  The image
is closed under tensor product and the embedding is a homomorphism.

Starting with the field~$\CC$ we have outlined the construction of a
commutative ring-like category~$\vect1$ (with involution) consisting of inner
product spaces over~$\CC$.  Now we iterate and consider inner product spaces
over~$\vect1$, which we call {\it complex 2-inner product
spaces\/}.\footnote{Since $\vect1$~is analogous to a ring, not a field, we
expect that not all of its modules are free.  The ones we consider in this
paper are sums of one dimensional cyclic modules, so are free.  A formal
development of this concept should probably demand freeness in the
definition~\cite{KV}.} So a complex 2-inner product space~$\Cal{W}$ is a
category with an abelian group law $\Cal{W}\times \Cal{W}\to \Cal{W}$, a
``scalar multiplication'' $\vect1\times \Cal{W}\to\Cal{W}$, and an ``inner
product'' $\Cal{W}\times \overline{\Cal{W}}\to\vect1$. There is a zero
complex 2-inner product space~$O$.  The dual, conjugate, direct sum, and
tensor product are defined.  The category~$\vect 1$ is a 2-inner product
space which is an identity element for the tensor product.  The collection of
all (virtual) complex 2-inner product spaces forms a commutative ring-like
2-category~$\vect2$ with involution.

Because a 2-inner product space is a category, and not a set, there is an
extra layer of structure (natural transformations) and so additional data as
part of the definition.  We do not claim to have a complete list, but mention
some additional structure related to the inner product.  Namely, for
all~$W_1,W_2\in \Cal{W}$ there is a specified map
  $$ (W_2,W_1)\cdot W_1\longrightarrow W_2. $$
The `$\cdot $' here is the scalar product.  We might further assume that
$\Mor(W_1,W_2)$~is isomorphic to the vector space~$(W_2,W_1)$; this holds in
the examples.  In addition, we postulate a preferred isometry
  $$ (W_1,W_2)\longrightarrow \overline{(W_2,W_1)} $$
whose ``square'' is the identity.  In particular, $(W,W)$~has a real
structure for all~$W\in \CW$, and we assume the existence of compatible maps
  $$ \CC\longrightarrow (W,W) \longrightarrow \CC. \tag{3.3} $$
The composition is then multiplication by a real number, which we call $\dim
W$.

A {\it linear map\/} of complex 2-inner product spaces $L\:\Cal{W}_1\to
\Cal{W}_2$ is a functor which preserves the addition and scalar
multiplication.  The space of all such linear maps is the 2-inner product
space $\Hom(\CW_1,\CW_2)\cong \CW_2\otimes \CW_1^*$.  If we assume some
freeness condition on 2-inner product spaces (see previous footnote), then we
can clearly generalize other standard notions of linear algebra.  For
example, we should be able to define linear independence and bases.  Then if
$P\:\Cal{W}\to\Cal{W}$ is a linear operator on~$\Cal{W}$, a matrix
representation relative to a basis of~$\CW$ is a matrix of inner product
spaces~$P^i_j\in \vect1$.  The {\it trace\/} $\Tr(P)=\bigoplus\nolimits
_iP^i_i$ is then an inner product space.  The {\it dimension\/} of~$\CW$ is
the trace of the identity map, which is $\dim\CW=\CC^n$ for some~$n$.  It
makes sense, then, to identify the dimension of~$\CW$ as~$n$.

If $\Cal{G}$~is a \T-gerbe, then we form the one dimensional complex 2-inner
product space
  $$ \aligned
     \CW_\Cal{G} &= \Cal{G}\times _{\tcat1}\vect 1\\
     &= \{\langle G,V  \rangle\in \Cal{G}\times \vect1\}\bigm/ \langle G\cdot
     T,V  \rangle \sim \langle G, L_T\otimes V  \rangle\endaligned
     \tag{3.4}$$
for all \T-torsors~$T$. Note that $\CW_{\tcat1}\cong \vect1$.  If we define
the inner product
  $$ (\Cal{W}_1,\Cal{W}_2) = \Cal{W}_1\otimes \overline{\Cal{W}_2}
     $$
on~$\vect2$, then we see that the image of the
embedding~$\tcat2\hookrightarrow \vect2$ determined by~\thetag{3.4} consists
of complex 2-inner product spaces of ``unit norm''.  The image is closed
under tensor product and the embedding is a homomorphism.

Here is a more concrete example of a nontrivial 2-inner product space which
is important in what follows.  Suppose $A$~is a finite group.  Let
$\ivr{\vect1}A$~denote the category of finite dimensional unitary
representations of~$A$.  The morphisms are required to commute with the
$A$~action.  Then $\ivr{\vect1}A$~is a 2-inner product space as follows.  If
$W\in \ivr{\vect1}A$ and~$V\in \vect1$ then we can ``scalar multiply''~$V$
by~$W$ using the ordinary vector space tensor product.  We obtain~$V\otimes
W$, which is a representation of~$A$.  The vector sum in~$\ivr{\vect1}A$ is
the usual direct sum of representations.  The inner product
on~$\ivr{\vect1}A$ is
  $$ (W_1,W_2) = (W_1\otimes \overline{W_2})^A, \tag{3.5}$$
where for any representation~$W\in \ivr{\vect1}A$ the inner product
space~$W^A\in \vect1$ is the subspace of {\it invariants\/}.  Note that if
$W$~is an {\it irreducible\/} unitary representation of~$A$, then
(cf.~\cite{FQ,Appendix~A})
  $$ (W,W)=\dim W\cdot \CC,  $$
since $\dim W$~is the norm square of the canonical element of~$W\otimes
\overline{W}$.  The composition~\thetag{3.3} is~$\dim W$ in the usual sense.
The dimension of~$\ivr{\vect1}A$ is the number of isomorphism classes of
irreducible representations of~$A$.

More generally, suppose that $\Cal{G}$~is a \T-gerbe with a nontrivial
$A$~action, which we denote by~$\rho $.  For any $G\in \Cal{G}$ let
  $$ L_G=\langle G,\CC  \rangle\in \CW_\CG. \tag{3.6}$$
(Recall the definition of~$\CW_\CG$ in~\thetag{3.4}.)  Note that for any line
$L\in \vect1$, the element $\langle G,L \rangle\in \CW_\CG$ is equivalent
to~$L_{G'}$ for some~$G'\in \CG$, and so any element of~$\CW_\CG$ is
isomorphic to a finite sum $L_{G_1}\oplus \dots \oplus L_{G_k}$.  Let $A$~act
on~$\WG$ by
  $$ a\cdot L_G = L_{a\cdot G}. \tag{3.7}$$
Finally, set\footnote{To make good sense of ``invariant'' we must identify
certain {\it canonically\/} isomorphic elements.  For example, we need to
identify different permutations of the sum $L_{G_1}\oplus \dots \oplus
L_{G_k}$.  Also, this definition is suspicious---the dimension of the
invariants is larger than the dimension of~$\CW_{\CG}$!}
  $$ \ivr{\WG}{A,\rho }=\spann\{W=L_{G_1}\oplus \dots \oplus L_{G_k}:
     \text{$W$ is invariant under the $A$ action} \}. \tag{3.8} $$
This is our sought-after 2-inner product space.  If $\CG=\tcat1$ is the
trivial \T-gerbe, then according to~\thetag{1.4} the action~$\rho $
determines a central extension~$\Atil$ of~$A$ by~$\TT$, and to each $a\in A$
corresponds a \T-torsor~$T_a$ which is the fiber of~$\Atil$ over~$a$.  We can
describe $\ivr{\CW_{\tcat1}}{A,\rho }=\ivr{\vect1}{A,\rho }$ as the category
of representations of~$\Atil$ such that the central~$\TT$ acts by standard
scalar multiplication.  For then an element of this category is of the form
$W=L_{T_1}\oplus \dots \oplus L_{T_k}$, where for each~$a\in A$ and each
index~$i$ there is an index~$j$ with $T_a\cdot T_i=T_j$.  Then
\thetag{3.7}~is an isometry $L_{T_a}\otimes L_{T_i}\to L_{T_j}$, and so
each~$\tilde{a}\in T_a$ induces an isometry $L_{T_i}\to L_{T_j}$.  This
describes the $\Atil$~action.  The dimension of~$\ivr{\CW_{\tcat1}}{A,\rho }$
as the number of isomorphism classes of such irreducible representations.
Since any \T~gerbe~$\CG$ is (noncanonically) isomorphic to~$\tcat1$, this is
also the dimension of~$\ivr{\WG}{A,\rho }$.  If $\rho $~is the trivial
$A$~action on~$\tcat1$, then $\ivr{\CW_{\tcat1}}{A,\rho }=\ivr{\vect1}{A,\rho
}$ is the 2-inner product space~$\ivr{\vect1}A$ we defined in the previous
paragraph.

Think of $\ivr{\CW_{\tcat1}}{A,\rho }$ as the space of $A$-invariants
in~$\WG$.  We can also consider invariants in the analogous situation ``one
dimension down''.  That is, if $A$~acts on a \T-torsor~$T$ through a
character $\mu \:A\to\TT$, then $A$~also acts on the hermitian line~$L_T$
through the same character.  We define
  $$ \ivr{L_T}{A,\mu } = \{\ell \in L_T : \text{$\ell$ is invariant under the
     $A$ action} \}. $$
But this is simple:
  $$ \ivr{L_T}{A,\mu } = \cases L_T ,&\text{if $\mu$ is trivial}
     ;\\0,&\text{otherwise} .\endcases $$

We remark that whereas $\ivr{\vect1}A$ has a natural {\it monoidal\/}
structure\footnote{A {\it monoidal category\/} is a category equipped with a
tensor product and an identity element.  In addition, an ``associator'' and
natural transformations related to the identity element must be specified
explicitly.  A monoidal category is the category-theoretic analogue of a
monoid, which is a set with an associative composition law and an identity
element.} given by the tensor product of representations, the category
$\ivr{\vect1}{A,\rho }$ for $\rho $~nontrivial do not: the tensor product of
representations of~$\Atil$ where $\TT$~acts as scalar multiplication is a
representation of~$\Atil$ where $\TT$~acts as the {\it square\/} of scalar
multiplication.  Also, if $\CG$~is a nontrivial gerbe, then $\ivr{\WG}{A,\rho
}$~is not monoidal in a natural way.

Finally, by forgetting the $A$~action we obtain an ``augmentation'' linear
map
  $$ \ivr{\WG}{A,\rho }\longrightarrow \WG. $$
If $\CG=\tcat1$ is trivial, it takes values in~$\CW_{\tcat1}=\vect1$.

Clearly these constructions have analogs in the higher complex inner
product spaces~\thetag{3.1}.

\newpage
\head
\S{4} Quantum Theory
\endhead
\comment
lasteqno 4@ 22
\endcomment

Now we are ready to quantize the classical $d+1$~dimensional classical field
theory described in~\S{2}.  We carry out the quantization on any compact
oriented manifold of dimension less than or equal to~$d+1$ by integrating the
classical action over the space of fields.  (We first use the constructions
in~\S{3} to convert the values of the classical action from an $n$-torsor to
an $n$-inner product space.)  Since there are symmetries of the fields, we
only integrate over {\it equivalence classes\/} of fields.  The residual
symmetry, that is, the automorphism groups of the fields, must also be taken
into account.  Since the gauge group is finite, the space of equivalence
classes of fields on a compact manifold is a {\it finite set\/}, so all we
need to perform the path integral is a measure on this finite set.  We also
need to define the product of a positive number~$\mu $ (the measure) by an
element~$\CW\in \vect n$.  This we denote as~$\mu \cdot \CW$ and interpret it
as~$\CW$ with the inner product multiplied by~$\mu $.  The rest is a
straightforward generalization of~\cite{FQ,\S2}, given the higher algebra
of~\S{3} and the classical theory of~\S{2}.  For a closed oriented
$(d+1-n)$-manifold~$Y$, $n>0$, the resulting quantum invariant is a complex
$n$-inner product space $E(Y)\in \vect n$.  If $Y=\emptyset $ is the empty
manifold, then $E(\emptyset )=\vect{n-1}$ is the trivial space.  The quantum
invariant of a compact oriented $(d+2-n)$-manifold~$X$, possibly with
boundary, is an element $Z_X\in E(\bX)$.  For~$n=1$ we recover the quantum
invariants of~\cite{FQ,\S2}---the ordinary path integral (partition function)
and the quantum Hilbert space.  For~$n=2$ the quantum invariant of a closed
oriented $(d-1)$-manifold~$S$ is a 2-inner product space~$E(S)$, and the
quantum invariant of a compact oriented $d$-manifold~$Y$ is an object~$Z_Y$
in the category~$E(\bY)$.  {\it Et cetera\/}.

We first introduce a measure~$\mu $ on the category of principal $\Gamma
$~bundles~$\fld M$ over any manifold~$M$.  For $P\in \fld M$ set
  $$ \mu _P = \frac{1}{\#\Aut P}. \tag{4.1}$$
Clearly $\mu_{P'}=\mu _P$ for equivalent bundles~$P'\cong P$, so $\mu
$~determines a measure on the set of equivalence classes~$\fldb M$.  This is
the assertion that the measure is invariant under the symmetries of the
fields.

If $M$~has a boundary, for each $Q\in \fld{\bM}$ set
  $$ \fld M(Q) = \{\langle P,\theta   \rangle: P\in \fld
     M,\;\text{$\theta\:\bP\to Q$ is an isomorphism} \}. \tag{4.2}$$
A morphism $\varphi \:\langle P',\theta ' \rangle\to\langle P,\theta \rangle$
is an isomorphism $\varphi \:P'\to P$ such that $\theta '=\theta \circ
\partial \varphi $.  The morphisms define an equivalence relation on~$\fld
M(Q)$, and we denote the set of equivalence classes by~$\fldb M(Q)$.
Equation~\thetag{4.1} determines a measure on~$\fldb M(Q)$.  Note that any
automorphism of $\langle P,\theta \rangle\in \fld M(Q) $ is the identity on
components of~$M$ with nontrivial boundary.  If $\psi \:Q'\to Q$ is an
isomorphism of $\Gamma $~bundles over~$\bM$, then $\psi $~induces a
measure-preserving map
  $$ \psi _*\:\fldb M(Q')\longrightarrow \fldb M(Q) $$
by $\psi _*(P,\theta ) = \langle P,\psi \theta   \rangle$.  In particular,
for $Q'=Q$ this gives a measure-preserving action of~$\Aut Q$ on~$\fldb
M(Q)$.

One important property of~$\mu $, which is an ingredient in the proof of the
gluing law~\thetag{4.17}, is its behavior under cutting and pasting.  Suppose
$N\hookrightarrow M$ is an oriented codimension one submanifold and $M\cut$
the manifold obtained by cutting~$M$ along~$N$.  For each $Q\in \fld N$,
\,$Q'\in \fld{\bM}$, we obtain a gluing map
  $$ \aligned
     g_Q\:\fldb{M\cut}(Q\sqcup Q\sqcup Q')&\longrightarrow \fldb M(Q')\\
     \langle P\cut;\theta _1,\theta _2,\theta   \rangle&\longmapsto \langle
     P\cut/(\theta _1=\theta _2) \,;\, \theta   \rangle.\endaligned
     $$
We refer to~\cite{FQ,\S2} for the proof of the following.

     \proclaim{\protag{4.3} {Lemma}}
 The gluing map~$g_Q$ satisfies:\newline
 \rom(a\rom)\ $g_Q$~maps onto the set of equivalence classes of bundles
over~$M$ whose restriction to~$N$ is isomorphic to~$Q$.\newline
 \rom(b\rom)\ Let $\phi \in \Aut Q$ act on $\langle P\cut;\theta _1,\theta _2,
\theta \rangle\in \fld{M\cut}(Q\sqcup Q)$ by
  $$ \phi \cdot \langle P\cut;\theta _1,\theta _2,\theta \rangle = \langle
     P\cut;\phi \circ \theta _1, \phi \circ \theta _2,\theta \rangle . $$
Then the stabilizer of this action at~$\langle P\cut;\theta _1,\theta
_2,\theta \rangle $ is the image $\Aut P\to\Aut Q$ determined by the~$\theta
_i$, where $P = g_Q(\langle P\cut;\theta _1,\theta _2,\theta \rangle )$.
\newline
 \rom(c\rom)\ There is an induced action on equivalence classes
$\fldb{M\cut}(Q\sqcup Q)$, and $\Aut Q$~acts transitively on~$g\inv _Q([P])$
for any~$[P]\in \fldb{M}$.  \newline
 \rom(d\rom)\ For any $[P]\in \fldb M(Q)$ we have
  $$ \mu _{[P]} = \voll\bigl( g\inv _Q([P]) \bigr)\cdot \mu
     ^{\vphantom{-1}}_{Q}.  \tag{4.4}$$
     \endproclaim

Now we are ready to carry out the quantization.  We treat all codimensions
simultaneously, but suggest that the reader first review the top dimensional
quantizations in~\cite{FQ,\S2}.  Again for clarity we first treat closed
manifolds and then arbitrary compact manifolds (possibly with boundary),
though the second case includes the first.

Suppose first that $Y$~is a closed oriented $(d+1-n)$-manifold, $n>0$.  The
classical action defined in~\S{2} is a map
  $$ T_Y\:\cat Y\longrightarrow \tcat n, $$
which we can think of as a bundle of ``$n$-torsors'' over~$\cat Y$.  By
\theprotag{2.5(a)} {Assertion} for each $Q\in \cat Y$ there is an
action~$\rho _Q$ of~$\Aut Q$ on $T_Y(Q)$.  Use the construction~\thetag{3.2},
{}~\thetag{3.4} to replace each~$T_Q$ by the one dimensional $n$-inner product
space
  $$ \CW_Q = \CW_{T_Y(Q)}. \tag{4.5} $$
\theprotag{2.5(a)} {Assertion} also implies that an isomorphism $\psi \:Q'\to
Q$ induces an isomorphism $\psi_*\: \CW_{Q'}\to \CW_{Q}$.  However, an
automorphism $\psi \in \Aut Q$ does not necessarily act trivially on~$\CW_Q$.
Rather, it only acts trivially on the subspace of invariants under the $\Aut
Q$~action (cf.~\thetag{3.8}).  More precisely, we construct a ``quotient''
complex $n$-inner product space~$\CW_{[Q]}$ associated to the {\it
equivalence class\/}~$[Q]\in \fldb Y$ as an inverse limit.  (The inverse
limit picks out the invariants under automorphisms.)  Consider the
category~$\Cal{C}_{[Q]}$ of bundles~$Q$ in the isomorphism class~$[Q]$, and
let $\Cal{F}_{[Q]}\:\Cal{C}_{[Q]}\to\vect n$ be the functor whose value
at~$Q$ is ~$\CW_Q$.  Set $\CW_{[Q]}$ to be the inverse limit
of~$\Cal{F}_{[Q]}$.  As $[Q]$~varies we then obtain a map
  $$ \CW_Y\:\fldb Y\longrightarrow \vect n. $$
The quantum space~$E(Y)$ is the integral of~$\CW_Y$ over~$\fldb Y$, which in
this case is a finite sum:
  $$ E(Y) = \int_{\fldb Y}d\mu ([Q])\,\CW_Y([Q]) = \bigoplus_{[Q]\in \fldb
     Y}\mu _{[Q]}\cdot \CW_{[Q]}\in \vect n. \tag{4.6} $$
If we think of~$\CW_Y$ as a bundle of $n$-inner product spaces over~$\fldb
Y$, then $E(Y)$~is the space of $L^2$~sections of that bundle.

Now suppose that $X$~is a compact oriented $(d+2-n)$-manifold, possibly with
boundary.  The classical action on the boundary~$\bX$ is a bundle of
$n$-torsors $T_{\bX}\to\cat{\bX}$, and the classical action~$e^{\tpi S_X}$
on~$X$ is a section of the pullback~$r^*T_{\bX}$, where $r$~is restriction to
the boundary:
  $$ \CD
     r^*T_{\bX} @>>> T_{\bX}\\
     @VVV @VVV\\
     \cat X @>r>> \cat{\bX}\endCD $$
By \theprotag{2.5(a)} {Assertion}
the action is invariant under the morphisms in~$\cat X$, that is, under
symmetries of the fields.  Now for each~$P\in \cat X$ we use the
construction~\thetag{3.6} to define an element
  $$ \lin XP = L_{e^{\tpi S_X(P)}}\in \CW_{\bP}=\CW_{\tors{\bX}{\bP}}.
     \tag{4.7} $$
Now $\lin XP$~is not necessarily invariant under~$\Aut P$; it transforms
under~$\psi \in \Aut P$ according to the action of the restricted
automorphism ~$\partial \psi \in \Aut(\bP)$ on~$\CW_{\tors{\bX}{\bP}}$.  We
only obtain invariance after integrating.  Thus fix~$Q\in \cat{\bX}$ and
consider~$\cat X(Q)$ as defined in~\thetag{4.2}.  If $\langle P,\theta
\rangle\in \cat X(Q)$ then using~$\theta $ to identify $\tors{\bX}{\bP}\cong
\tors{\bX}Q$ we have the action $\eac X{P,\theta }\in \tors{\bX}Q$ and the
associated $L_X(P,\theta )\in \CW_Q$, as in~\thetag{4.7}.  If $\langle
P,\theta \rangle\cong \langle P',\theta ' \rangle$ then there is an
isomorphism between the values of the actions on these fields as elements
of~$\tors{\bX}Q$.  By another inverse limit construction we define $\lin
X{[P,\theta ]}\in \CW_Q$.  Set
  $$ Z_X(Q) = \int_{\fldb X(Q)}d\mu ([P,\theta ])\,\lin X{[P,\theta ]} =
     \bigoplus _{[P,\theta ]\in \fldb X(Q)}\mu _{[P,\theta ]}\cdot \lin
     X{[P]}\in \CW_{Q}. \tag{4.8} $$
Now we claim that $Z_X(Q)$ is invariant under the $\Aut Q$~action
on~$\CW_{Q}$, and so
  $$ Z_X(Q)\in \ivr{\CW_{\tors{\bX}Q}}{\Aut Q,\rho _Q} \tag{4.9} $$
More generally, we check that for an isomorphism $\psi \:Q'\to Q$ we have
  $$ \spreadlines{6pt}\split
      \psi _*Z_X(Q') &= \bigoplus_{[\langle P',\theta' \rangle]} \mu
     _{[P']}\cdot \psi _*\lin X{[P',\theta ']}\\
      &\cong \bigoplus_{[\langle P',\theta '\rangle]} \mu _{[P']}\cdot \lin
     X{[P',\psi \theta ']}\\
      &= Z_X(Q),\endsplit \tag{4.10} $$
since $\langle P',\psi \theta '\rangle$~runs over a set of equivalence
classes in~$\fld X(Q)$ as $\langle P',\theta '\rangle$~runs over a set of
equivalence classes in~$\fld X(Q')$.  Using the definition~\thetag{3.8}
of~$\ivr{\CW_{\tors{\bX}Q}}{\Aut Q,\rho _Q}$ we deduce~\thetag{4.9}, and
furthermore \thetag{4.10}~shows that $\{Z_X(Q):Q\in [Q]\}$ is a collection of
elements in $\{\CW_Q:Q\in [Q]\}$ invariant under symmetries.  In other words,
it is an element of the inverse limit~$\CW_{[Q]}$:
  $$ Z_X([Q])\in \CW_{[Q]}.  $$
Finally, then,
  $$ Z_X = \bigoplus_{[Q]\in \fldb{\bX}} Z_X([Q])\;\in \;\bigoplus_{[Q]\in
     \fldb {\bX}} \mu _{[Q]}\cdot \CW_{[Q]} = E(\bX) \tag{4.11} $$
is the desired quantum invariant.

The basic properties of these quantum invariants, which we might term
``higher quantum Hilbert spaces'' and ``higher path integrals'', are listed
in the following.

     \proclaim{\protag{4.12} {Assertion}}
  Let $\Gamma $~be a finite group and $\alpha\in C^{d+1}(B\Gamma ;\RZ)$ a
cocycle.  Then the assignments\footnote{Again the notation is awkward, and
possibly it is best to use~$Z_X$ for all~$X$ and write $Z_X\in Z_{\bX}$.}
  $$ \aligned
       Y &\longmapsto E(Y) \in \vect n,\\
       X &\longmapsto Z_X\in E(\bX),\endaligned  $$
defined above for closed oriented $(d+1-n)$-manifolds~$Y$ and compact
oriented $(d+2-n)$-manifolds~$X$ satisfy:\newline
 \rom(a\rom)\ \rom({\it Functoriality\/}\rom)\ Suppose $f\:Y'\to Y$ is an
orientation preserving diffeomorphism.  Then there is an induced isometry
  $$ f_*\:E(Y') \longrightarrow E(Y) \tag{4.13} $$
and these compose properly.  If $F\:X'\to X$ is an orientation preserving
diffeomorphism, then there is an induced isometry\footnote{If $n=1$ this is
an equality, as are~\thetag{4.15}, \thetag{4.16}, and~\thetag{4.17}.}
  $$ (\partial F)_*(Z_{X'}) \longrightarrow Z_X, \tag{4.14} $$
where $\partial F\:\partial X'\to\partial X$ is the induced map over the
boundary.\newline
 \rom(b\rom)\ \rom({\it Orientation\/}\rom)\ There are natural isometries
  $$ E(-Y) \cong \overline{E(Y)},  $$
and
  $$ Z_{-X} \cong  \overline{Z_X}. \tag{4.15} $$
 \rom(c\rom)\ \rom({\it Multiplicativity\/}\rom)\ If $Y=Y_1\sqcup Y_2$ is a
disjoint union, then there is a natural isometry
  $$ E(Y _1\sqcup Y _2) \cong E(Y _1)\otimes E(Y _2). $$
If $X=X_1\sqcup X_2$ is a disjoint union, then there is a natural isometry
  $$ Z_{X_1\sqcup X_2} \cong Z_{X_1} \otimes Z_{X_2}. \tag{4.16} $$
 \rom(d\rom)\ \rom({\it Gluing\/}\rom)\ Suppose $Y\hookrightarrow X$ is a
closed oriented codimension one submanifold and $X\cut$~is the manifold
obtained by cutting~$X$ along~$Y$.  Write $\partial X\cut = \partial X\sqcup
Y \sqcup -Y$.  Then there is a natural isometry
  $$ \Tr_Y ( Z_{X\cut})\longrightarrow  Z_ X , \tag{4.17} $$
where $\Tr_Y $ is the contraction
  $$ \Tr_Y \:E(\partial X \cut) \cong E(\partial X )\otimes E(Y) \otimes
     \overline{E(Y)}\longrightarrow E(\partial X ) \tag{4.18} $$
using the inner product on~$E(Y) $.
     \endproclaim

\flushpar
 Just as on the classical level, iterated gluings commute.

     \demo{Proof}
 We only comment on the gluing law~(d).  The proof is formally the same as the
one in~\cite{FQ,\S2}, but we repeat it here anyway.  Recall that for a
field~$P$ over a compact oriented $(d+2-n)$-manifold~$X$ we have the action
$\eac XP\in \tors{\bX}{\bP}$ which lives in an $n$-torsor, and the associated
$\lin XP\in \CW_{\tors{\bX}{\bP}}$ which lives in an $n$-vector space
(cf.~\thetag{2.4} and~\thetag{4.7}.)  Fix a bundle $Q'\to\bX$.  Then for
each $Q\to Y$ and each $P\cut\in \fld{X\cut}(Q'\sqcup Q\sqcup Q)$ we have an
isometry
  $$ \lin X{g_Q(P\cut)}\cong \Tr_Q\bigl(\lin{X\cut}{P\cut}\bigr) \tag{4.19} $$
by~\thetag{2.12}, where now $Tr_Q$~is the contraction
  $$ \Tr_Q\:\CW_{\tors{\bX\cut}{\bP\cut}} \cong \CW_{\tors{\bX}{\bP}}\otimes
     \CW_{\tors YQ} \otimes \overline{\CW_{\tors YQ}}\longrightarrow
     \CW_{\tors{\bX}{\bP}}  $$
using the inner product on~$\CW_{\tors YQ}$, and $g_Q$~is the gluing map
  $$ g_Q\: \fldb{X\cut}(Q'\sqcup Q\sqcup Q)\longrightarrow \fldb X(Q').
     \tag{4.20} $$
Fix $\Pb\in \fldb X(Q')$ and consider~$g_Q\inv (\Pb)$.  By \theprotag{4.3(c)}
{Lemma} the group~$\Aut Q$ acts transitively on~$g_Q\inv (\Pb)$.  This means
that the invariants in the representation
  $$ \bigoplus_{[P\cut]\in g_Q\inv (\Pb)} \lin{X\cut}{[P\cut]} \tag{4.21} $$
of~$\Aut Q$ by its diagonal action on $\CW_{T_Y(Q)}\times \CW_{T_{-Y}(Q)}$
via $\rho _Q\times \rho _Q$ are the ``constant functions'' under the
isomorphism~\thetag{4.19}.  Then the inner product~\thetag{3.5} in
$\ivr{\CW_{T_Y(Q)}}{\Aut Q,\rho _Q}$ applied to~\thetag{4.21} gives
  $$ \( \bigoplus_{[P\cut]\in g_Q\inv (\Pb)} \lin{X\cut}{[P\cut]} \)^{\Aut
     Q}\!\!\!\!\!\cong \#g_Q\inv ([P])\cdot \lin X{\Pb}. \tag{4.22} $$
Fix a set of representatives~$\{Q\}$ for~$\fldb Y$.  Let $\fldb
X(Q')_Q$~denote the equivalence classes of bundles over~$X$ whose restriction
to~$\bX$ is~$Q'$ and to ~$Y$ is~$Q$ (with given isomorphisms as
in~\thetag{4.2}).  Thus using equation~\thetag{4.4} on the measure and the
isometry~\thetag{4.22} we calculate
  $$ \spreadlines{6pt} \split
      Z_X(Q') &= \int_{\fldb X(Q')} d\mu \,(\Pb)\, \lin X{[P]}\\
      &= \sum\limits_{Q\in \{Q\}}\int_{\fldb X(Q')_Q}d\mu\, (\Pb) \,\lin
     X{[P]}\\
      &\cong \sum\limits_{Q\in \{Q\}}\mu _Q\cdot\[ \int_{\fldb
     {X\cut}(Q'\sqcup Q\sqcup Q)}d\mu ([P\cut])
     \,\Tr_Q\(\lin{X\cut}{[P\cut]}\)\]^{\Aut Q}\\
      &= \sum\limits_{Q\in \{Q\}}\mu _Q\cdot \Tr_Q \bigl( Z_{X\cut}(Q'\sqcup
     Q\sqcup Q) \bigr)^{\Aut Q}\\
      &= \Tr_Y(Z_{X\cut}(Q')).\endsplit  $$
     \enddemo

\newpage
\head
\S{5} Product Structures
\endhead
\comment
lasteqno 5@ 29
\endcomment

Some form of the following assertion holds: In a $d+1$~dimensional
topological quantum field theory the $d$-inner product space~$E(\cir)$ has
the structure of a ``higher commutative associative algebra with identity and
compatible real structure and inner product''.  In this section we only
discuss the cases $d=1$ and~$d=2$.  For~$d=1$ we obtain an ordinary algebra
structure on the vector space~$E(\cir)$, together with a compatible real
structure.  The inner product on~$E(\cir)$ is compatible with all of these
structures.  This is a standard argument, which we repeat here as a warmup.
For~$d=2$ the quantum space~$E(\cir)$ is a 2-inner product space, which in
particular is a category.  The algebra structure we discuss gives it the
structure of a {\it braided monoidal category\/}~\cite{JS}.\footnote{In fact,
we obtain what some refer to as a {\it tortile category\/}.
See~\cite{Y1,\S1}, ~\cite{Y2} for a precise definition and more thorough
discussion.  The notion of a tortile category is due to Shum~\cite{Sh}.} Here
the commutativity and associativity conditions give additional data (rather
than being conditions on the multiplication, as in an ordinary algebra), and
there is an additional piece of data coming from nontrivial loops of
diffeomorphisms of the circle (a {\it balancing\/}).  All of the arguments in
this section proceed directly from the axioms in \theprotag{4.12}
{Assertion}.  So they hold for any theory which obeys these axioms, not just
for a gauge theory with finite gauge group.

We begin with some standard deductions about arbitrary $d+1$~dimensional
theories.  First, a deduction about the classical theory.  Suppose $Y$~is a
closed oriented manifold and $Q\in \fld Y$ a $\Gamma $~bundle.  Consider the
product $[0,1]\times Q\in \fld{[0,1]\times Y}$, which is a bundle over the
``cylinder'' $[0,1]\times Y$.  The classical action\footnote{We use the
notation~$T_X(P)$ instead of~$\eac XP$, even though $X=[0,1]\times Y$ is not
closed.}~$\tors{\zot Y}{\zot Q}$ is an automorphism of~$\tors YQ$.  Now glue
two copies of~$\zot Q$ end to end and apply the gluing law~\thetag{2.12} to
construct an isomorphism
  $$ \tors{\zot Y}{\zot Q}\cdot \tors{\zot Y}{\zot Q}\longmapsto \tors{\zot
     Y}{\zot Q}. \tag{5.1} $$
This implies that there is a canonical element
  $$ t\in \tors{\zot Y}{\zot Q} \tag{5.2} $$
which satisfies $t\cdot t=t$.  In other words, the classical action of a
product field is trivialized.  If $\dim Y=d$ the classical action is the
identity map of~$\tors YQ$.  The quantum version of~\thetag{5.1}, obtained
from the quantum gluing law~\thetag{4.17}, asserts that
  $$ Z_{[0,1]\times Y}\:E(Y)\longrightarrow E(Y) \tag{5.3} $$
is an idempotent.  In other words, there is an isometry
  $$ (Z_{[0,1]\times Y})^2\longrightarrow Z_{[0,1]\times Y}.  \tag{5.4} $$
We may as well assume that $Z_{[0,1]\times Y}$~is isometric to the identity,
since in any case we can replace~$E(Y)$ by the image of~\thetag{5.3} to
obtain a new theory with this property.  Similarly, gluing the ends
of~$[0,1]\times Y$ together we deduce the existence of an isometry
  $$ Z_{\cir\times Y}\cong \dim E(Y). \tag{5.5} $$
Here the dimension of an $n$-inner product space is an $(n-1)$-inner product
space, as discussed in~\S{3}.  More generally, if $f\:Y\to Y$ is an
orientation preserving diffeomorphism, we can glue with a twist by~$f$ to
form the mapping torus~$\cir\times _fY$.  The axioms now imply the existence
of an isometry
  $$ Z_{\cir\times _fY}\cong \Tr_{E(Y)} (f_*),  $$
where $f_*\:E(Y)\to E(Y)$ is the isometry~\thetag{4.13}.

Another easily deduced property also relates to the
functoriality~\thetag{4.13}.  Suppose that $f_0,f_1\:Y'\to Y$ are isotopic
orientation preserving diffeomorphisms, and that $f_t\:Y'\to Y$ is an
isotopy.  Form the map
  $$ \aligned
     F\:\zo\times Y'&\longrightarrow \zo\times Y\\
     \langle t,y'  \rangle&\longmapsto \langle t,f_t(y')  \rangle.\endaligned
      $$
(More generally, our considerations apply to {\it pseudoisotopies\/}~$F$,
that is, to arbitrary diffeomorphisms~$F$ which restrict on the ends to~$f_0$
and~$f_1$.)  Now apply the functoriality axiom~\thetag{4.14} as follows.  The
partition functions~$Z_{\zo\times Y'}$ and~$Z_{\zo\times Y}$ are the
identity, according to~\thetag{5.3}.  The boundary maps $f_0$ and
$f_1$~induce isometries $(f_i)_*\:E(Y')\to E(Y)$.  The functoriality axiom
asserts that $F$~induces an isometry between $(f_1)_*\circ (f_0)_*\inv $
and~$\id_{E(Y)}$, or equivalently that
  $$ \text{$F$ induces an isometry $F_*\:(f_0)_*\to(f_1)_*$.} \tag{5.6} $$
The proper interpretation of~\thetag{5.6} depends on the dimension of~$Y$.
For example, if $\dim Y=d$ then $E(Y)$~is an ordinary inner product space and
\thetag{5.6}~asserts an equality $(f_0)_*=(f_1)_*$.  This implies in
particular that the action of~$\Diff^+(Y)$ on~$E(Y)$ factors through an
action of orientation-preserving diffeomorphisms~$\pi _0\Diff^+(Y)$
on~$E(Y)$.  If $\dim Y=d-1$, then $E(Y)$~is a 2-inner product space, which is
a category, and \thetag{5.6}~asserts that $F$~induces a natural
transformation~$F_*$ between the functors~$(f_0)_*$ and~$(f_1)_*$.  A further
argument shows that isotopic maps~$F$ induce the same natural transformation.
In the particular case where $f_0=f_1=\id$, this shows that $\pi
_1\Diff^+(Y)$ acts on~$E(Y)$ by automorphisms of the identity
functor.\footnote{An automorphism of the identity functor (i.e., a natural
transformation from the identity functor to itself) on a category~$\Cal{C}$
is for each object~$W\in \Cal{C}$ a choice of morphism $\theta _W\:W\to W$
such that if $W@>f>> W'$ is any morphism in~$\Cal{C}$, then
  $$ f\circ \theta _W = \theta _{W'}\circ f.   $$
}  This discussion generalizes to higher codimensions.

Now fix a $1+1$~dimensional theory and denote
  $$ E=E(\cir).  $$
Since any orientation-preserving diffeomorphism of~$\cir$ is isotopic to the
identity, \thetag{5.6}~implies that we can uniquely identify~$E(S)$ with~$E$
for any connected closed oriented 1-manifold~$S$.  Also, any two
orientation-reversing diffeomorphisms of~$\cir$ are isotopic, so there is a
well-determined isometry
  $$ c\:E\longrightarrow \overline{E}.  $$
Since the composite of two orientation-reversing diffeomorphisms is
orientation-preserving, $\bar{c}c=\id$.  Thus $c$~defines a real structure
on~$E$:
  $$ E_\RR = \{e\in E:c(e)=e\}. \tag{5.7} $$
Since $c$~is an isometry, $E_\RR$~is a real inner product space.  The inner
product identifies $E\mstrut _\RR\cong \ER^*$ as usual.  Since any compact
oriented 2-manifold has an orientation-reversing diffeomorphism, the
generalized partition function of any such manifold is real, by
\thetag{4.14}.

Next, we observe that the generalized partition function of the disk
  $$ \bo=Z_{D^2}\in \ER  $$
is a special element of~$\ER$.

The partition function of the ``pair of pants''~$P$, which is a disk with two
smaller disks removed (Figure~2), is an element
  $$ Z_P\in \ER\otimes \ER\otimes \ER. \tag{5.8} $$
Equation~\thetag{4.14} applied to diffeomorphisms of~$P$ which permute the
boundary circles (as in Figure~5) implies that $Z_P$ lives in the {\it
symmetric\/} triple tensor product of~$\ER$.  Identifying $\ER\mstrut \cong
\ER^*$ with the inner product, this defines a commutative multiplication
$\ER\otimes \ER\to\ER$.  In fact, the trilinear form
  $$ x\otimes y\otimes z\longmapsto \bigl(x\cdot y,z\bigr)_{\ER},\qquad
     x,y,z\in \ER,  $$
dual to~\thetag{5.8} is totally symmetric.  This symmetry is a compatibility
condition between the inner product and the multiplication.  For the complex
vector space~$E=E(\cir)$ we have the analogous statement that
  $$ x\otimes y\otimes z\longmapsto (x\cdot y,c(z))_{E},\qquad x,y,z\in E,
      \tag{5.9} $$
is totally symmetric.  Gluing a disk~$D^2$ onto~$P$ and
applying~\thetag{4.17} and~\thetag{5.3} we deduce that $\bo$~acts as the
identity map for the multiplication.  Finally, a standard gluing argument
that we do not repeat here shows that the multiplication is associative.

We summarize this discussion in the following.

        \proclaim{\protag{5.10} {Proposition}}
 In a $1+1$~dimensional topological quantum field theory \rom(which satisfies
the axioms of \theprotag{4.12} {Assertion}\rom) the inner product
space~$E(\cir)$ has a compatible real algebra structure which is commutative,
associative, and has an identity.  In addition, the map~\thetag{5.9} is
totally symmetric.
        \endproclaim

\flushpar
 It is not too hard to see that $E=E(\cir)$~contains no nilpotents.  For
if~$x\not= 0$, then since $(xc(x),\bo)=(x,x)\not= 0$, we see that~$xc(x)\not=
0$.  Iterating we find $x^{2^n}c(x)^{2^n}\not= 0$ and $(x^{2^n}c(x)^{2^n},
\bo) = (x^{2^n},x^{2^n})\not= 0$ for all~$n$.  Standard theorems in algebra
imply that $E$~contains a basis of idempotents $e_1,\dots ,e_N$, unique up to
permutation, with $e_ie_j=0$ for~$i\not= j$, and that $E$~is a product of one
dimensional algebras.\footnote{We need the {\it complex\/} algebra since
there exists a nontrivial commutative algebra over~$\RR$, namely~$\CC$.  Note
too that the conjugation $\langle z,w \rangle\mapsto\langle \bar{w},\bar{z}
\rangle$ on~$E=\CC\times \CC$ produces $\ER\cong \CC$ as an algebra
over~$\RR$.  So it is not true in general that the idempotents belong
to~$\ER$.} It is easy to express the partition function of a closed oriented
surface~$\Sigma _g$ of genus~$g$ in terms of the norms~$\lambda
_i^2=|e_i|^2$:
  $$ Z_{\Sigma _g} = \sum\limits_{i}(\lambda _i^2)^{1-g}.  $$

Now consider a $2+1$~dimensional theory, and as before denote $E=E(\cir)$.
Here $E$~is a 2-inner product space, so in particular is a category.  If
$f\:S\to \cir$ is an orientation-preserving diffeomorphism, then there is an
induced linear isometry $f_*\:E(S)\to E$.  Furthermore, any two such
$f_0,f_1\:S\to\cir$ are homotopic, and a homotopy $F\:f_0\to f_1$ induces an
isometry $F_*\:(f_0)_*\to(f_1)_*$, as in~\thetag{5.6}, but now $F_*$~depends
on the choice of~$F$.  (In the $1+1$~dimensional theory $F_*$~is an
equality.)  In fact, the positive generator of~$\pi _1\Diff^+(\cir)\cong \ZZ$
induces an automorphism of the identity functor on~$E$, that is, a morphism
  $$ \theta _W\:W\longrightarrow W \tag{5.11} $$
for each object $W\in \Obj(E)$.  So we cannot assert that $E(S)$~and $E$~are
uniquely isomorphic.

We do need, however, to identify the spaces~$E(S)$ for different circles~$S$
to derive the ``algebra'' structure on~$E$, so we resort to the following
device in what follows.  We use circles~$S$ which lie in~$\CC$.  There is a
unique composition of translations and homotheties which maps any such
circle~$S$ to the standard circle~$\cir=\TT\subset \CC$.  We use this to
uniquely identify $E(S)\cong E$ for any such~$S$.

As for the automorphism of the identity~$\theta $, we can compute it from the
diffeomorphism of the cylinder
  $$ \aligned
     \tau \:[0,1]\times \cir&\longrightarrow [0,1]\times S^1\\
     \langle t,s  \rangle&\longmapsto \langle t,s+t  \rangle,\endaligned
     \tag{5.12} $$
where here we write $\cir=\RR/\ZZ$ additively.  This glues to a
diffeomorphism of the torus~$\cir\times \cir$ described by the matrix
  $$ T=\pmatrix 1&0\\1&1  \endpmatrix. \tag{5.13} $$
By~\thetag{5.5} we have an isomorphism
  $$ E(\cir\times \cir)\cong \dim E,  $$
where $\dim E$~is understood as an inner product space, and in some sense the
action of~\thetag{5.13} on~$E(\cir\times \cir)$ is the action of~$\theta $ on
the identity endomorphism of~$E$.

The reflection $s\mapsto -s$ of the circle~$\cir=\RR/\ZZ$ induces an isometry
  $$ c\:E\longrightarrow \overline{E}. \tag{5.14} $$
On the underlying category~$E$ determines an involution on the objects.
Denote
  $$ c(W)=W^*,\qquad W\in \Obj(E).  $$
This is the {\it definition\/} of~`${}^*$'.  As in~\thetag{5.7} we can
consider the invariants~$\ER$.  For any 2-manifold~$Y$ there is an isometry
$Z\mstrut _Y\cong Z_Y^*$ determined by any orientation-reversing
diffeomorphism of~$Y$ which restricts to~$r$ on~$\bY$.  Of course, this
isometry depends on the choice of diffeomorphism, which we will standardize
in what follows.  Namely, our figures will sit in~$\CC$, symmetrically about
the real axis, and the boundary circles will have centers on that axis.  Then
reflection about the real axis is our standard orientation-reversing
diffeomorphism.

To compute the relationship between~$c$ and~$\theta $, consider the
cylinder~$C$ as shown in Figure~1.  The cylinder sits in~$\CC$, the boundary
circles have centers on the real axis, and $C$~is symmetric about the real
axis.  Now the diffeomorphism~\thetag{5.12} does not commute with reflection
in the real axis, but rather the reflection conjugates it to the
diffeomorphism $\langle t,s  \rangle\mapsto \langle t,s-t  \rangle$.
However, since the orientation of the boundary circles are reversed under
reflection, this conjugated diffeomorphism represents the positive generator
of $\pi _1\Diff^+\cir$ for the reflected circle.  Thus we conclude that for
any $W\in \Obj(E)$,
  $$ \theta \mstrut _{W^*} = \theta _W^*. \tag{5.15} $$
Here $\theta _W^*$~denotes the image of~$\theta _W$ under the
functor~\thetag{5.14}.

\midinsert
\bigskip
\centerline{\epsffile{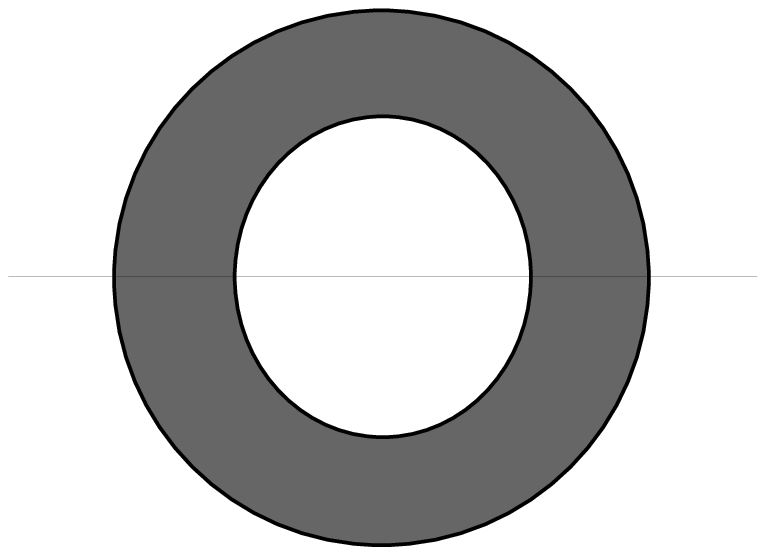}}
\nobreak
\centerline{Figure~1: The cylinder~$C$}
\bigskip
\endinsert

Let $D^2$ be the unit disk in~$\CC$.  Then
  $$ \bo=Z_{D^2}\in E \tag{5.16} $$
is a distinguished element of~$E$, and reflection in the real axis determines
an isometry
  $$ \bo\cong \bo^*. \tag{5.17} $$

Fix a standard pair of pants~$P$ as shown in Figure~2.  (The ordering of the
boundary circles is motivated by Figure~8.)  As with all of our figures it is
symmetric about the real axis and the boundary circles have centers on that
axis.  Any other $P'$~with the same properties is isotopic to~$P$ by an
isotopy which moves the boundary circles only by translations along the real
axis and by homotheties.  Furthermore, any two such isotopies are isotopic,
since any self-diffeomorphism of~$P$ which is the identity on~$\partial P$ is
isotopic to the identity.  This means that there is a uniquely defined
isotopy $Z_{P'}\cong Z_P$.  Now the partition function is
  $$ Z_P\in E\otimes E\otimes E,  $$
and reflection about the real axis determines an isometry
  $$ Z\mstrut _P\cong Z_P^*. \tag{5.18} $$
By duality $Z_P$~determines a map
  $$ m\:E\otimes E\longrightarrow E. \tag{5.19} $$
In particular, $m$~is a functor $E\times E\to E$, but it has linearity
properties as well.  Denote
  $$ m(W_1,W_2)=W_1\mytimes W_2,\qquad W_1,W_2\in \Obj(E).  $$
This is the {\it definition\/} of~`$\mytimes $'.  The isometry~\thetag{5.18}
translates into a natural isometry
  $$ (W_1\mytimes W_2)^* \cong W_1^*\mytimes W_2^*,\qquad W_1,W_2\in \Obj(E).
     \tag{5.20} $$
Glue a disk to the inner boundary circles in~$P$ to obtain natural isometries
  $$ \aligned
     \bo\mytimes W&\cong W,\\
     W\mytimes \bo&\cong W,\endaligned \tag{5.21} $$
for all~$W\in \Obj(E)$.\footnote{There should also be natural transformations
$W\mytimes W^*\to\bo$ and $\bo\to W\mytimes W^*$ which we did not succeed in
finding.}

\midinsert
\bigskip
\centerline{\epsffile{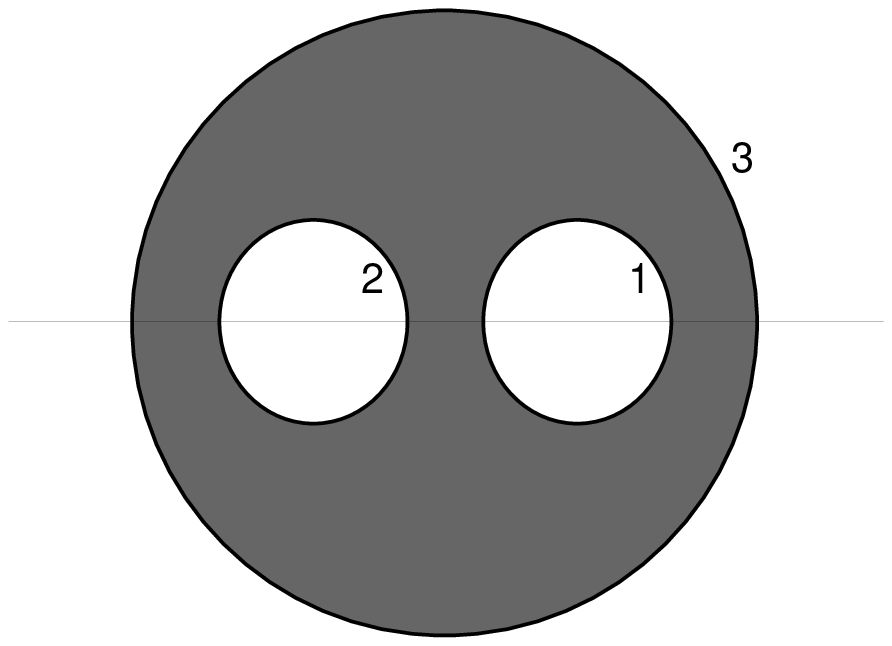}}
\nobreak
\centerline{Figure~2: The pair of pants~$P$}
\bigskip
\endinsert

\midinsert
\bigskip
\centerline{\epsffile{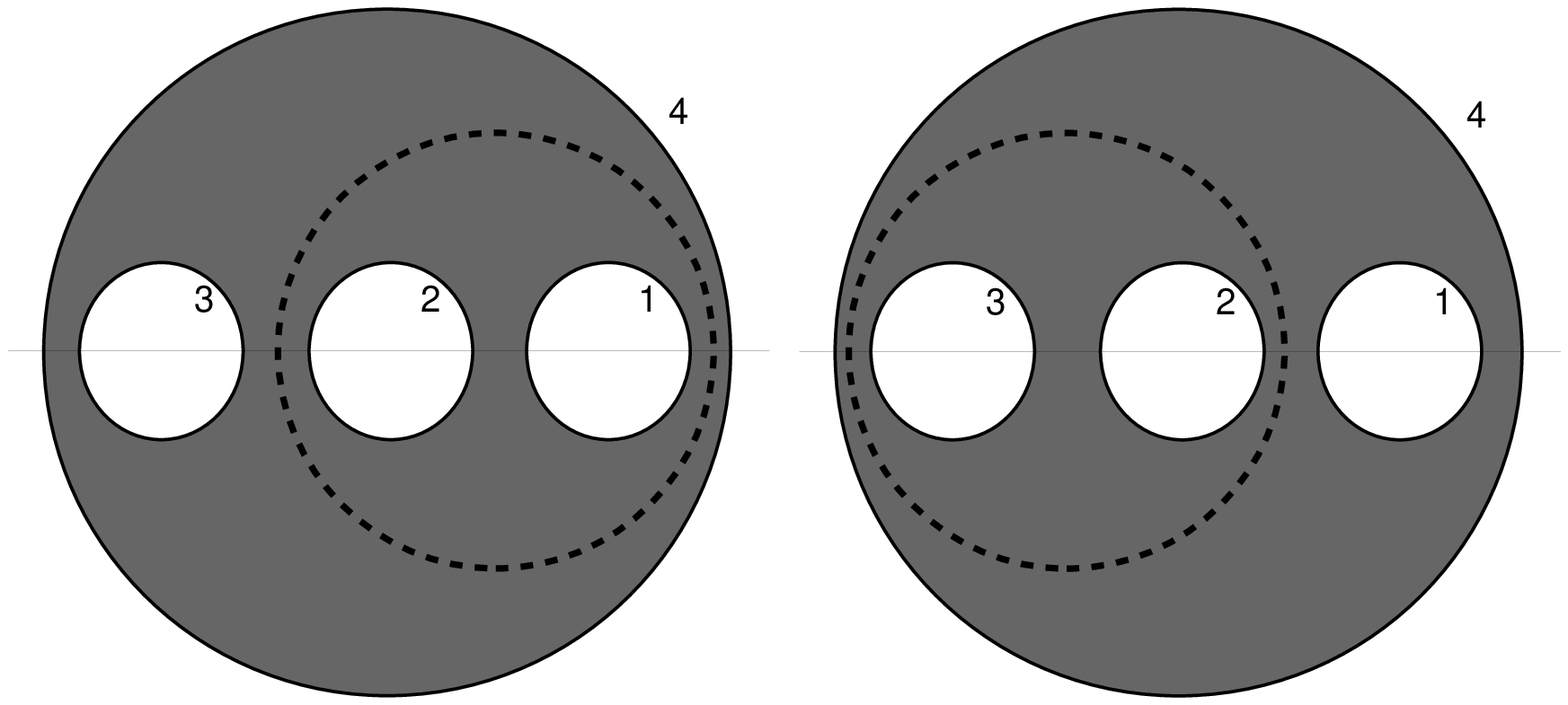}}
\nobreak
\centerline{Figure~3: Associativity}
\bigskip
\endinsert

It remains to discuss associativity and commutativity.  Whereas in the
$1+1$~dimensional theory these are constraints on the multiplication, here
they are new structures which satisfy ``higher order'' constraints.  The
associative law is a natural isometry
  $$ \varphi _{W_1,W_2,W_3}\:(W_1\mytimes W_2)\mytimes W_3\longrightarrow
     W_1\mytimes (W_2\mytimes W_3),\qquad W_1,W_2,W_3\in \Obj(E),
     \tag{5.22} $$
obtained from the obvious diffeomorphism indicated in Figure~3.  This figure
indicates an isometry between two different contractions of~$Z_P\otimes Z_P$,
which is equivalent to~\thetag{5.22}.  One can think of~\thetag{5.22} as
obtained by gluing and ungluing according to the dashed lines in Figure~3.
Performing such gluings and ungluings in Figure~4 makes obvious the
commutativity of the usual pentagon diagram
  $$ \CD
      ((W_1\mytimes W_2)\mytimes W_3)\mytimes W_4 @>>>
     (W_1\mytimes W_2)\mytimes (W_3\mytimes W_4) @>>>
      W_1\mytimes(W_2\mytimes (W_3\mytimes W_4))\\
      @VVV\\
      (W_1\mytimes (W_2\mytimes W_3))\mytimes W_4 @>>> W_1\mytimes
     ((W_2\mytimes W_3)\mytimes W_4) \endCD \tag{5.23} $$
A similar check shows that
  $$ \CD
      (W_1\mytimes \bo)\mytimes W_2 @>>> W_1\mytimes
     (\bo\mytimes W_2))\\
      @VVV\\
      W_1\mytimes W_2
      \endCD \tag{5.24} $$
commutes.

\midinsert
\bigskip
\centerline{\epsffile{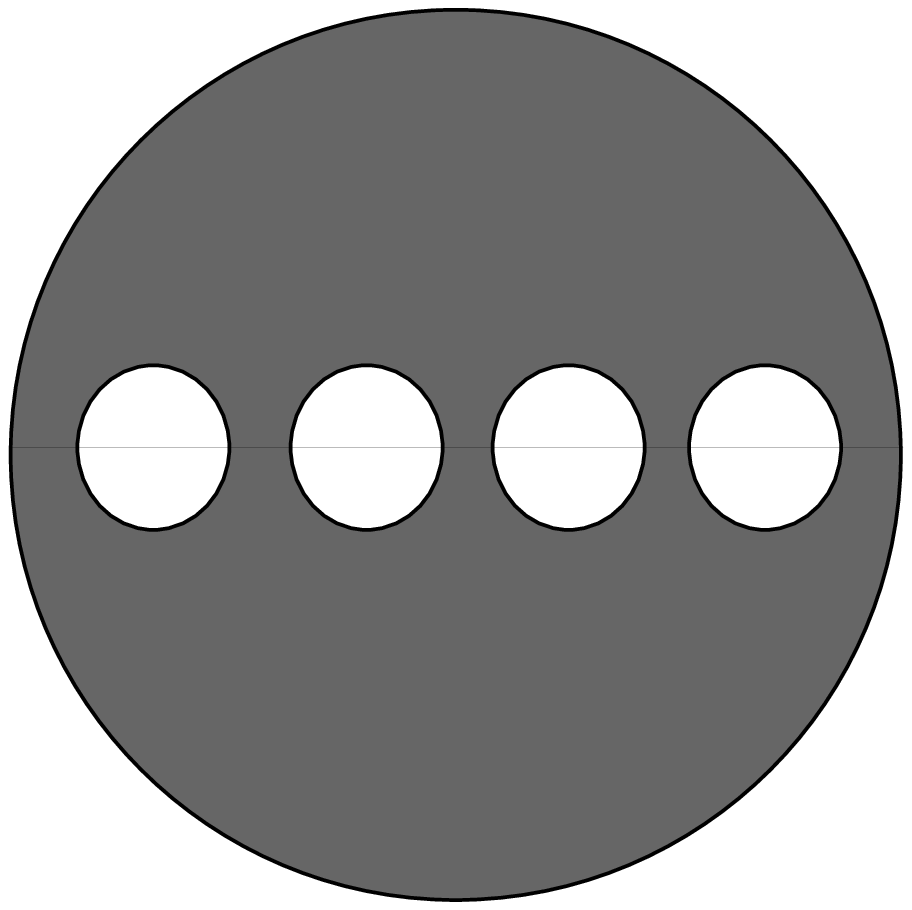}}
\nobreak
\centerline{Figure~4: Gluings and ungluings of pieces of this surface prove
 the
pentagon}
\bigskip
\endinsert

\midinsert
\bigskip
\centerline{\epsffile{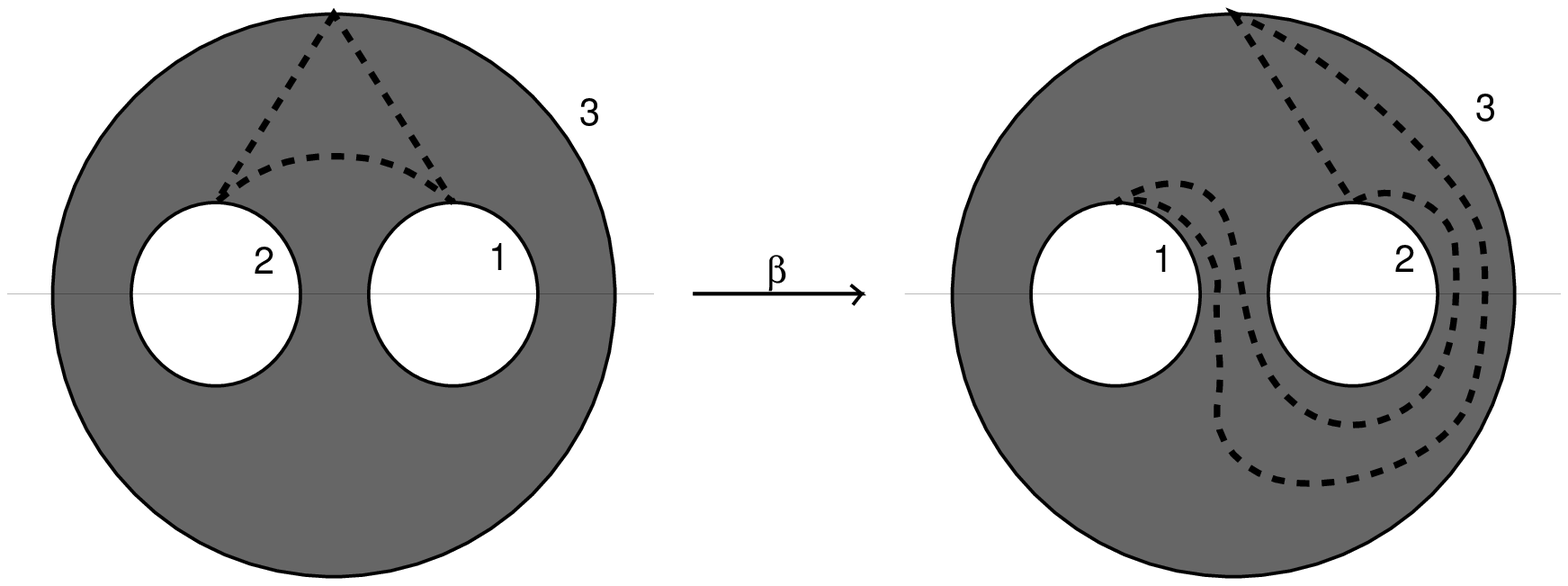}}
\nobreak
\centerline{Figure~5: The braiding diffeomorphism~$\beta $}
\bigskip
\endinsert

It does not make sense to say that the multiplication~\thetag{5.19} is
commutative.  Rather, there is a natural {\it braiding\/} isometry
  $$ R_{W_1,W_2}\:W_1\mytimes W_2\longrightarrow W_2\mytimes W_1 \tag{5.25} $$
obtained from the self-diffeomorphism $\beta \:P\to P$ indicated in Figure~5.
The auxiliary dashed lines indicate the motion of the boundary circle
labeled~$2$ over that labeled~$1$.  There is a compatibility between the
braiding~$R$ and the automorphism~$\theta $: the diagram
  $$ \CD
      W_1\mytimes W_2 @>R_{W_1,W_2}>> W_2\mytimes W_1\\ @V{\theta _{W_1\mytimes
     W_2}}VV @VV{\theta _{W_2}\mytimes \theta _{W_1}}V\\
      W_1\mytimes W_2 @>R_{W_2,W_1}\inv >> W_2\mytimes W_1\endCD \tag{5.26} $$
commutes for $W_1,W_2\in \Obj(E)$.  (Thus $\theta $~is termed ``balanced''.)
This follows from an equation in~$\Diff^+(P)$.  Namely, let $\tau _i$ denote
a positive Dehn twist around the boundary labeled~$i$.  Then the desired
equation is
  $$ \tau _2 \tau _1 \beta = \beta \inv  \tau _3,  $$
which is easily checked using pictures like those in Figure~5.  Similar
computations using Figure~6 show that the hexagon diagrams
  $$ \CD (W_1\mytimes W_2)\mytimes W_3 @>{R_{W_1,W_2}\mytimes \id}>>
     (W_2\mytimes W_1)\mytimes W_3 @>{\varphi _{W_2,W_1,W_3}}>> W_2\mytimes
     (W_1\mytimes W_3)\\
      @V{\varphi _{W_1,W_2,W_3}}VV @. @VV{\id\mytimes R_{W_1,W_3}}V\\
      W_1\mytimes (W_2\mytimes W_3) @>{R_{W_1,W_2\mytimes W_3}}>>
     (W_2\mytimes W_3)\mytimes W_1 @>{\varphi _{W_2,W_3,W_1}}>> W_2\mytimes
     (W_3\mytimes W_1)\endCD \tag{5.27} $$
and
  $$ \CD W_1\mytimes (W_2\mytimes W_3) @>{\id\mytimes R_{W_2,W_3}}>>
     W_1\mytimes (W_3\mytimes W_2) @>{\varphi _{W_1,W_3,W_2}\inv }>>
     (W_1\mytimes W_3)\mytimes W_2\\
      @V{\varphi _{W_1,W_2,W_3}\inv }VV @. @VV{R_{W_1,W_3}\mytimes\id }V\\
      (W_1\mytimes W_2)\mytimes W_3 @>{R_{W_1\mytimes W_2,W_3}}>>
     W_3\mytimes (W_1\mytimes W_2) @>{\varphi _{W_3,W_1,W_2}\inv }>>
     (W_3\mytimes W_1)\mytimes W_2\endCD \tag{5.28} $$
commute.  Each of \thetag{5.27}~and \thetag{5.28}~follows from an equation in
the diffeomorphism group of the surface pictured in Figure~6.  The
diffeomorphisms are formed from the braiding~$\beta $ shown in Figure~5.  The
associators are formed from gluings and ungluings, so do not enter.

\midinsert
\bigskip
\centerline{\epsffile{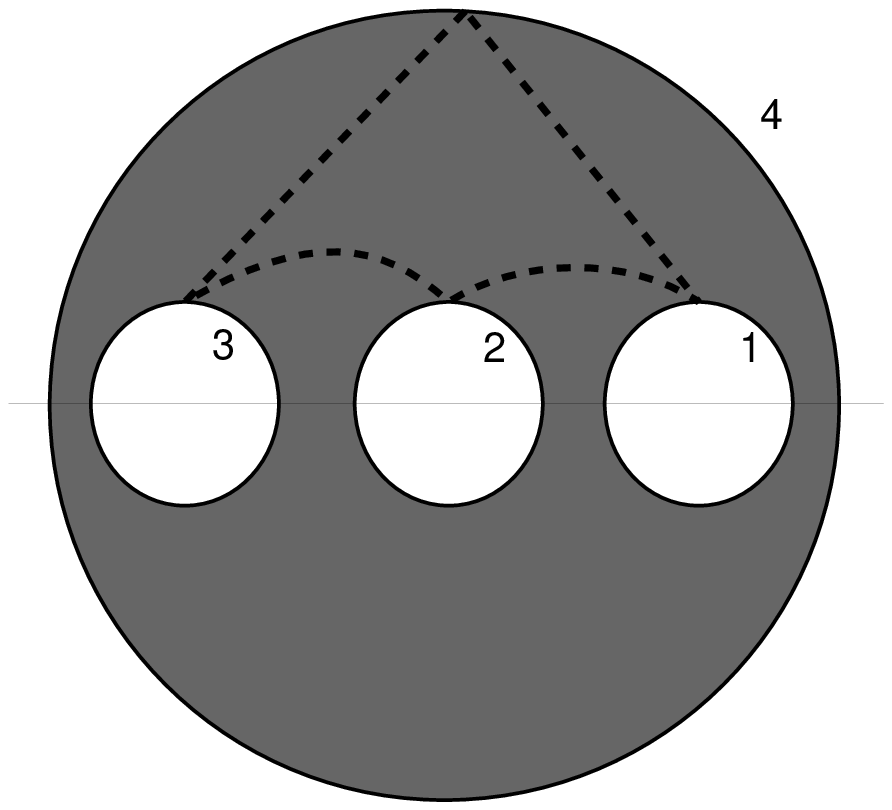}}
\nobreak
\centerline{Figure~6: Surface used to prove hexagon diagrams~\thetag{5.27}
 and~\thetag{5.28}}
\bigskip
\endinsert

We summarize this discussion in the following.

        \proclaim{\protag{5.29} {Proposition}}
 In a $2+1$~dimensional topological quantum field theory \rom(which satisfies
the axioms of \theprotag{4.12} {Assertion}\rom) the 2-inner product
space~$E(\cir)$ is a braided monoidal category with a compatible balanced
automorphism of the identity and compatible duality.\footnote{As mentioned
earlier, this is sometimes termed a {\it tortile category\/}.  Also, there is
a gap here in that we did not find the natural transformations mentioned in
the footnote following~\thetag{5.21}.}
        \endproclaim

\flushpar
 There is a notion of semisimplicity for such categories~\cite{Y2}, and it is
desirable to prove that $E$~is semisimple using the inner product, as we
indicated for the $1+1$~dimensional case after \theprotag{5.10} {Proposition}.
Surely one should think of the 2-inner product space structure together with
the monoidal structure.  In other words, one should think of~$E$ as a higher
version of the algebra encountered in \theprotag{5.10} {Proposition}.

There are {\it reconstruction theorems\/} in category theory which recover
certain algebraic objects from certain types of categories.  For example,
in~\cite{DM} it is shown how to recover a group from its category of
representations.  The structure in \theprotag{5.29} {Proposition} is almost
enough to reconstruct a {\it quasitriangular quasi-Hopf
algebra\/}~\cite{Ma1}.  (This is often termed a {\it quasi-quantum group\/}.
Probably there is a ribbon element as well~\cite{RT}, ~\cite{AC}
corresponding to the automorphism of the identity.)  Missing is a functor
from~$E$ to the category of vector spaces, though more abstract
reconstructions are possible~\cite{Ma2}.  We remark that there are examples
where no such ``fiber functor'' exists; the simplest is $\vect1\times
\vect1$.  (This arises from a three dimensional $\sigma $-model into a space
consisting of two points.)  But it seems that we can always decompose into a
product of spaces where reconstruction is possible.  For the finite gauge
theory we carry out the reconstruction in~\S\S{7--9}.  There we choose
various trivializations to construct a functor from~$E$ to the category of
vector spaces, and this allows the reconstruction of the quasi-quantum group.

Finally, we remark that we can take products with any closed oriented~$Y$ in
all of these constructions to obtain a higher algebra structure
on~$E(\cir\times Y)$.  In particular, the generalized quantum Hilbert space
of any torus~$\cir\times \dots \times \cir$ has a higher algebra structure.

\newpage
\head
\S{6} The $1+1$~Dimensional Theory
\endhead
\comment
lasteqno 6@ 11
\endcomment

We resume our discussion of the finite group gauge theory of~\S{2} and~\S{4}.
In this section we examine the $d=1$ case.  We know from \theprotag{5.10}
{Assertion} that $E(\cir)$~is an algebra, the algebra of central
functions~$\Cal{F}_{\text{cent}}(\Gamma )$ under convolution, as was computed
in~\cite{FQ,\S5}.  The new point is to compute~$E(\pt)$ and~$Z_{\zo}$.  The
results are fairly trivial, but they illustrate the definitions and
constructions of the previous sections and are a good warmup to the
$d=2$~case we discuss in~\S\S{7--9}.

Recall that the lagrangian is specified by a cocycle $\alpha \in C^2(B\Gamma
;\RZ)$.  We first consider the simplest case (the ``untwisted theory'')
where~$\alpha =0$.  Obviously, $\fldb{\pt}$~has a single element, the
equivalence class of the trivial bundle $\Qtriv=\pt\times \Gamma $.  The
value of the classical action $\ca{\pt}{\Qtriv}\in \tcat2$ is the trivial
\T-gerbe~$\tcat1$.  We identify the automorphism group of~$\Qtriv$
with~$\Gamma $, acting by {\it left\/} multiplication, and it acts trivially
on~$\ca{\pt}{\Qtriv}$.  Hence the associated 2-vector space~$\CW_{\Qtriv}$
in~\thetag{4.5} is~$\ivr{\vect1}\Gamma $, the category of representations
of~$\Gamma $.  Now $\Ep$~is computed by the path integral~\thetag{4.6} as an
inverse limit over the category of trivial bundles $Q\to\pt$.  The
automorphism groups~$\Aut Q$ which enter~\thetag{4.5} are not {\it
canonically\/} isomorphic to~$\Gamma $.  Rather, we use the distinguished
bundle~$\Qtriv\to\pt$ to trivialize the inverse limit:
  $$ E(\pt)\cong \frac{1}{\#\Gamma }\cdot \ivr{\vect1}\Gamma . \tag{6.1} $$
(Recall that the prefactor is $1/(\#\Aut\Qtriv)$.)  We use this
trivialization in what follows.

According to~\thetag{5.4} the generalized partition function~$Z_{\zo}$ is
isometric to the identity operator on~$E(\pt)$.  It is instructive to compute
this isometry directly from the definition of the path
integral~\thetag{4.8}.  There is a bijection
  $$ \fldb{\zo}(\Qtriv\sqcup \Qtriv)\longleftrightarrow \Gamma \tag{6.2} $$
by comparing the trivializations of a bundle $P\to\zo$ over the the two
endpoints of~$\zo$.  More explicitly, fix a basepoint in~$\Qtriv$ and let
$p_0\in P_0,\;p_1\in P_1$ be the corresponding basepoints in~$P$ using the
trivializations.  Parallel transport along~$\zo$ is an isomorphism $\psi
\:P_0\to P_1$.  Define $g\in \Gamma $ by $\psi (p_0)=p_1\cdot g$.  Then
$g$~is the element of~$\Gamma $ corresponding to~$P$ under the
correspondence~\thetag{6.2}.  The action of $\langle h_0,h_1 \rangle\in
\Gamma \times \Gamma \cong \Aut(\Qtriv)\times \Aut(\Qtriv)$ on the left hand
side of~\thetag{6.2} corresponds to the action
  $$ \langle h_0,h_1 \rangle\cdot g=h\mstrut _1gh_0\inv ,\qquad g\in \Gamma ,
      $$
on the right hand side.  The classical action~\thetag{2.4} is trivial, so
in~\thetag{4.7} we obtain $\lin{\zo}g=\CC$ for all~$g$ in~\thetag{6.2}.
Since $\zo$~has nonempty boundary the measure~$\mu $ in~\thetag{4.1} is
identically equal to~1.  Hence the path integral~\thetag{4.8} gives
  $$ Z_{\zo} = \bigoplus_{g\in \Gamma }\CC. \tag{6.3} $$
We identify this as the set of complex-valued functions~$\fug$ on~$\Gamma $,
with $\Gamma \times \Gamma $~acting as
  $$ \bigl(\langle h_0,h_1 \rangle\cdot f\bigr)(g)=f(h_1gh_0\inv ),\qquad
     f\in \fug,\quad g\in \Gamma , \tag{6.4} $$
with the standard inner product
  $$ (f_1,f_2) = \sum\limits_{g\in \Gamma }\,f_1(g)\overline{f_2(g)},\qquad
     f_1,f_2\in \fug. \tag{6.5} $$
View $\fug$ as an element in~$\overline{E(\pt)}\otimes E(\pt)$, or using the
inner product on~$\Ep$ as an element in $\Ep^*\otimes \Ep\cong \Hom\bigl(\Ep
\bigr)$.  Call this endomorphism~$K$.  Suppose that $W\in \Ep$ is a unitary
representation of~$\Gamma $ with action $\rho \:\Gamma \to\Aut(W)$.
According to the inner product~\thetag{3.5} and the factor~$1/\#\Gamma $
in~\thetag{6.1}, the action of~$K$ on~$W$ is
  $$ K(W)= \frac{1}{\#\Gamma }\cdot \bigl(\fug\otimes W \bigr)^\Gamma
      $$
Here we take $\Gamma $-invariants under the action of~$h\in \Gamma $
by~$\langle h,1 \rangle$ on~$\fug$ and $\rho (h)$ on~$W$; then $h\in \Gamma
$~acts on~$\bigl(\fug\otimes W \bigr)^\Gamma $ through the action of~$\langle
1,h \rangle$ on~$\fug$.

Now by~\thetag{5.4} we can derive from the gluing law an isometry~$K^2\to
K$.  (The underlying map of categories is a natural transformation.)  We
compute it by analyzing the gluing map~\thetag{4.20} for the gluing of two
intervals.  We find that the desired isometry is
  $$ \aligned
      \ooGd\bigl(\fug\otimes \fug \bigr)^\Gamma \cong \ooGd\Cal{F}(\Gamma
     \times \Gamma )^{\Gamma }&\longrightarrow \fug\\
      f(\cdot ,\cdot )&\longmapsto f(e,\cdot ).\endaligned \tag{6.6} $$
(The $\Gamma $~invariance in~\thetag{6.6} refers to the action $(h\cdot
f)(g_1,g_2)=f(g_1h,h\inv g_2)$ for~$h\in \Gamma $.)  This yields the desired
isometry~$K^2\to K$ which on~$W\in E(pt)$ is
  $$ \aligned
      K^2(W) = \frac{1}{(\#\Gamma )^2}\cdot \bigl(\Cal{F}(\Gamma \times
     \Gamma )\otimes W \bigr)^{\Gamma \times \Gamma }&\longrightarrow
     \ooGd\bigl(\fug\otimes W \bigr)^\Gamma = K(W)\\
       f^i\otimes w_i \qquad &\longmapsto \qquad \bigl(g\mapsto f^i(e,g)w_i
     \bigr).\endaligned  $$
(These expressions are summed over~$i$.)  This is an isometry $K\to \id$ on
the image of~$K$, and is compatible with the isometry~$K\to \id$ which on
$W\in E(pt)$ is
  $$ \aligned
      \ooGd\bigl(\fug\otimes W \bigr)^\Gamma &\longrightarrow W\\
      f^i\otimes w_i&\longmapsto f^i(e)w_i.\endaligned \tag{6.7} $$

We can also check the gluing which leads to~\thetag{5.5}.  That is to say we
can check the gluing law~\thetag{4.17} when we glue the two ends of~$\zo$
together.  Now $\fldb\cir$~can be identified with the set of conjugacy
classes in~$\Gamma $, and the gluing map~\thetag{4.20} with~$Q=\Qtriv$ sends
an element in~$\Gamma $ to its equivalence class.  The map~$\Tr_{\pt}$
in~\thetag{4.18} is $1/\nG$~times the $\Gamma $-invariants under the diagonal
action in~\thetag{6.4}, and applied to~$Z_{\zo}=\fug$ this gives
  $$ \Tr_{\pt}\bigl(\fug \bigr) = \frac{1}{\nG}\cdot
     \Cal{F}_{\text{cent}}(\Gamma ) \tag{6.8} $$
where $\Cal{F}_{\text{cent}} (\Gamma )$~is the space of central functions
with inner product~\thetag{6.5}.  This is~$E(\cir)$, as follows easily
from~\thetag{4.6} (cf.~\cite{FQ,\S5}).

If the lagrangian $\alpha \in C^2(B\Gamma ;\RZ)$ is nonzero (the ``twisted
theory''), then the classical action also enters in a nontrivial way.  We
compute the classical action on the trivial bundle $\Qtriv\to\pt$.  Since
there is a unique cycle in~$C_0(\pt)$ which represents the fundamental class
$[\pt]\in H_0(\pt)$, the integration theory in the appendix gives
  $$  \exp(\tpi\int_{\pt}\bar{f}^*\alpha ) =\tcat1  $$
for any $\bar{f}\:\pt\to B\Gamma $.  (This is what we must compute
in~\thetag{2.1}.)  In other words, we can think of~$\alpha $ as defining the
trivial \T-gerbe bundle over~$B\Gamma $, which then lifts to the trivial
\T-gerbe bundle over~$E\Gamma $.  The nontrivial part comes from homotopies
between classifying maps of~$\Qtriv$, which we identify with paths
in~$E\Gamma $.  The integral in~\thetag{2.2} is then a \T-torsor.  The
classical action~$\cao$ is a nontrivial \T-gerbe computed by an inverse limit
over the ``path category'' of~$E\Gamma $.  The value of the classical action
on a field $P\to\zo$, whose boundary we assume trivialized by an isomorphism
$\bP\cong \Qtriv\times \Qtriv$, is then an automorphism of~$\cao$.
By~\thetag{6.2} we identify the equivalence class of~$P$ with an
element~$g\in \Gamma $, and by~\thetag{2.8} the classical action is
well-defined on the equivalence class. Taking an inverse limit over all such
bundles in the equivalence class we obtain for each $g\in \Gamma $ an
automorphism~$T_g$ of~$\cao$.  Furthermore, there are isomorphisms
  $$ T_{gh}\longrightarrow T_g\cdot T_h  $$
from the gluing law~\thetag{2.12} applied to the gluing of intervals.

As in the $\alpha =0$ case we compute the quantum space~$\Eap$ by taking an
inverse limit over the category of all trivial bundles.  We use the
distinguished object~$\Qtriv$ to trivialize the inverse limit
(cf.~\thetag{4.5} and~\thetag{3.8}):
  $$ \Eap\cong \frac{1}{\nG}\cdot \ivr{\CW_{\cao}}{\Gamma ,\rho }.
     \tag{6.9} $$
Here $\rho $~is the action of~$\Gamma $ on~$\cao$ via the torsors~$T_g$.  If
we trivialize the \T-gerbe~$\cao$, for example by choosing a basepoint
in~$E\Gamma $, then we obtain an isomorphism $\cao\cong \tcat1$, and so the
$T_g$~are identified with \T-torsors.  As in~\thetag{1.4} these torsors
define a central extension
  $$ 1 @>>> \TT @>>> \tilde{\Gamma} @>>> \Gamma @>>> 1.   $$
Incidentally, they are isomorphic to the torsors and central extension which
come from the action of~$\Aut(\Qtriv)\cong \Gamma $ on~$\cao\cong \tcat1$.
This assertion follows from the fact~\thetag{5.2} that the classical action
of the product bundle $[0,1]\times \Qtriv\to[0,1]$ is trivial.  With the
trivialization of~$\cao$ the isometry~\thetag{6.9} becomes
  $$ \Eap\cong \frac{1}{\nG}\cdot \ivr{\vect1}{\Gamma ,\rho }. \tag{6.10} $$
Recall from the paragraph following~\thetag{3.8} that $\ivr{\vect1}{\Gamma
,\rho }$~is the category of representations of~$\tilde{\Gamma }$ where the
central~$\TT$ acts by scalar multiplication.  We emphasize that
\thetag{6.10}~requires two choices of trivialization (of two inverse limits).

Computing with the trivialization~\thetag{6.10} we find analogous
to~\thetag{6.3} that
  $$ Z_{\zo}\cong \bigoplus_{g\in \Gamma }L_g, \tag{6.11} $$
where $L_g$~is the hermitian line obtained from the torsor~$T_g$ as
in~\thetag{3.2}.  We leave the reader to modify the verification
of~\thetag{6.7} above to show that \thetag{6.11}~acts isometrically to the
identity map.  The twisted version of~\thetag{6.8} is also easy to check.

\newpage
\head
\S{7} The $2+1$ dimensional Chern-Simons theory and quasi-quantum groups:
Untwisted
Case
\endhead
\comment
lasteqno 7@ 28
\endcomment

We turn to the $2+1$~dimensional case of gauge theory with finite gauge
group, which can be considered as a Chern-Simons theory.  Our goal is to
derive the quasi-Hopf algebras of~\cite{DPR} directly from the path
integral~\thetag{4.6}.  We already investigated several features of this
theory in~\cite{FQ}.  The new point is an investigation of the 2-inner
product space~$E(\cir)$, which according to \theprotag{5.29} {Assertion} is a
certain type of braided monoidal category.  With suitable trivializations we
claim that it is isomorphic to the category of representations\footnote{It is
probably more natural to use corepresentations here, but in any case we have
enough finiteness to switch back and forth between representations and
corepresentations.  Also, this will reconstruct the algebras in~\cite{DPR}
rather than their duals.  Our convention here differs from~\cite{FQ,\S3},
where we use corepresentations.  Note also that in~\cite{FQ,\S3} we use {\it
right\/} comodules whereas here we use {\it left\/} modules.  Thus the
groupoid~\thetag{7.5} is opposite that in~\cite{FQ,\S3}.} of the
quasitriangular quasi-Hopf algebra constructed in~\cite{DPR}.  We also
recover the results of~\cite{FQ,\S\S3--4}, including Segal's modular
functor~\cite{S1}, from our approach here.  In this section we treat the
untwisted case where the lagrangian $\alpha \in C^3(B\Gamma ;\RZ)$ vanishes.
In sections~\S\S{8--9} we generalize to the twisted case~$\alpha \not= 0$.

The holonomy of a bundle around the circle induces a bijection
  $$ \fldb{\cir} \longleftrightarrow \text{conjugacy classes in $\Gamma$} .
     \tag{7.1} $$
The classical action in the $\alpha =0$ theory is trivial.  Choose a bundle
$\Qx\to\cir$ representing each conjugacy class~$[x]$ in~$\Gamma $ under the
correspondence~\thetag{7.1}.  Then following the same steps as
in~\thetag{6.1}, this choice of bundles leads to an isometry
  $$ E=E(\cir)\cong \bigoplus_{[x]} \frac{1}{\#\Aut\Qx}\cdot
     \ivr{\vect1}{\Aut\Qx}. \tag{7.2} $$

It is convenient to use a more concrete description of~$E$ directly in terms
of the group~$\Gamma $, and this requires a choice of some basepoints.
(Compare with the choice of basepoints in~\cite{FQ,\S3}.)  Fix a conjugacy
class~$[x]$ and consider the fiber~$F_{[x]}$ of $\Qx\to\cir$ over the
basepoint $1\in \cir=\TT$.  A point in~$F_{[x]}$ determines a particular
value of the holonomy of~$\Qx$, which is an element of the conjugacy
class~$[x]$.  Choose a (base)point~$f_x$ in the fiber of the holonomy map
$F_{[x]}\to[x]$ for each~$x\in [x]$.  Then $f_x$~induces an isomorphism
  $$ \Aut\Qx\longrightarrow  C_x \tag{7.3} $$
by assigning to $\psi \in \Aut\Qx$ the element $g\in C_x$ which satisfies
$\psi (f_x)=f_x\cdot g$.  Thus if $W$~is a representation of~$\Aut\Qx$, then
under this isomorphism $W$~is also a representation of~$C_x$.  Let
$\Wb$~denote the trivial vector bundle over~$[x]$ whose fiber at each~$x\in
[x]$ is~$W$.  Now $\Gamma $~acts on~$[x]$ on the left by conjugation
($g\:x\mapsto gxg\inv $), and we want to lift this action to~$\Wb$.  For
each~$x\in [x]$ the stabilizer~$C_x$ already acts on the fiber~$\Wb_x=W$.
For $x,x'\in [x]$ there is a unique~$g_{x,x'}\in \Gamma $ with
$f_{x}=f_{x'}\cdot g_{x,x'}$.  Then~$x'=g\mstrut _{x,x'}xg_{x,x'}\inv $.
Lift $g_{x,x'}\:x\mapsto x'$ to the identity map $\id\:\Wb_x\to\Wb_{x'}$.
There is then a unique extension of the $C_x$~action and the action of
the~$g_{x,x'}$ on~$W$ to a $\Gamma $~action on~$W$ which lifts the
conjugation action on~$[x]$.

Summarizing, the choice of basepoints in the bundles~$\Qx$ leads to an
isometry
  $$ E\cong \frac{1}{\#\Gamma }\cdot \Vect_\Gamma (\Gamma ), \tag{7.4} $$
where $\Vect_\Gamma (\Gamma )$~is the 2-inner product space of hermitian
vector bundles over~$\Gamma $ with a unitary lift of the left $\Gamma
$~action on~$\Gamma $ by conjugation.  We write an element of~$\Vect_\Gamma
(\Gamma )$ as $W=\oplus _{x\in \Gamma }W_x$.  If $W_1,W_2\in \Vect_\Gamma
(\Gamma )$, then the inner product is defined as
  $$ (W_1,W_2)_{\Vect_\Gamma (\Gamma )} = \( \bigoplus_x (W_1)_x\otimes
     \overline{(W_2)_x}\) ^{\tsize\Gamma} .  $$
It is easy to check that $1/\#\Gamma $ times this inner product is the inner
product in~\thetag{7.2}.

There is another description of~$E$ which is useful.  Let $\goid$~denote the
groupoid which is the set~$G\times G$ with the composition law
  $$ \langle x_2,g_2 \rangle\circ \langle x_1,g_1 \rangle = \langle
     x_1,g_2g_1 \rangle,\qquad \text{if $x\mstrut _2=g\mstrut _1x\mstrut
     _1g_1\inv $} . \tag{7.5} $$
Composition is not defined if~$x\mstrut _2\not= g\mstrut _1x\mstrut _1g_1\inv
$.  Then
  $$ E\cong \frac{1}{\#\Gamma }\cdot \ivr{\vect1}{\goid} , \tag{7.6} $$
where $\ivr{\vect1}{\goid} $~is the 2-inner product space of finite
dimensional unitary representations of~$\goid$.  What we mean by a
representation of the groupoid~$\goid$ amounts exactly to a $\Gamma $-bundle
over~$\Gamma $, so \thetag{7.6}~is essentially identical to~\thetag{7.4}.
More precisely, these are representations ({\it left\/} modules) of the
``groupoid algebra''
  $$ \CC[\goid] = \bigoplus_{x,g}\CC\langle x,g \rangle, \tag{7.7} $$
with multiplication
  $$ \langle x_2,g_2 \rangle\cdot \langle x_1,g_1 \rangle = \cases \langle
     x_1,g_2g_1 \rangle ,&x\mstrut _2=g\mstrut _1x\mstrut _1g_1\inv
     ;\\0,&\text{otherwise} .\endcases  $$
The unit element is
  $$ 1=\sum\limits_{x}\langle x,e  \rangle.  $$
If $W\in \Vect_\Gamma (\Gamma ) =\ivr{\vect1}{\goid} $ we use the notation
  $$ A^W_g=A\mstrut _g\:W_x\longrightarrow W_{gxg\inv }  $$
for the action of~$\langle x,g  \rangle\in \goid$.  In terms of the
$\goid$~action we have
  $$ W_x = \langle x,e  \rangle\cdot W. \tag{7.8} $$

We use the trivialization~\thetag{7.4}, or equivalently~\thetag{7.6}, in what
follows.

An irreducible element~$W\in E$ is supported on some equivalence class~$[x]$,
and the fiber~$W_x$ is an irreducible representation~$\rho $ of~$C_x$.  Since
the various $C_x,\;x\in [x]$ are identified up to inner automorphisms, the
equivalence class~$[\rho ] $ of the representation is well-defined.  Up to
isomorphism the irreducible elements of~$E$ are labelled by the pair~$\langle
[x],[\rho ] \rangle$.  These labels appear in all treatments of this
theory~\cite{DVVV}, ~\cite{DPR}, ~\cite{DW}, ~\cite{FQ}.

It is convenient to use the isomorphism~\thetag{7.6} to identify the path
integral~\thetag{4.11} over a compact oriented 2-manifold~$X$, which is an
element of~$E(\bX)$, as an element in tensor products of~$E$.  Recall our
convention stated after~\thetag{5.11} for identifying~$\bX$ as a disjoint
union of copies of the standard circle~$\cir$.  For this we restrict to
surfaces~$X$ which are subsets of~$\CC$.  Under these identifications each
component of~$\bX$ has a basepoint corresponding to the standard
basepoint~$1\in \cir$.  Let $\bfld X$ denote the category of principal
$\Gamma $~bundles $P\to X$ endowed with a basepoint in the fiber over each
basepoint in~$\bX$.  Morphisms are required to preserve the basepoints. Let
$\bfldb X$~denote the set of equivalence classes.  For the cylinder~$C$ the
holonomy and parallel transport define a bijection
  $$ \bfldb C\longleftrightarrow\goid \tag{7.9} $$
as illustrated in Figure~7.  (Compare with~\thetag{6.2}.)  Now for a
surface~$X$ we can glue~$C$ to any component of~$\bX$ using the basepoints.
This induces a $\goid$~action on~$\bfldb X$ for each component of~$\bX$.

\midinsert
\bigskip
\centerline{\epsffile{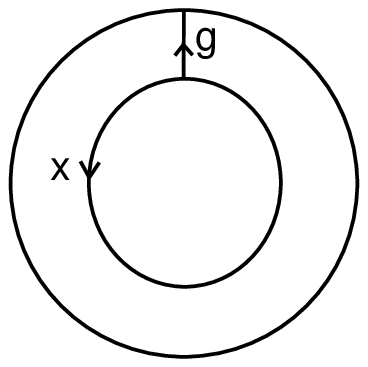}}
\nobreak
\centerline{Figure~7: The bundle over~$C$ corresponding to~$\langle x,g
\rangle\in \CG$ }
\bigskip
\endinsert

        \proclaim{\protag{7.10} {Proposition}}
 Let $X\subset \CC$ be a compact oriented 2-manifold.\footnote{The same
arguments apply to arbitrary surfaces with {\it parametrized\/} boundary.  If
the surface has closed components, then we must modify the inner product
in~\thetag{7.11}.}  Then under the isomorphism~\thetag{7.6} the path integral
over~$X$ is
  $$ Z_X \cong L^2(\bfldb X) \tag{7.11} $$
with the $\goid$~actions induced by gluing cylinders onto components
of~$\bX$.
        \endproclaim

        \demo{Proof}
 Let $P\in \bfld X$ and fix a component~$S$ of~$\bX$.  The basepoint
determines an isomorphism $P\res S\to \Qx$ for some~$[x]$.  If the holonomy
around~$S$ is~$x$, then the basepoint maps to~$f_x$.  Apply this to a pointed
bundle $P\in \bfld C$ over the cylinder~$C$ which corresponds
under~\thetag{7.9} to an element~$\langle x,g \rangle\in \goid$.  Using
parallel transport along the axis of~$C$, this bundle also determines an
element of~$\Aut\Qx$.  If $g\in C_x$ then the correspondence between the
automorphism of~$\Qx$ and~$g$ agrees with~\thetag{7.3}.  (This follows
from~\thetag{5.2}.)  Also, the bundle labeled by~$\langle x,g_{x,x'}
\rangle$ corresponds to the identity in~$\Aut\Qx$ for all $x'\in [x]$.  Thus
the action of~$\goid\approx\bfldb C$ on the quantization~\thetag{7.11}
induced by gluing is the action described in the text leading to~\thetag{7.4}
and~\thetag{7.6}.
        \enddemo

The 2-inner product space~$E$ has extra structure determined by the path
integral over special surfaces and special diffeomorphisms, as described
in~\S{5}.

        \proclaim{\protag{7.12} {Proposition}}
 The finite gauge theory described in \theprotag{4.12} {Assertion}
with~$\alpha =0$ determines the following structure on~$E$.
\medskip
 \noindent\rom(a\rom)\ \rom(Automorphism of the identity~\thetag{5.11}\rom)\
For $W\in E$ we have
  $$ \theta _W\res{W_x} = A_x\:W_x\longrightarrow W_x. \tag{7.13} $$
 \noindent\rom(b\rom)\ \rom(Involution~\thetag{5.14}\rom)\ For~$W\in E$ the
dual $W^*\in \overline{E}$ is defined by $(W^*)\mstrut _x=W_{x\inv }^*$ and
$A^{W^*}_g = (A^W_{g\inv })^*$.
 \smallskip
 \noindent\rom(c\rom)\ \rom(Identity~\thetag{5.16}\rom)\ The identity~$\bo$ is
  $$ \bo_x = \cases \CC ,&x=e;\\0,&x\not= e,\endcases \tag{7.14} $$
with $C_e=\Gamma $~acting trivially on~$\bo_e$.
 \smallskip
 \noindent\rom(d\rom)\ \rom(Multiplication~\thetag{5.19}\rom)\ The tensor
product of $W_1,W_2\in E$ is
  $$ (W_1\mytimes W_2)_x = \bigoplus_{x_1x_2=x}(W_1)_{x_1}\otimes (W_2)_{x_2}
     \tag{7.15} $$
with the $\Gamma $~action
  $$ A^{W_1\mytimes W_2}_g = A^{W_1}_g \otimes A^{W_2}_g. \tag{7.16} $$
 \noindent\rom(e\rom)\ \rom(Associator~\thetag{5.22}\rom)\ The
associator~$\varphi$ is induced from the standard associator of tensor
products of vector spaces.
 \smallskip
 \noindent\rom(f\rom)\ \rom($R$-matrix~\thetag{5.25}\rom)\ For $W_1,W_2\in E$
we have
  $$ \aligned
       R_{W_1,W_2}\: (W_1)\mstrut _{x_1}\otimes (W_2)\mstrut
     _{x_2}&\longrightarrow (W_2)\mstrut _{x\mstrut _1x\mstrut _2x_1\inv
     }\otimes (W_1)\mstrut _{x_1} \\
       w_1\otimes w_2 &\longmapsto A_{x_1}^{W_2}(w_2)\otimes w_1\endaligned
     \tag{7.17} $$
and all other components are zero.
        \endproclaim

\flushpar
 A few remarks are in order.  First, since $x$~is a central element of~$C_x$,
the transformation~\thetag{7.13} is a scalar on each irreducible component
of~$W_x$.  (We decompose~$W_x$ under the $C_x$~action.)  If $W$~is an
irreducible element of~$E$ labelled by~$\langle [x],[\rho]   \rangle$, then
the scalar transformation~$A_x$ is independent of~$x\in [x]$.  The {\it
conformal weight\/}~$h_{\langle [x],[\rho]   \rangle}$ is defined up to an
integer by the equation
  $$ A_x = e^{\tpi h_{\langle [x],[\rho]   \rangle}}. \tag{7.18} $$
This  agrees with the results of~\cite{FQ,\S5}, where we calculated
the conformal weight from the action of~\thetag{5.13} on the torus.  Notice
that $\theta _W$~can also be described as the action of
  $$ v=\sum\limits_{x}\langle x,x  \rangle \tag{7.19} $$
on~$W$, where $v$~is a special element\footnote{This element plays the role
of the {\it inverse\/} of the ribbon element of Reshetikhin/Turaev~\cite{RT}.
The quasitriangular quasi-Hopf algebras we encounter have a ribbon structure
(cf.~\cite{AC}).} of~$\CC[\goid]$.  The identity element~$\bo$ corresponds to
the label~$\langle [e],\text{trivial} \rangle$.  Another description of the
multiplication~\thetag{7.15}, \thetag{7.16} is
  $$ W_1\mytimes W_2 = \mu _*(W_1\boxtimes W_2),  $$
where $\mu \:\Gamma \times \Gamma \to\Gamma $ is group multiplication and
$W_1\boxtimes W_2\to\Gamma \times \Gamma $ is the external tensor product.
Finally, we invite the reader to verify~\thetag{5.15}, \thetag{5.17},
\thetag{5.20}, \thetag{5.21}, \thetag{5.23}, \thetag{5.24}, \thetag{5.26},
\thetag{5.27}, and~\thetag{5.28} directly from the data listed in
\theprotag{7.12} {Proposition}.\footnote{The natural transformations
$W\mytimes W^*\to\bo$ and $\bo\to W\mytimes W^*$ mentioned in the footnote
following~\thetag{5.21} are evidently the duality pairing $\bigoplus_x
W\mstrut _x\otimes W_x^*\to\CC$ and its dual.}

        \demo{Proof}
 We use \theprotag{7.10} {Proposition} to compute the path integrals over the
various surfaces.

\smallskip

(a)\ We compute the action of the diffeomorphism~\thetag{5.12} on the
cylinder~$C$.  From~\thetag{7.9} and~\thetag{7.11} we obtain an isomorphism
  $$ Z_{C}\cong \fugo. \tag{7.20} $$
An argument similar to that in~\S{6} (see~\thetag{6.7}) shows that
$Z_C$~acts isometrically to the identity on~$E$ via the isometry
  $$ \aligned
      \ooGd\bigl(\fugo\otimes W \bigr)^{\goid} &\longrightarrow W\\
      f^i\otimes w_i&\longmapsto f^i\bigl(\langle \pi (w_i),e
     \rangle\bigr)w_i,\endaligned \tag{7.21} $$
where $\pi \:W\to\Gamma $ is an element of~$\Vect_\Gamma (\Gamma )$.  On the
left hand side of~\thetag{7.21} we take $\goid$-invariants under the action
$a\:f^i(\cdot )\otimes w_i\mapsto f^i(a\inv \cdot )\otimes aw_i$, and then
$a\in \goid$ acts on the invariants by $a\:f^i(\cdot )\otimes w_i\mapsto
f^i(\cdot a)\otimes w_i$.  Here `$\cdot $'~indicates the argument of the
function.  Now the diffeomorphism~$\tau $ in~\thetag{5.12} induces by
pullback the map
  $$ \tau ^*\langle x,g  \rangle = \langle x,gx  \rangle,\qquad \langle x,g
     \rangle\in \goid \tag{7.22} $$
on fields~\thetag{7.9}, and so the map
  $$ \bigl(\tau _*f \bigr)\bigl(\langle x,g  \rangle \bigr) = f\bigl(\langle
     x,gx  \rangle \bigr),\qquad f\in \fugo,  $$
on the quantization~\thetag{7.20}.  In terms of the element $v\in \CC[\goid]$
in~\thetag{7.19}, this is
  $$ (\tau _*f)(\cdot ) = f(\cdot \;v).  $$
Thus on the left hand side of~\thetag{7.21} the diffeomorphism $\tau
$~induces the action
  $$ f^i(\cdot )\otimes w_i\longmapsto (\tau _*f^i)(\cdot )\otimes w_i =
     f^i(\cdot \;v)\otimes w_i,  $$
which corresponds to the action $w\mapsto vw$ on the right hand side
of~\thetag{7.21}.  This is~\thetag{7.13}.

\smallskip

(b)\ We first calculate that the reflection of~$\cir$ induces the map
$\Qx\mapsto Q_{[x\inv ]}$ on fields by pullback.  (Actually, this is the map
on equivalence classes of fields written using our distinguished
representatives.)  Since the reflection reverses orientation, this induces a
map $\tcat1\mapsto\tcat1\inv $ on the classical action, and in the
quantization leads us to use the dual space.  Under the
identification~\thetag{7.4} this gives $(W^*)\mstrut _x = W_{x\inv }^*$.
Then the induced representation of~$\Aut\Qx\cong \Aut Q_{[x\inv ]}$ is
$A^{W^*} _g = (A^W_{g\inv })^*$.

\smallskip

(c)\ It is easy to see that $\bfldb{D^2}$~consists of one element, and the
restriciton of any representative bundle to~$\partial D^2=\cir$ is~$Q_{[e]}$.
Furthermore, $\Aut Q_{[e]}\cong \Gamma $ acts trivially.

\midinsert
\bigskip
\centerline{\epsffile{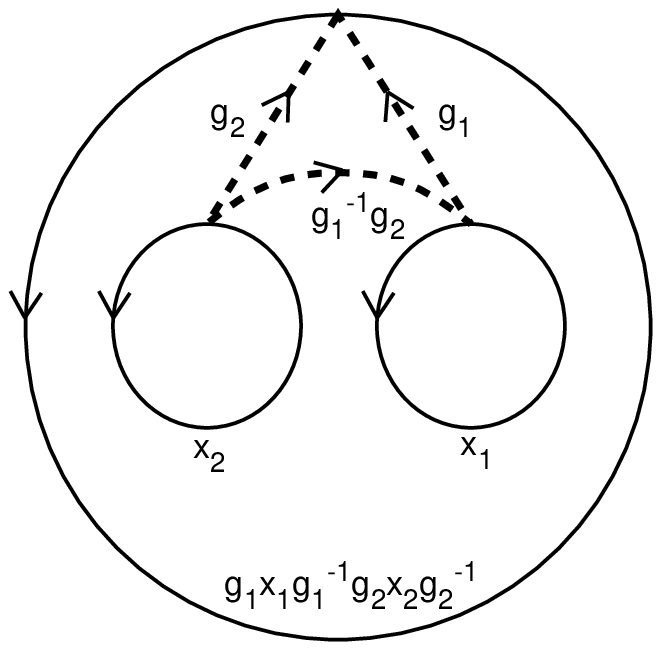}}
\nobreak
\centerline{Figure~8: The bundle over~$P$ corresponding to $\langle x_1,g_1
 \rangle\times \langle x_2,g_2  \rangle\in \goid\times \goid$}
\bigskip
\endinsert

(d)\ For the pair of pants~$P$ we identify
  $$ \bfldb P\longleftrightarrow \goid\times \goid \tag{7.23} $$
using the parallel transports and holonomies indicated in Figure~8.  This
leads to an isometry
  $$ Z_P\cong \Cal{F}(\goid\times \goid). \tag{7.24} $$
The actions of $\langle x,g  \rangle\in \goid$ corresponding to the two inner
components of~$\partial P$ are
  $$ \aligned
     f(\cdot ,\cdot )&\longmapsto f(\cdot \langle x,g  \rangle\inv ,\cdot ),\\
     f(\cdot ,\cdot )&\longmapsto f(\cdot,\cdot\langle x,g  \rangle\inv
     ).\endaligned \tag{7.25} $$
The action of~$\langle x,g  \rangle\in \goid$ corresponding to the outer
component is
  $$ \bigl(\langle x,g \rangle\cdot f \bigr)\bigl(\langle x_1,g_1
     \rangle,\langle x_2,g_2 \rangle \bigr) = \cases f\bigl(\langle x_1,gg_1
     \rangle,\langle x_2,gg_2 \rangle \bigr) ,&\text{if $x = g\mstrut
     _1x\mstrut _1g_1\inv g\mstrut _2x\mstrut _2g_2\inv$}
     ;\\0,&\text{otherwise} .\endcases \tag{7.26} $$
Using the inner product~\thetag{7.6} on~$E$ we see that the
multiplication~\thetag{5.19} is the map
  $$ W_1\otimes W_2\longmapsto \frac{1}{(\#\Gamma )^2}\cdot
     \bigl(\Cal{F}(\goid\times \goid)\otimes W_1\otimes W_2
     \bigr)^{\goid\times \goid},  $$
where $\goid\times \goid$ acts on~$\Cal{F}(\goid\times \goid)$
via~\thetag{7.25}.  The $\goid$~action on the right hand side is
via~\thetag{7.26}.  Then a routine check shows that
  $$ \aligned
     \frac{1}{(\#\Gamma )^2}\cdot
     \bigl(\Cal{F}(\goid\times \goid)\otimes W_1\otimes W_2
     \bigr)^{\goid\times \goid}&\longrightarrow W_1\mytimes W_2\\
     f^{ij}\otimes w^{(1)}_i\otimes w^{(2)}_j&\longmapsto f^{ij}\bigl(\langle
     \pi (w_1),e  \rangle,\langle \pi (w_2),e  \rangle \bigr)
     w^{(1)}_i\otimes w^{(2)}_j\endaligned \tag{7.27} $$
is an isometry, where $W_1\otimes W_2$~is defined by~\thetag{7.15}
and~\thetag{7.16}.

\smallskip

(e)\ This is immediate from the definition of the associator.

\smallskip

(f)\ We compute the action of the braiding diffeomorphism~$\beta $ (Figure~5)
on the fields~\thetag{7.23} by pullback as
  $$ \langle x_1,g_1  \rangle\times \langle x_2,g_2  \rangle\longmapsto
     \langle x_2\mstrut ,g\mstrut _1x\mstrut _1g_1\inv g\mstrut _2
     \rangle\times \langle x_1,g_1  \rangle. \tag{7.28} $$
So the action on the quantization~\thetag{7.24} by pushforward is
  $$ \bigl(\beta _*f \bigr)\bigl(\langle x_1,g_1 \rangle,\langle x_2,g_2
     \rangle \bigr) = f\bigl(\langle x\mstrut _2,g\mstrut _1x\mstrut
     _1g_1\inv g\mstrut _2 \rangle,\langle x\mstrut _1,g\mstrut _1 \rangle
     \bigr).   $$
Under the isometry~\thetag{7.27} this corresponds to~\thetag{7.17}, as
desired.
        \enddemo

Reconstruction theorems in category theory assert that $E$~is (equivalent to)
the category of representations of a Hopf algebra~$H$.  In fact, since $E$~is
braided $H$~is a {\it quasitriangular\/} Hopf algebra~\cite{Dr}.  We do not
need the general arguments from category theory to carry out the
reconstruction, as the Hopf algebra~$H$ is apparent from our explicit
descriptions of~$E$ in~\thetag{7.4} and~\thetag{7.6}, and from the formulas
in \theprotag{7.12} {Proposition}.

Indeed, as an algebra $H$~is the ``groupoid algebra'' $H=\CC[\goid]$ defined
in~\thetag{7.7}.  We have already seen in~\thetag{7.6} that $E$~is
isomorphic to the category of representations of the algebra~$H$.
Explicitly, if $\rho \:H\to\End(W)$ is a representation of~$H$, set $W_x =
\rho (\langle x,e \rangle)(W)$ as in~\thetag{7.8} and set $A_g\:W_x\to
W_{gxg\inv }$ equal to~$\rho (\langle x,g \rangle)$.  The quasitriangular
Hopf structure on~$H$ is easily deduced from \theprotag{7.12} {Proposition}.
{}From~\thetag{7.15} and~\thetag{7.16} we see that the coproduct $\Delta \:H\to
H\otimes H$ is
  $$ \Delta \bigl(\langle x,g \rangle\bigr) = \sum\limits_{x_1x_2=x}\langle
     x_1,g \rangle\otimes \langle x_2,g \rangle.  $$
The counit $\epsilon \:H\to\CC$ is
  $$ \epsilon \bigl(\langle x,g \rangle\bigr) = \cases 1
     ,&x=e;\\0,&\text{otherwise} ,\endcases  $$
as we see from the action of~$H$ on~$\bo$~\thetag{7.14}.  The antipode
$S\:H\to H$ is implemented on the dual (\theprotag{7.12(b)} {Proposition}), so
is
  $$ S\bigl(\langle x,g \rangle\bigr) = \langle gx\inv g\inv ,g\inv \rangle.
      $$
The quasitriangular structure is an element $R\in H\otimes H$ such that for
every pair of representations $(W_1,\rho _1)$, $(W_2,\rho _2)$ of~$H$, we
have
  $$ R_{W_1,W_2} = \tau _{W_1,W_2}\circ (\rho _1\otimes \rho _2)(R),
      $$
where $\tau _{W_1,W_2}\:W_1\otimes W_2\to W_2\otimes W_1$ is the
transposition.  Hence from~\thetag{7.17} we deduce
  $$ R = \sum\limits_{x_1,x_2} \langle x_1,e \rangle\otimes \langle x_2,x_1
     \rangle.  $$
Since the associator~$\varphi $ is the standard associator on vector spaces
(\theprotag{7.12(e)} {Proposition}), we obtain a Hopf algebra (as opposed to
a quasi-Hopf algebra).  Finally, we have already observed that the
automorphism of the identity~$\theta $ in~\thetag{7.13} is implemented by the
element~$v$ in~\thetag{7.19}:
  $$ v=\sum\limits_{x}\langle x,x  \rangle .  $$
This special element in~$H$ is the {\it inverse\/} of the ribbon element of
Reshetikhin/Turaev~\cite{RT}. We interpret it here in terms of the
``balancing'' of the category of representations.

The quasitriangular Hopf algebra~$H$ is identified in~\cite{DPR} as the
``quantum double'' of~$\fug$.

Finally, we indicate how to recover the ``modular functor''~\cite{S1},
{}~\cite{FQ,\S4}.  Once and for all fix a basis~$\{W_\lambda \}$ of the 2-inner
product space~$E=E(\cir)$.  Here $\lambda $~runs over the labeling set~$\Phi
$ mentioned earlier.  Now suppose $X$~is a compact oriented 2-manifold with
each boundary component parametrized.  The parametrizations identify~$E(\bX)$
with a tensor product of copies of~$E$ and~$\overline{E}$.  Thus we can
decompose~$Z_X$ according to the chosen basis for~$E$:
  $$ Z_X\cong \bigoplus_{\bal} E(X,\bal)\otimes W_{\bal},  $$
where $\bal=\langle \lambda _1,\dots ,\lambda _k  \rangle$ runs over
labelings of the boundary components and
  $$ W_{\bal}=W_{\lambda _1}^{\pm1}\otimes \dots \otimes W_{\lambda
     _k}^{\pm1},  $$
the signs chosen according to the orientation.  The inner product
spaces~$E(X,\bal)$ define the modular functor.  The gluing law for the
modular functor follows directly from \theprotag{4.12(d)} {Assertion}.

\newpage
\head
\S{8} The $2+1$ dimensional Chern-Simons theory and quasi-quantum groups:
Twisted
Case
\endhead
\comment
lasteqno 8@ 36
\endcomment

In this section we extend the results of~\S{7} to the $2+1$~dimensional
finite gauge theory with nontrivial lagrangian $\alpha \in C^3(B\Gamma
;\RZ)$.  The classical theory is nontrivial, and this leads to corresponding
modifications of the quantum theory.  We must choose additional
trivializations (of gerbes) to express the theory in terms of familiar
objects, and in particular to construct a quasi-Hopf algebra.  (Recall the
remarks following \theprotag{5.29} {Proposition}.)  Such trivializations
appear more naturally in~\S{9}, where we cut open the circle and make
calculations on the interval.  We rely here on the exposition in~\S{7} and
only indicate the necessary modifications.  The cocycle~$\alpha \in
C^3(B\Gamma ;\RZ)$ is fixed throughout.  We often omit it from the notation.

We use the choices made in~\S{7} of representative bundles $Q_{[x]}\to\cir$
and basepoints~$f_x$.  The classical action~$T\ua_{\cir}(\Qx)$ is a \T-gerbe,
which we denote~$\CG_{[x]}$.
The automorphism group $\Aut\Qx$~acts on this gerbe, and the action is a
homomorphism
  $$ \rho \ua_{[x]}=\rho \mstrut _{[x]}\:\Aut\Qx\longrightarrow
     \Aut(\CG_{[x]}).  \tag{8.1} $$
Fix a trivializing element
  $$ G_{[x]}\in \CG_{[x]}=T\ua_{\cir}(\Qx), \tag{8.2} $$
and so an isomorphism $\CG_{[x]}\cong \tcat1$.  This can be done as
in~\cite{FQ,\S3} by fixing a representative cycle~$s\in C_1(\cir)$ for the
fundamental class $[\cir]\in H_1(\cir)$, and by fixing classifying maps
$\Qx\to E\Gamma $.  With these trivializations the action~\thetag{8.1}
determines a central extension of~$\Aut\Qx$ by~$\TT$, as in~\thetag{1.4}.
There is an induced isometry (see~\thetag{4.6}, \thetag{6.10})
  $$ E\ua=E\ua(\cir)\cong \bigoplus_{[x]} \frac{1}{\#\Aut\Qx}\cdot
     \ivr{\vect1}{\Aut\Qx,\rho \ua_{[x]}}.  $$

We want to express this directly in terms of~$\Gamma $, using the
basepoints~$f_x$ as in~\S{7}.  The central extensions of~$\Aut(\Qx)$ lead via
the isomorphism~\thetag{7.3} to central extensions $\Ctil_x$ of the
centralizer subgroup of any~$x\in \Gamma $.  That is, for each~$g\in C_x$ we
have a \T-torsor~$\ttor xg$ together with appropriate isomorphisms under
composition.  Note that there are trivializations
  $$ \ttor xe\cong \TT \tag{8.3} $$
since $\ttor xe\cdot \ttor xe\cong \ttor xe$.  (This is~\thetag{5.2}.)
Extend to a central extension of the groupoid~$\goid$ in~\thetag{7.5} as
follows.  First, for any two elements~$x,x'$ in the same conjugacy class let
  $$ \ttor x{g_{x,x'}}=\TT,\qquad x'\in [x]. \tag{8.4} $$
(The element~$g_{x,x'}\in \Gamma $ was defined following~\thetag{7.3}.)  Then
for any $x,g\in \Gamma $ we have $\langle x,g \rangle=\langle x,g_{x,gxg\inv
}\rangle\circ \langle x,h \rangle$ for some unique~$h\in C_x$.  Set
$T(x,g)=T(x,g_{x,gxg\inv })\cdot T(x,h)$.  This determines the desired
central extension
  $$ 1 @>>> \TT @>>> \tilde{\goid}\ua @>>> \goid @>>> 1,  $$
where for $\langle x,g \rangle\in \goid$ the \T-torsor~$\ttor xg$ is the
preimage of~$\langle x,g \rangle$ in~$\goidtilde\ua$.  There are appropriate
isomorphisms under composition.  Let $\hl xg$~be the hermitian line
corresponding to the \T-torsor~$\ttor xg$, and
  $$ \te xe\in \hl xe \tag{8.5} $$
the trivializing element derived from~\thetag{8.3}.  We ignore the
trivializations~\thetag{8.4}, which are artifacts of our definitions.

With this understood an element of~$E\ua$ corresponds to a vector bundle
$W=\bigoplus_{x\in \Gamma }W_x$ over~$\Gamma $ with isomorphisms
  $$ A^W_g=A\mstrut _g\:\hl xg\otimes W\mstrut _x\longrightarrow W_{gxg\inv }
      $$
which compose properly.  Set
  $$ H\ua = \bigoplus_{x,g} \hl xg. \tag{8.6} $$
Define an algebra structure\footnote{In~\cite{FQ} we defined a coalgebra
structure instead.} using the multiplication in~$\goidtilde$:
  $$ \hl{x_2}{g_2}\otimes \hl{x_1}{g_1}\longrightarrow \cases
     \hl{x_1}{g_2g_1},&x\mstrut _2=g\mstrut _1x\mstrut _1g_1\inv
     ;\\0,&\text{otherwise}. \endcases \tag{8.7} $$
The identity element in~$H\ua$ is
  $$ 1=\sum\limits_{x}\te xe. \tag{8.8} $$
We can view~$E\ua$ as the 2-inner product space of representations of~$H\ua$,
with the natural inner product multiplied by~$1/\#\Gamma $.  Or, by analogy
with~\thetag{7.6}, we write
  $$ E\ua\cong \frac{1}{\#\Gamma }\cdot \ivr{\vect1}{\goidtilde},
     \tag{8.9} $$
where we only take representations in which the central circles $\ttor
xe\cong \TT$ act as scalar multiplication.

Now suppose $X$~is a compact oriented surface, either with a given
parametrization of the components of~$\bX$, or with an embedding~$X\subset
\CC$ which induces such parametrizations according to our conventions.
Suppose $P\in \bfld X$ is a $\Gamma $~bundle over~$X$ with basepoints on the
boundary.  Let $Y$~be a component of~$\bX$ and suppose the holonomy of~$P\res
Y$ is~$x$.  Then the basepoint in~$P\res Y$ and the parametrization of~$Y$
determine an isomorphism $P\res Y\cong Q_{[x]}$, and so an isomorphism
$T\ua_{\bX}(\partial P)\cong T\ua_{\cir}(\Qx)=\CG_{[x]}$.  This \T-gerbe is
trivialized by our choice in~\thetag{8.2}.  Hence the classical
action~\thetag{2.6} of~$P$ can be identified with a \T-torsor $T\ua_X(P)$,
using this trivialization.  As in~\thetag{4.7} this \T-torsor determines a
hermitian line, and by taking an inverse limit we obtain a line~$L\ua_X([P])$
depending only on the equivalence class of~$P$.  (This line could degenerate
to~$0$ if $X$~has a closed component.)  Let
  $$ L\ua_X\longrightarrow \bfldb X  $$
denote the resulting line bundle over the finite set~$\bfldb X$.  The
following  generalizes \theprotag{7.10} {Proposition}.

        \proclaim{\protag{8.10} {Proposition}}
 Let $X\subset \CC$ be a compact oriented 2-manifold.\footnote{The same
arguments apply to arbitrary surfaces with {\it parametrized\/} boundary.  If
the surface has closed components, then we must modify the inner product
in~\thetag{8.11}.}  Then under the isomorphism~\thetag{8.9} the path integral
over~$X$ is space of $L^2$~sections
  $$ Z\ua_X \cong L^2(\bfldb X,L\ua_X), \tag{8.11} $$
with the $\goidtilde$~action induced by gluing cylinders onto components
of~$\bX$.
        \endproclaim

        \demo{Proof}
 The only new point is an isometry
  $$ L_C(x,g)\cong \hl xg, \tag{8.12} $$
where $L_C(x,g)=L_C([P_{\langle x,g \rangle}])$ for $P_{\langle x,g
\rangle}\to C$ a pointed bundle over the cylinder corresponding to~$\langle
x,g \rangle\in \goid$ under~\thetag{7.9}.  Recall the proof of
\theprotag{7.10} {Proposition}, where we show that the basepoints determine
an isomorphism $\partial P_{\langle x,g\rangle}\cong \Qx\sqcup \Qx$, and so
$P_{\langle x,g \rangle}$~determines an element of~$\Aut\Qx$.  The classical
action of~$P_{\langle x,g \rangle}$ is then an element of~$\Aut
(\CG_{[x]})\cong \tcat1$.  But by~\thetag{5.2} the classical action
$T_C([0,1]\times \Qx)$ of a product bundle is trivial, and then the desired
isometry~\thetag{8.12} follows easily.
        \enddemo

We adopt the notation
  $$ \lce xe=\te xe  $$
for the element in~\thetag{8.5}.

We need a few preliminaries to generalize \theprotag{7.12} {Proposition}.
For any~$x\in \Gamma $ there is a trivialization
  $$ \lce xx\in \lc xx \tag{8.13} $$
as follows.  By~\thetag{7.22} the diffeomorphism $\tau \:C\to C$ satisfies
$\tau ^*\langle x,e \rangle=\langle x,x \rangle$.  Notice that $\tau $~is the
identity on~$\partial C$, so it respects the trivializations~\thetag{8.2}.
By the functoriality of the classical action~\thetag{2.7} the
diffeomorphism~$\tau $ induces an isomorphism~$T_C(x,x)\cong T_C(x,e)$, and
so an isometry $L_C(x,x)\cong L_C(x,e)$.  Then $\lce xx$~corresponds to $\lce
xe\in L_C(x,e)$ (cf.~\thetag{8.5}).

Next, consider the diffeomorphism of the cylinder~$C$
  $$ \aligned
      \iota \:[0,1]\times \cir&\longrightarrow [0,1]\times S^1\\
       \langle t,s \rangle&\longmapsto \langle -t,-s \rangle.\endaligned
      $$
It is not the identity on~$\partial C$.  Rather, $\partial \iota $ swaps the
two boundary components, and if we identify them in the obvious way,
$\partial \iota $ is the reflection~$s\mapsto -s$.  By the
functoriality~\thetag{2.7} and the orientation axiom~\thetag{2.9} this
reflection induces an isomorphism $T_{\cir}\bigl(\Qx \bigr)\inv \to
T_{\cir}\bigl(Q_{[x\inv ]} \bigr)$, and so we can compare the trivializations
in~\thetag{8.2}.  Use this isomorphism to define the \T-torsor
  $$ T_{[x]}=G_{[x]}\cdot G_{[x\inv ]}. \tag{8.14} $$
Let $L_{[x]}$~be the hermitian line corresponding to the \T-torsor~$T_{[x]}$.
Then since $\iota $~induces the map $\iota ^*\langle x,g  \rangle=\langle
gx\inv g\inv ,g\inv   \rangle$ on fields, the induced isometry on the
classical action is
  $$ \iota _*\:\lc{gx\inv g\inv }{g\inv }\otimes L_{[x]}\longrightarrow \lc
     xg \otimes L_{[gxg\inv ]}. \tag{8.15} $$
Of course, $L_{[gxg\inv ]}=L_{[x]}$, so we can cancel these terms
from~\thetag{8.15}.

\midinsert
\bigskip
\centerline{\epsffile{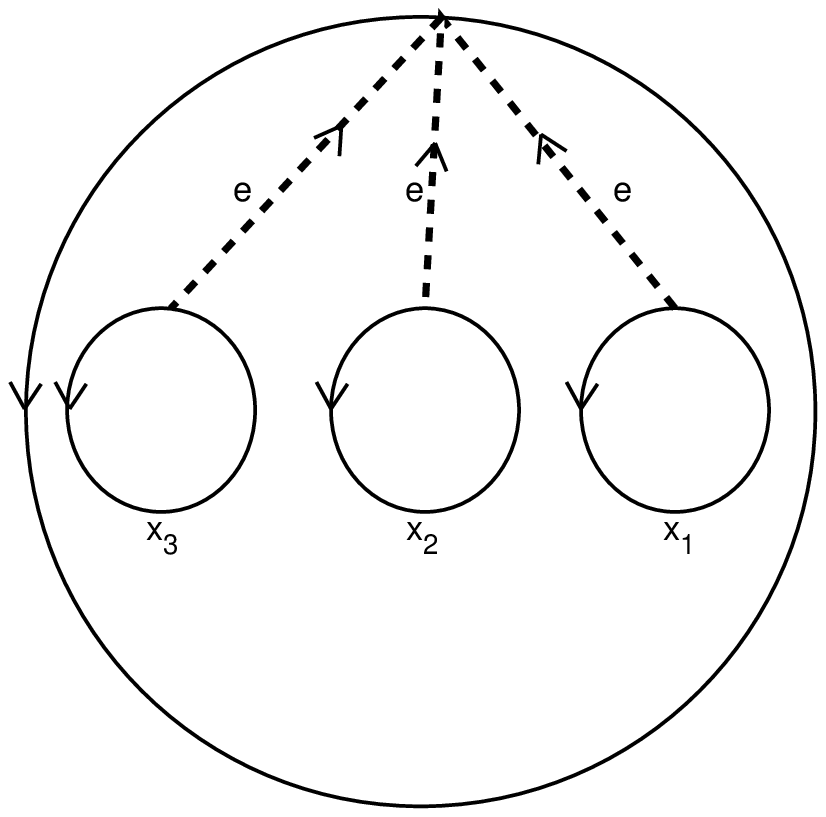}}
\nobreak
\centerline{Figure~9: Field used in the proof of~\thetag{8.16}}
\bigskip
\endinsert

\midinsert
\bigskip
\centerline{\epsffile{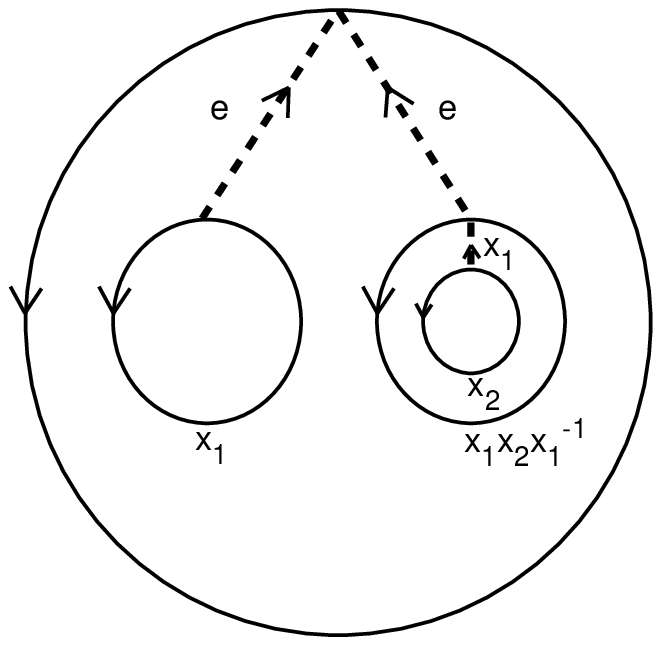}}
\nobreak
\centerline{Figure~10: The isometry~\thetag{8.19}}
\bigskip
\endinsert

We use~\thetag{7.23} to identify an equivalence class of pointed bundles over
the pair of pants~$P$ with an element in~$\goid\times \goid$ (see Figure~8).
Let
  $$ \lt {x_1}{x_2} = L_P(x_1,e;x_2,e)  $$
denote the hermitian line obtained from the classical action on the
equivalence class corresponding to $\langle x_1,e \rangle\times \langle x_2,e
\rangle$.  We claim that for any $x_1,x_2,x_3,g\in \Gamma $ there are
isometries
  $$ \gather
      \phi _{x_1,x_2,x_3}\: \lt{x_1x_2}{x_3} \otimes \lt{x_1}{x_2}
     \longrightarrow \lt{x_1}{x_2x_3} \otimes \lt{x_2}{x_3}, \tag{8.16}\\
      \sigma _{x_1,x_2}\: \lt{x_1}{x_2}\longrightarrow \lt{x\mstrut
     _1x\mstrut _2x_1\inv }{x\mstrut _1}\otimes L_C(x_2,x_1), \tag{8.17}
     \endgather $$
and
  $$ \gamma _{x_1,x_2,g}\: L_C(x_1x_2,g)\otimes \lt{x_1}{x_2} \longrightarrow
     \lt{gx_1g\inv }{gx_2g\inv }\otimes L_C(x_1,g)\otimes L_C(x_2,g).
     \tag{8.18} $$
For~\thetag{8.16} we use the gluings in Figure~3 to see that both sides are
isomorphic to the bundle~$L(x_1,e;x_2,e;x_3,e)$ indicated in Figure~9.  The
isometry~\thetag{8.17} is constructed from the braiding diffeomorphism~$\beta
$, which by~\thetag{7.28} induces an isometry
  $$ \beta ^*\:L_P(x_1,e;x_2,e)\longrightarrow L_P(x_2,x_1;x_1,e),
      $$
and from the gluing in Figure~10, which induces an isometry
  $$ L_P(x_2,x_1;x_1,e)\longrightarrow L_P(x\mstrut _1x\mstrut _2x_1\inv,e;
     x\mstrut _1,e)\otimes L_C(x_2,x_1). \tag{8.19} $$
The isometry~\thetag{8.18} is constructed from the gluing in Figure~11 and
the duality
  $$ L_C(x_i,g)\otimes L_C(x_i,g\inv )\longrightarrow L_C(x_i,e)\cong \CC,
     \tag{8.20} $$
which follows from Figure~12 and~\thetag{8.5}.

\midinsert
\bigskip
\centerline{\epsffile{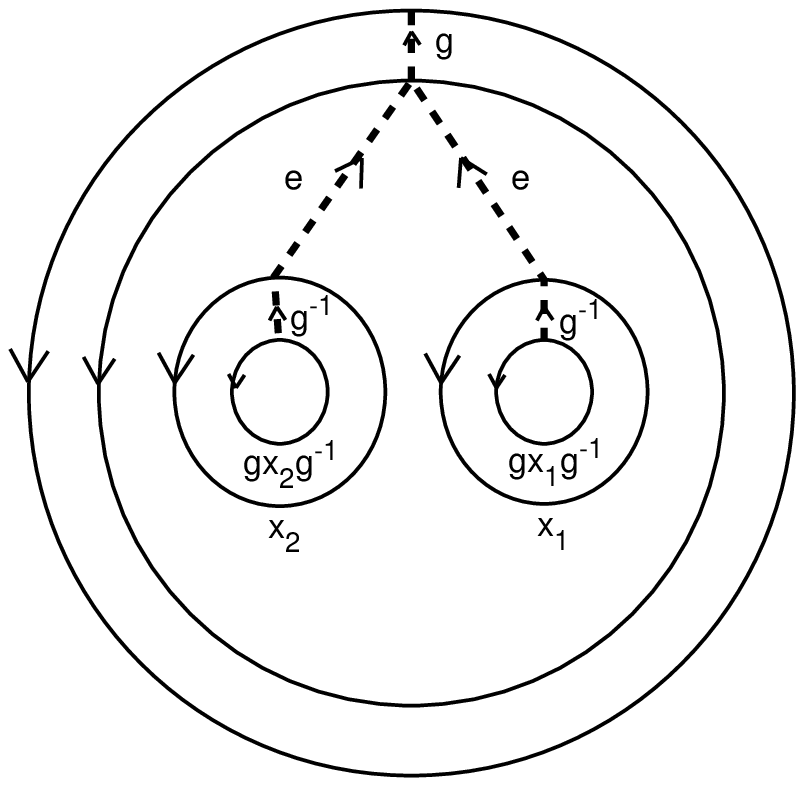}}
\nobreak
\centerline{Figure~11: The isometry~\thetag{8.18}}
\bigskip
\endinsert

\midinsert
\bigskip
\centerline{\epsffile{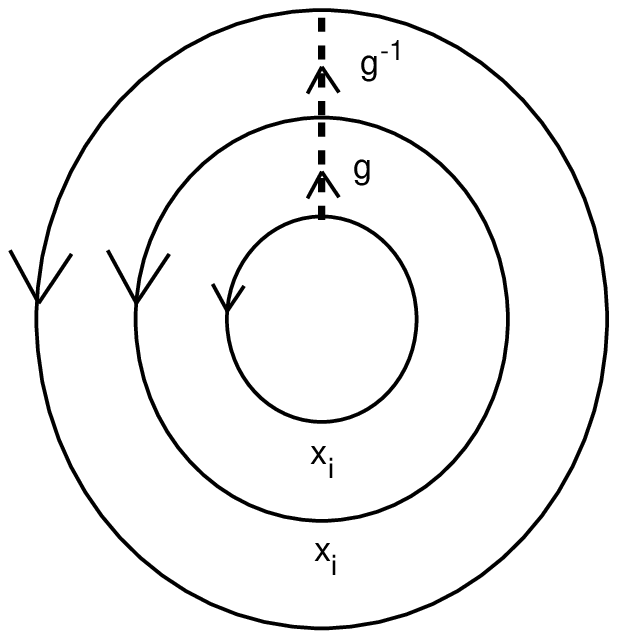}}
\nobreak
\centerline{Figure~12: The duality~\thetag{8.20}}
\bigskip
\endinsert

        \proclaim{\protag{8.21} {Proposition}}
 Consider the finite gauge theory described in \theprotag{4.12} {Assertion}
with lagrangian $\alpha \in C^3(B\Gamma ;\RZ)$.  This field theory and the
trivializations chosen in~\thetag{8.2} determine the following structure
on~$E\ua$.
 \medskip
 \noindent\rom(a\rom)\ \rom(Automorphism of the identity~\thetag{5.11}\rom)\
For $W\in E$ we have
  $$ \theta _W\res{W_x} = A_x(\lce xx)\:W_x\longrightarrow W_x.  $$
 \noindent\rom(b\rom)\ \rom(Involution~\thetag{5.14}\rom)\ For~$W\in E$ the
dual $W^*\in \overline{E}$ is defined by $(W^*)\mstrut _x=W_{x\inv }^*\otimes
L_{[x]}^*$ and $A^{W^*}_g = (A^W_{g\inv })^*$.
 \smallskip
 \noindent\rom(c\rom)\ \rom(Identity~\thetag{5.16}\rom)\ The identity~$\bo$ is
  $$ \bo_x = \cases L_{D^2}([P_{\text{triv} }]) ,&x=e;\\0,&x\not=
     e,\endcases \tag{8.22} $$
with the action of the central extension $\Ctil_e $~ on~$\bo_e$ determined by
gluing a cylinder~$C$ to a disk~$D^2$.
 \smallskip
 \noindent\rom(d\rom)\ \rom(Multiplication~\thetag{5.19}\rom)\ The tensor
product of $W_1,W_2\in E$ is
  $$ (W_1\mytimes W_2)_x = \bigoplus_{x_1x_2=x} \lt{x_1}{x_2}\otimes
     (W_1)_{x_1} \otimes (W_2)_{x_2} \tag{8.23} $$
with the $\Gamma $~action
  $$ A^{W_1\mytimes W_2}_g = \(\id\otimes A^{W_1}_g \otimes A^{W_2}_g\)\circ
     (\gamma _{x_1,x_2,g}\otimes \id) \tag{8.24} $$
on $\lt{x_1}{x_2}\otimes (W_1)_{x_1} \otimes (W_2)_{x_2}$.
 \smallskip
 \noindent\rom(e\rom)\ \rom(Associator~\thetag{5.22}\rom)\ For $W_1,W_2,W_3\in
E$ the associator is
  $$ \varphi _{W_1,W_2,W_3} = \phi _{x_1,x_2,x_3}\otimes \id  $$
on $\lt{x_1}{x_2}\otimes \lt{x_1x_2}{x_3}\otimes (W_1)\mstrut _{x_1}\otimes
(W_2)\mstrut _{x_2}\otimes (W_3)\mstrut _{x_3}$.
 \smallskip
 \noindent\rom(f\rom)\ \rom($R$-matrix~\thetag{5.25}\rom)\ For $W_1,W_2\in E$
we have
  $$ \aligned
       R_{W_1,W_2}\: \lt{x_1}{x_2}\otimes (W_1)\mstrut _{x_1}\otimes
     (W_2)\mstrut _{x_2}
      &\longrightarrow \lt{x\mstrut _1x\mstrut _2x_1\inv}{x\mstrut _1}\otimes
     (W_2)\mstrut _{x\mstrut _1x\mstrut _2x_1\inv }\otimes (W_1)\mstrut
     _{x_1} \\
       \ell \otimes w_1\otimes w_2 &\longmapsto (\id\otimes
     A_{x_1}^{W_2})(\sigma _{x_1,x_2} (\ell )\otimes (w_2))\otimes
     w_1\endaligned \tag{8.25} $$
and all other components are zero.
        \endproclaim

\flushpar
 A few remarks.  First, we omitted transposition of ordinary tensor products
of vector spaces from the notation in~\thetag{8.24} and~\thetag{8.25}.  Also,
the conformal weight is defined by~\thetag{7.18} with $A_x(\lce xx)$
replacing~$A_x$ on the left hand side.  The special (inverse ribbon) element
of~$H\ua$ replacing~\thetag{7.19} is
  $$ v\ua = \sum\limits_{x}\lce xx. \tag{8.26} $$
In~(b) the isometry~\thetag{8.15} is implicit in the equation $A^{W^*}_g =
(A^W_{g\inv })^*$.  In~\thetag{8.22}, $[P_{\text{triv} }]$~is the equivalence
class of the trivial bundle over the disk, and gluing a cylinder gives
isometries
  $$ \lc eg\otimes L_{D^2}([P_{\text{triv} }]) \longrightarrow
     L_{D^2}([P_{\text{triv} }]) , \tag{8.27} $$
which is the required action of~$\Ctil _e$.  Of course, \thetag{8.27} is
equivalent to a linear map
  $$ \epsilon \:\bigoplus_g \lc eg\longrightarrow \CC. \tag{8.28} $$
The verifications of ~\thetag{5.15}, \thetag{5.17}, \thetag{5.20},
\thetag{5.21}, \thetag{5.23}, \thetag{5.24}, \thetag{5.26}, \thetag{5.27},
and~\thetag{5.28} directly from the data listed in \theprotag{8.21}
{Proposition} require some additional identities in the classical theory
easily derived from simple gluings of the type already considered.

The proof of \theprotag{8.21} {Proposition} is a straightforward extension of
the proof of \theprotag{7.12} {Proposition}, so we omit it.

It remains to deduce a quasi-Hopf algebra structure on~$H\ua$.  For this we
need to choose trivializing elements\footnote{From the point of view of the
reconstruction theorems, the reason we need to choose these elements is to
obtain a functor from~$E$ to the category of vector spaces which preserves
the tensor product.  Hence the line which appears in~\thetag{8.23} must be
trivialized.}
  $$ \lte{x_1}{x_2}\in \lt{x_1}{x_2}. \tag{8.29} $$
Define
  $$ \omega (x_1,x_2,x_3) = \frac{\lte{x_1}{x_2x_3}\otimes
     \lte{x_2}{x_3}}{\phi _{x_1,x_2,x_3}\bigl(\lte{x_1x_2}{x_3} \otimes
     \lte{x_1}{x_2}\bigr)}\in \TT. \tag{8.30} $$
An argument with gluings and ungluings of the four times punctured disk shows
that $\omega $~satisfies the cocycle identity
  $$ \frac{\omega (x_1\,,\,x_2\,,\,x_3)\,\omega
     (x_1\,,\,x_2x_3\,,\,x_4)\,\omega (x_2\,,\,x_3\,,\,x_4)}{\omega
     (x_1\,,\,x_2\,,\,x_3x_4)\,\omega (x_1x_2\,,\,x_3\,,\,x_4)} =1,\qquad
     x_1,x_2,x_3,x_4\in \Gamma . \tag{8.31} $$
In a sense this is the classical analog of the pentagon diagram\thetag{5.23}.
So $\omega $~defines a class $[\omega ]\in H^3(\Gamma ;\RZ)$ in group
cohomology.  The following proposition is analogous
to~\cite{FQ,Proposition~3.14}.  We state it without proof.

        \proclaim{\protag{8.32} {Proposition}}
 Under the isomorphism $H^\bullet(\Gamma )\cong H^\bullet(B\Gamma )$ the
group cohomology class~$[\omega ]$ corresponds to the singular cohomology
class~$[\alpha ]$.
        \endproclaim

Now we write the quasitriangular quasi-Hopf structure on~$H\ua$ induced from
the data in \theprotag{8.21} {Proposition}.  The coproduct is
  $$ \Delta \ua(\ell ) = \sum\limits_{x_1x_2=x} \frac{\gamma
     _{x_1,x_2,g}\bigl(\ell \otimes \lte{x_1}{x_2} \bigr)}{\lte{gx_1g\inv
     }{gx_2g\inv }},\qquad \ell \in \lc xg. \tag{8.33} $$
The counit is the linear map defined in~\thetag{8.28}; it maps~$\lc xg$
to~$0$ if~$x\not= 0$.  The antipode is computed from \theprotag{8.21(b)}
{Proposition} as the inverse
  $$ S\:\lc xg\longrightarrow \lc{gx\inv g\inv }{g\inv } \tag{8.34} $$
of~\thetag{8.15}.  The quasitriangular element $R\ua\in H\ua\otimes H\ua$ is
  $$ R\ua = \sum\limits_{x_1,x_2}\lce{x_1}e \otimes \frac{\sigma
     _{x_1,x_2}\bigl(\lte{x_1}{x_2} \bigr)}{\lte{x\mstrut _1x\mstrut
     _2x_1\inv }{x\mstrut _1}}. \tag{8.35} $$
Finally, there is an invertible element $\varphi \ua\in H\ua\otimes
H\ua\otimes H\ua$ which implements the quasiassociativity condition
  $$ (\id\otimes \Delta\ua )\Delta\ua (\ell ) = \bigl(\varphi\ua
     \bigr)(\Delta\ua \otimes \id)\Delta\ua (\ell )\bigl(\varphi\ua\bigr)\inv
     ,\qquad \ell \in H\ua. $$
This is the element
  $$ \varphi \ua = \sum\limits_{x_1,x_2,x_3}\omega (x_1,x_2,x_3)\inv\;
     \lce{x_1}e\otimes \lce{x_2}e\otimes \lce{x_3}e. \tag{8.36} $$
A routine check shows that the modular tensor category described in
\theprotag{8.21} {Proposition} is the category of representations of the
quasitriangular quasi-Hopf algebra~$H\ua$.

The quasi-Hopf algebra in~\cite{DPR,\S3.2} looks similar to~$H\ua$, but is
expressed in terms of a basis.  We will choose this basis geometrically in
the next section, and so construct an isomorphism between~$H\ua$ and the
algebra in~\cite{DPR,\S3.2}.

\newpage
\head
\S{9} Higher Gluing and Good Trivializations
\endhead
\comment
lasteqno 9@ 26
\endcomment

In this section we introduce a ``higher order gluing law'' for gluing
manifolds with corners.  The corners we use are in codimension two; clearly
there are generalizations of this gluing law to higher codimension.  Also the
gluing law we use here pertains to the classical theory; there are quantum
versions as well.  While the formulation of this gluing law is rather
abstract, the computations which follow should make its meaning clear.  We
study the classical theory over the inteval~$[0,1]$.  We choose
trivializations~\thetag{9.4} which replace the trivializations~\thetag{8.2}
we chose in the last section.  The procedure here is more natural than that
of~\S{8}.  Furthermore, the trivializations~\thetag{9.4} induce
trivializations of the lines~$\lt{x_1}{x_2}$ which we previously chose
separately in~\thetag{8.29}, and they also induce trivializations of the
lines $L(x,g)\cong L_C(x,g)$.  The latter amount to a basis of the
algebra~$H\ua$ in~\thetag{8.6}.  In terms of this basis the quasitriangular
quasi-Hopf structure we computed in~\S{8} is exactly the one constructed
in~\cite{DPR,\S3.2}, as we verify.  The reader may wish to consider
analogous, but simpler, computations in the $1+1$~dimensional theory.

\midinsert
\bigskip
\centerline{\epsffile{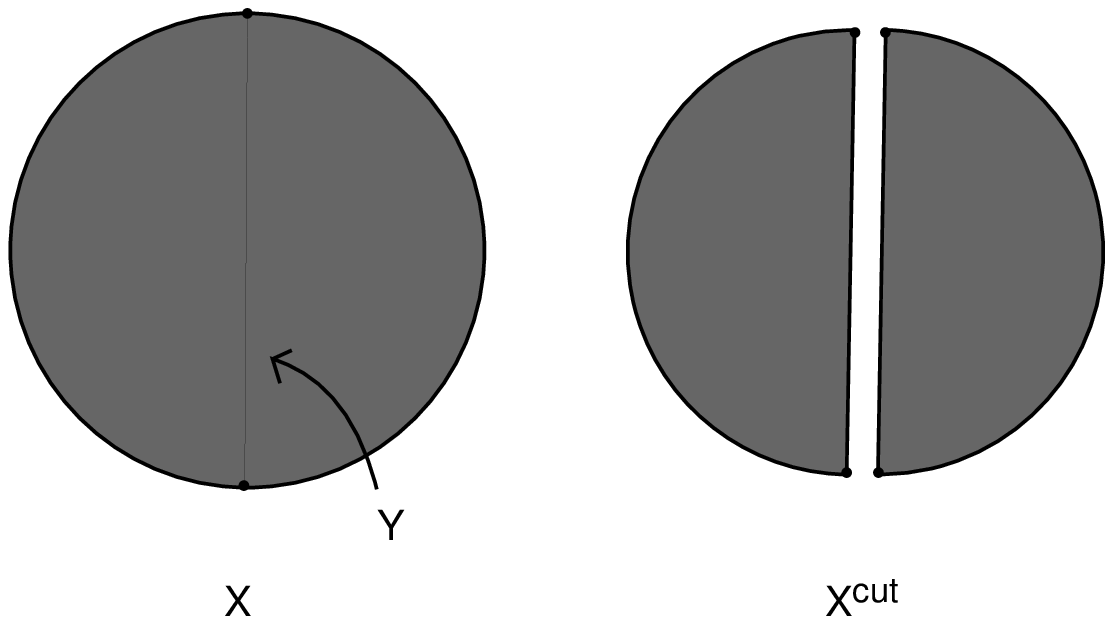}}
\nobreak
\centerline{Figure~13: Gluing manifolds with corners}
\bigskip
\endinsert

We begin with a statement of the gluing law which should hold in any
classical field theory, but for our purposes we consider the classical
$d+1$~dimensional theory of \theprotag{2.5} {Assertion}.  Suppose $X$~is a
compact oriented $(d+2-n)$-manifold and $Y\hookrightarrow X$ a {\it neat\/}
oriented codimension one submanifold~(Figure~13), that is, $\bY=Y\cap\bX$ and
$Y$~intersects $\bX$ transversely.  Then $\bY\hookrightarrow \bX$ is a closed
oriented codimension one submanifold, and
  $$ \aligned
     \bX\cut &= Y \cup_{\bY} (\bX)\cut \cup_{-\bY} -Y,\\
     \partial (\bX)\cut &= -\bY\sqcup \bY.\endaligned  $$
Suppose $P\to X$ is a $\Gamma $~bundle and $Q\to Y$ its restriction to~$Y$.
Then the usual gluing law \theprotag{2.5(d)} {Assertion} implies that there
is an isomorphism\footnote{We use the notation $T_Y(Q)$ for the classical
action, even though $Y$~is not closed.}
  $$ \Tr_{12,34}\:\ca YQ\cdot T_{(\bX)\cut}\bigl((\partial P)\cut \bigr)
     \cdot \ca YQ\inv \longrightarrow \ca{\bX\cut}{\partial P\cut}. \tag{9.1}
     $$
Note that the left hand side of~\thetag{9.1} is an element of
  $$ \ca{\bY}{\partial Q}\cdot \ca{\bY}{\partial Q}\inv \cdot
     \ca{\bY}{\partial Q}\cdot \ca{\bY}{\partial Q}\inv .  $$
Also, there is an isomorphism
  $$ \Tr_{14,23}\:\ca YQ\cdot T_{(\bX)\cut}\bigl((\partial P)\cut \bigr)
     \cdot \ca YQ\inv \longrightarrow \ca{\bX}{\partial P},   $$
and so finally an isomorphism
  $$ \Tr_Q = \Tr_{14,23}\circ \Tr_{12,34}\inv \:\ca{\bX\cut}{\partial
     P\cut}\longrightarrow \ca{\bX}{\partial P}.  $$

        \proclaim{\protag{9.2} {Assertion}}
 In the situation described, there is a natural isomorphism
  $$ \Tr_Q \( \eac{X\cut}{P \cut}\)\longrightarrow \eac XP . \tag{9.3} $$
        \endproclaim

Now we resume our work from~\S{8}, retaining the notations there.  As
in~\S{6} fix a trivial bundle $\Rtriv= pt\times \Gamma $ over a point.  Use
the correspondence~\thetag{6.2} to identify equivalence classes of fields
over~$[0,1]$ trivialized over the endpoints with elements of~$\Gamma $.  Then
the classical action of the equivalence class~$[Q_x]$ corresponding to~$x\in
\Gamma $ is a \T-gerbe $\CG_x=T\ua_{[0,1]}([Q_x])$.  Choose trivializing
elements
  $$ G_x\in \CG_x = T\ua_{[0,1]}([Q_x]),\qquad x\in \Gamma . \tag{9.4} $$
Now for $x_1,x_2\in \Gamma $ we glue $[Q_{x_2}]$ and~$[Q_{x_1}]$ to
obtain~$[Q_{x_1x_2}]$.  Hence the isomorphism~\thetag{2.12} implies that
there is an isomorphism
  $$ \CG_{x_1}\cdot \CG_{x_2}\longrightarrow \CG_{x_1x_2}. \tag{9.5} $$
In particular, \thetag{9.5}~implies that $\CG_e$~has a trivialization
compatible with gluing, and we assume that $G_e$~is that trivialization.  In
other words,
  $$ G_e\cdot G_e = G_e.  \tag{9.6} $$
Define the \T-torsor~$T_{x_1,x_2}$ by the equation
  $$ G_{x_1}\cdot G_{x_2} = G_{x_1x_2}\cdot T_{x_1,x_2},\qquad x_1,x_2\in
     \Gamma , \tag{9.7} $$
where we implicitly use the isomorphism~\thetag{9.5} to compare the two
sides.  Equation~\thetag{9.6} implies that~$T_{e,e}=\TT$.  Three intervals
can be glued together in two different ways to obtain a single interval.  The
behavior of the classical action under iterated gluings, which we did not
explicitly state in \theprotag{2.5(d)} {Assertion}, implies that for
any~$x_1,x_2,x_3\in \Gamma $ the diagram
  $$ \CD
      (\Gb1\cdot \Gb2)\cdot \Gb3 @>>> \CG_{x_1x_2}\cdot
      \Gb3 @>>>\CG_{x_1x_2x_3}\\
	@VVV\\
      \Gb1\cdot (\Gb2\cdot \Gb3)@>>> \Gb1\cdot \CG_{x_2x_3}
      \endCD  $$
commutes up to a natural transformation.  Using~\thetag{9.7} this natural
transformation amounts to an isomorphism
  $$ T_{x_1,x_2}\cdot T_{x_1x_2,x_3}\cong T_{x_2,x_3}\cdot T_{x_1,x_2x_3},
     \qquad x_1,x_2,x_3\in \Gamma . \tag{9.8} $$
In particular, taking two of~$x_1,x_2,x_3$ to be~$e$ we deduce isomorphisms
  $$ T_{x,e} \cong T_{e,x} \cong \TT,\qquad x\in \Gamma . \tag{9.9} $$

Now we explain the relationship of the choices~\thetag{9.4} to the
choices~\thetag{8.2} made in the last section.  Fix~$x\in \Gamma $ and
consider the bundle $Q_{[x]}\to\cir$ with basepoint~$f_x$, as chosen
in~\S\S{7--8}.  Cutting the circle at its basepoint, and using the
basepoint~$f_x$ to identify~$\partial Q\cut_{[x]}$ with~$\Rtriv\sqcup
\Rtriv$, we obtain from the gluing law~\thetag{2.12} an isomorphism
  $$ \CG_x\longrightarrow \CG_{[x]}. \tag{9.10} $$
It is {\it not\/} necessarily true that the trivializations of~$\CG_{x'}$
in~\thetag{9.4} for different~$x'\in [x]$ lead to the same trivialization
of~$\CG_{[x]}$.\footnote{In this connection notice that whereas
$T(x,g_{x,x'})$~was chosen to be~$\TT$ in~\thetag{8.4}, this torsor is
nontrivial with our current set of choices (cf.~\thetag{9.16}).} Now let
$X$~be a compact oriented 2-manifold with parametrized boundary and $P\in
\bfld X$ a bundle with basepoints on the boundary.  Suppose $Y$~is a
component of~$\bX$ and $P\res Y$~has holonomy~$x$.  The basepoint and
parametrization induce an identification $P\res Y\cong \Qx$, and so
by~\thetag{9.10} an isomorphism $T\ua_Y(P\res Y)\cong \CG_x$.  We trivialize
this \T-gerbe using~\thetag{9.4}.  Then as in the argument preceding
\theprotag{8.10} {Proposition} the classical action of~$P$ is a \T-torsor.
It is not the same \T-torsor obtained in~\S{8}, since we use different
trivializations.  None of the subsequent arguments are affected by this
change, and we use these new trivializations in what follows.

As a first application of \theprotag{9.2} {Assertion} we claim that the
classical action of the trivial bundle over the disk is
  $$ T_{D^2}([P_{\text{triv} }])=G_e. \tag{9.11} $$
This can be deduced from the gluing in Figure~13 and~\thetag{9.6}.

Next, choose trivializing elements
  $$ t_{x_1,x_2}\in T_{x_1,x_2},\qquad x_1,x_2\in \Gamma . \tag{9.12} $$
We assume that
  $$ t_{x,e}=t_{e,x}=1,\qquad x\in \Gamma , \tag{9.13} $$
under the isomorphism~\thetag{9.9}.  Define $\omega (x_1,x_2,x_3)\in \TT$
by the equation
  $$ t_{x_1x_2,x_3}\cdot t_{x_1,x_2}\cdot \omega (x_1,x_2,x_3) =
     t_{x_1,x_2x_3}\cdot t_{x_2,x_3}, \qquad x_1,x_2,x_3\in \Gamma ,
     \tag{9.14} $$
where the equality refers to the isomorphism~\thetag{9.8}.  The behavior of
the classical action under iterated gluings of four intervals shows that
$\omega $~satisfies the cocycle identity~\thetag{8.31}.

\midinsert
\bigskip
\centerline{\epsffile{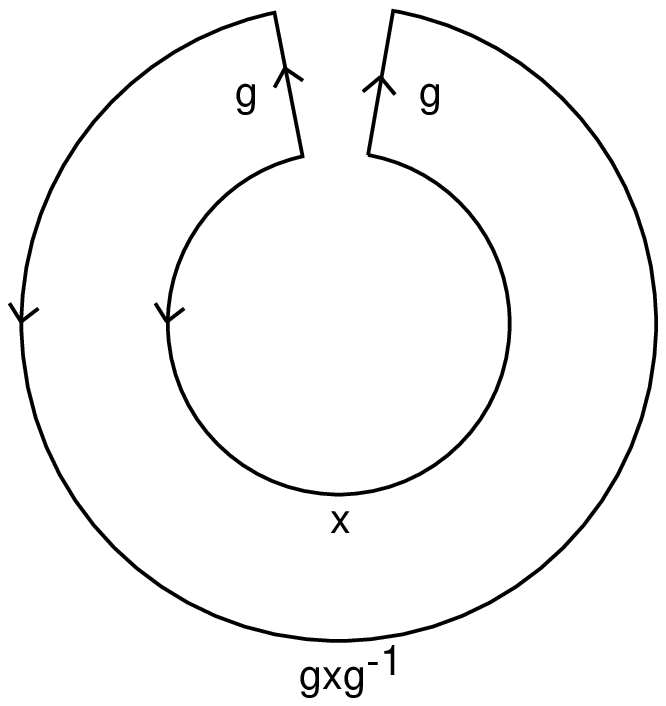}}
\nobreak
\centerline{Figure~14: The isomorphism~\thetag{9.16}}
\bigskip
\endinsert

Let~$L_{x_1,x_2}$ be the hermitian line corresponding to the
\T-torsor~$T_{x_1,x_2}$ and
  $$ \nte{x_1}{x_2}\in L_{x_1,x_2} \tag{9.15} $$
the element of unit norm corresponding to~$t_{x_1,x_2}$.  We claim that with
the choices of trivializations we have made, the higher gluing
law~\thetag{9.3} constructs isometries
  $$ \alignat2
      \lc xg&\cong \frac{L_{g,x}}{L_{gxg\inv ,g}},\qquad &x,g&\in \Gamma ,
     \tag{9.16}\\
      \lt{x_1}{x_2} &\cong L_{x_1,x_2},\qquad &x_1,x_2&\in \Gamma . \tag{9.17}
      \endalignat $$
The isomorphism~\thetag{9.16} is derived from the gluing in Figure~14, where
we obtain the cylinder~$C$ by gluing a disk~$D^2$ along part of its boundary.
The usual gluing law~\thetag{2.12} applied to~$\partial D^2$ yields an
isomorphism
  $$ \CG_e \cong  \CG_{gxg\inv }\cdot \CG_g\cdot \CG_x\inv \cdot \CG_g\inv ,
      $$
and a short computation with~\thetag{9.7} shows that under this isomorphism
we have
  $$ G_e = G_{gxg\inv }\cdot G_g\cdot G_x\inv \cdot G_g\inv \cdot
     \frac{T_{g,x}}{T_{gxg\inv ,g}}\;.  $$
Now \thetag{9.16}~follows from~\thetag{9.11} and the gluing law.  The
isomorphism~\thetag{9.17} is derived in a similar manner from Figure~15.  In
that figure
  $$ \CG_e\cong \CG_{x_1x_2}\cdot \Gb2\inv \cdot \Gb1\inv ,  $$
and under this isomorphism
  $$ G_e = G_{x_1x_2}\cdot G_{x_2}\inv \cdot G_{x_1}\inv \cdot T_{x_1,x_2}.
      $$
The gluing law and~\thetag{9.11} imply~\thetag{9.17}.

\midinsert
\bigskip
\centerline{\epsffile{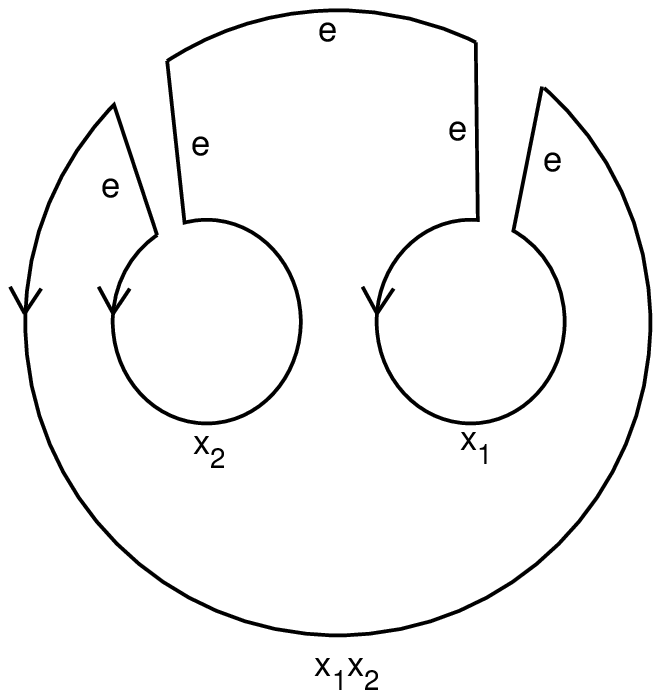}}
\nobreak
\centerline{Figure~15: The isomorphism~\thetag{9.17}}
\bigskip
\endinsert

We use~\thetag{9.15} to trivialize the lines~$\lc xg$ and~$\lt{x_1}{x_2}$.
Namely, set
  $$ \lce xg = \frac{\nte gx}{\nte{gxg\inv }g} \tag{9.18} $$
and
  $$ \lte{x_1}{x_2} = \nte{x_1}{x_2}. \tag{9.19} $$
The elements in~\thetag{9.19} replace the arbitrary choice~\thetag{8.29} we
made in~\S{8}.  We now define the quasi-Hopf quasitriangular structure
on~$H\ua$ in terms of the choices~\thetag{9.19}.  The elements
in~\thetag{9.18} form a basis of~$H\ua$, and our last task is to compute the
quasi-Hopf quasitriangular structure in terms of this basis.  Observe also
that \thetag{9.18}~agrees with the special trivializations~\thetag{8.5}
and~\thetag{8.13}.

First, we compute the isomorphisms \thetag{8.16}--\thetag{8.18} in terms
of~\thetag{9.18} and~\thetag{9.19}.  We make the obvious computations and
leave the justification to the reader.  (This involves the compatibility of
various gluings and diffeomorphisms.)  The isomorphism~$\phi _{x_1,x_2,x_3}$
is still expressed by~\thetag{8.30}, which follows directly
from~\thetag{9.14}.  For~$\sigma _{x_1,x_2}$ we compute
  $$ \frac{\sigma _{x_1,x_2}\bigl(\lte{x_1}{x_2} \bigr)}{\lte{x\mstrut
     _1x\mstrut _2x_1\inv }{x\mstrut _1}\otimes \lce{x_2}{x_1}} \;=\;
     \frac{\nte{x_1}{x_2}}{\nte{x\mstrut _1x\mstrut _2x_1\inv }{x\mstrut
     _1}\otimes \dfrac{\nte{x_1}{x_2}}{\nte{x\mstrut _1x\mstrut _2x_1\inv
     }{x\mstrut _1}}} \;=\; 1.  $$
A direct computation yields
  $$ \frac{\gamma _{x_1,x_2,g}\bigl(\lce{x_1x_2}{g} \otimes \lte{x_1}{x_2}
     \bigr)}{\lte{gx_1g\inv }{gx_2g\inv }\otimes \lce{x_1}g\otimes
     \lce{x_2}g} \;=\; \frac{\omega (g\,,\,x_1\,,\,x_2)\,\omega (gx_1g\inv
     \,,\,gx_2g\inv ,\,g)}{\omega (gx_1g\inv ,\,g\,,\,x_2)}.  $$

Now for the structure on~$H\ua$.  A short computation shows that the
multiplication~\thetag{8.7} is
  $$ \lce{g\mstrut _1xg_1\inv }{g_2}\cdot \lce x{g_1} = \frac{\omega
     (g\mstrut _2\,,\,g\mstrut _1\,,\,x)\,\omega (g\mstrut _2g\mstrut
     _1xg_1\inv g_2\inv ,\,g\mstrut _2\,,\,g\mstrut _1)}{\omega (g\mstrut
     _2\,,\,g\mstrut _1xg_1\inv ,\,g\mstrut _1)} \; \lce x{g_2g_1}.
     \tag{9.20} $$ 
The identity element is~\thetag{8.8}:
  $$ 1 = \sum\limits_{x}\lce xe. \tag{9.21} $$ 
The coproduct~\thetag{8.33} is
  $$ \Delta \ua(\lce xg ) = \sum\limits_{x_1x_2=x} \frac{\omega
     (g\,,\,x_1\,,\,x_2)\,\omega (gx_1g\inv ,\,gx_2g\inv
     \,,\,g)}{\omega (gx_1g\inv ,\,g\,,\,x_2)}\;\lce{x_1}g\otimes
     \lce{x_2}g. \tag{9.22} $$ 
The counit~\thetag{8.28} is
  $$ \epsilon \bigl(\lce xg \bigr) = \cases 1 ,&\text{if $x=e$}
     ;\\0,&\text{otherwise} .\endcases \tag{9.23} $$  
The quasitriangular element~\thetag{8.35} is
  $$ R\ua = \sum\limits_{x_1,x_2}\lce{x_1}e\otimes \lce{x_2}{x_1}.
     \tag{9.24} $$
The element~$\varphi \ua$ which measures the deviation from coassociativity
is~\thetag{8.36}:
  $$ \varphi \ua = \sum\limits_{x_1,x_2,x_3}\omega (x_1,x_2,x_3)\inv\;
     \lce{x_1}e\otimes \lce{x_2}e\otimes \lce{x_3}e. \tag{9.25} $$ 
 Recall that the antipode~\thetag{8.34} is the inverse of~\thetag{8.15}.  With
the trivializations of this section equation~\thetag{8.14} is replaced by the
equation
  $$ G\mstrut _x\cdot G_{x\inv } = G\mstrut _e\cdot T_{x,x\inv },
     $$   
and so \thetag{8.15}~by a map
  $$ \iota _*\:\lc{gx\inv g\inv }{g\inv }\otimes L_{x,x\inv }\longrightarrow
     \lc xg \otimes L_{gxg\inv ,gx\inv g\inv }. $$
The ratio
  $$ \frac{\lce xg\otimes \nte{gxg\inv }{gx\inv g\inv }}{\iota
     _*\bigl(\lce{gx\inv g\inv }{g\inv }\otimes \nte x{x\inv } \bigr)} $$
is the numerical factor in the expression
  $$ \spreadlines{6pt}\multline
      S\bigl(\lce xg \bigr)\\
      = \frac{\omega (g\inv ,\,gx\inv g\inv ,\,g)\,\omega (gxg\inv
     ,\,g\,,\,x\inv )}{\omega (g\inv ,\,g\,,\,x\inv )\,\omega (x\inv ,\,g\inv
     ,\,g)\omega (g\,,\,x\,,\,x\inv )\,\omega (gxg\inv ,\,gx\inv g\inv ,\,g)}
     \; \lce{gx\inv g\inv }{g\inv } \endmultline\tag{9.26} $$ 
for the antipode.  The inverse ribbon element is~\thetag{8.26}:
  $$ v\ua = \sum\limits_{x}\lce xx.   $$

Equations \thetag{9.20}--\thetag{9.26} are exactly the equations
in~\cite{DPR,\S3.2}, up to some changes in notation.

Suppose we replace the trivializations~$t_{x_1,x_2}$ in~\thetag{9.12} with
$\beta (x_1,x_2)t_{x_1,x_2}$ for some~$\beta (x_1,x_2)\in \TT$.  We assume
that $\beta (x,e)=\beta (e,x)=1$ for all~$x\in \Gamma $ so that
\thetag{9.13}~is respected.  Then this change of basis has the effect of {\it
twisting\/} (cf.~\cite{Dr}) the formulas \thetag{9.20}--\thetag{9.26} by the
element
  $$ \sum\limits_{x_1,x_2}\beta (x_1,x_2)\;\lce{x_1}e\otimes \lce{x_2}e. $$

We conclude with some brief general remarks about gluing.  The first should
be valid for arbitrary topological theories in any dimension.  Consider
$Y\hookrightarrow X$ a closed oriented codimension one submanifold and
$X\cut$ the cut manifold as in \theprotag{2.5(d)} {Assertion}.  For a new
manifold~$W$ by identifying the two pieces in the boundary of $[0,1]\times
X\cut$ which correspond to $[\frac 12,1]\times Y$, as illustrated in
Figure~16.  Then
  $$ \partial W=X\,\sqcup \,-X\cut\,\sqcup \,[0,\frac 12]\times Y\,\sqcup
     \,-[0,\frac 12]\times Y. $$
In the classical theory we also are given a field~$P$ on~$X$ and the
corresponding~$P\cut$ on~$X\cut$.  We claim that the gluing~\thetag{2.12} of
the classical action (resp.~the gluing~\thetag{4.17} of the path integral) is
computed by the classical action (resp.~path integral) over~$W$.  For this we
trivialize the classical action (resp.~path integral) over $[0,\frac 12]$
using~\thetag{5.2} (resp.~\thetag{5.4}).  Such pictures help compute the
gluing isometries.

\midinsert
\bigskip
\centerline{\epsffile{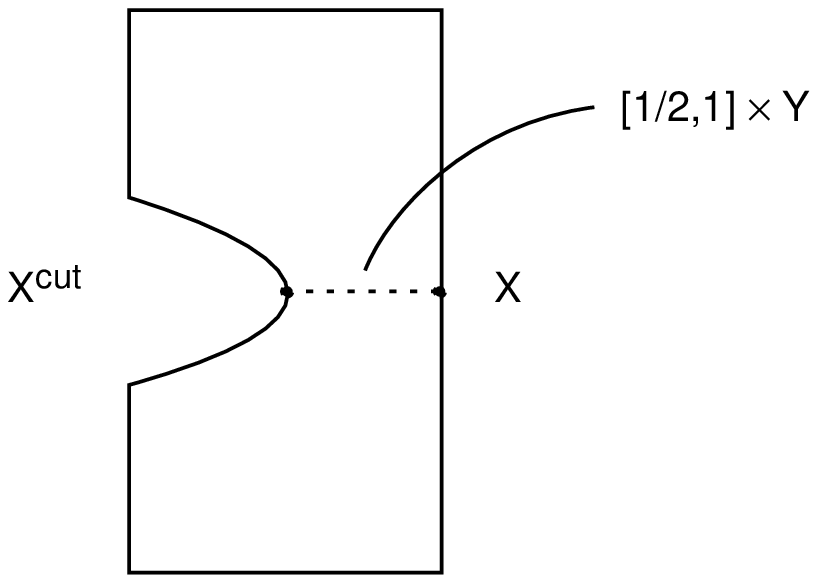}}
\nobreak
\centerline{Figure~16: Gluing along a closed submanifold}
\bigskip
\endinsert

Figure~16 is a schematic for arbitrary dimensions as well as an exact picture
of the gluing of two intervals.  The reader may wish to contemplate various
gluings of this figure and relate the computations in~\S{8} to those in~\S{9}.

There should also be refined gluing laws of the following sort.  Recall from
\theprotag{5.29} {Proposition} that in a $2+1$~dimensional theory
$E(\cir)$~is a ``higher commutative associative algebra with compatible real
structure'' which presumably is semisimple (in a unitary theory).  In
particular, it is a braided monoidal category, or better a tortile category.
For such categories one can apparently define a ``Grothendieck ring''~$\grot$
(see~\cite{Y2,Prop.~26}).  If $E(\cir)$~is the category of representations of
a quasi-Hopf algebra~$H$, then the Grothendieck ring is the ring of
equivalence classes of representations, the multiplication given by the
tensor product.  Equation~\thetag{5.5} is a gluing law on the level of inner
product spaces, and in this case surely there is an extension to an
isomorphism
  $$ E(\cir\times \cir)\cong \grot $$
of algebras.  ($E(\cir\times \cir)$~is an algebra by the remark at the end
of~\S{5}.  It is commonly called the {\it Verlinde algebra\/}.)  The
Grothendieck ring is the ``dimension'' of~$E(\cir)$ from the point of view
of~\thetag{5.5}.  Notice that $\grot$ has a distinguished basis of
irreducible representations.  These are the ``labels'' mentioned in~\S{7}.

\newpage
\head
Appendix: Integration of Singular Cocycles Revisited
\endhead
\comment
lasteqno A@ 12
\endcomment

In ~\cite{FQ,Appendix~B} we describe some elements of an integration theory
for singular cocycles with coefficients in~$\RZ$.  Here we describe an
extension of that theory to higher codimensions in terms of the higher
algebra discussed in~\S{1}.  Notice that we do not introduce any basepoints
or special choices, as in~\cite{FQ,Proposition~B.5}.  Instead, we extend the
integration theory in a more intrinsic manner to all codimensions.  The
higher algebra of~\S{1} is a prerequisite to this appendix.

Our goal is to integrate a singular $(d+1)$-cocycle~$\alpha $ over compact
oriented manifolds of any dimension less than or equal to~$(d+1)$.
In~\cite{FQ} we described the integral of~$\alpha $ over closed oriented
$(d+1)$-manifolds, compact oriented $(d+1)$-manifolds (possibly with
boundary), and closed oriented $d$-manifolds.  In the easiest case $\alpha
$~is a $(d+1)$-cocycle on a closed oriented $(d+1)$-manifold~$X$.  Then if
$x\in C_{d+1}(X)$ is an oriented cycle which represents the fundamental class
$[X]\in H_{d+1}(X)$, we form the pairing~$\etpi{\alpha (x)}\in \RZ$.  If
$x'$~is another representative, then $x'-x=\partial w$ for some $w\in
C_{d+1}(X)$.  Hence $\alpha (x')-\alpha (x)= \alpha (\partial w) = \delta
\alpha (w)=0$ since $\alpha $~is a cocycle.  This is the usual argument which
shows that the integral
  $$ \exp(\tpi\int_{X}\alpha )\in \tcat0=\TT \tag{A.1}$$
is well-defined.  In fact, \thetag{A.1}~can be viewed as the pairing between
the cohomology class $[\alpha ]\in H^{d+1}(X;\RZ)$ and the homology
class~$[X]\in H_{d+1}(X)$.  This is the only one of the integrations we
discuss which has cohomological meaning.

Now suppose $\alpha $~is a $(d+1)$-cocycle on a closed oriented
$d$-manifold~$Y$.  Then we claim that there is a well-defined integral
  $$ I_{Y,\alpha }=\exp(\tpi\int_{Y}\alpha )\in \tcat1 \tag{A.2}$$ 
which is a \T-torsor.  The following is a slight modification of what appears
in~\cite{FQ,Appendix~B}.  The justification for terming this an `integral'
are the properties listed in \theprotag{A.4} {Assertion}.  Let $\cat Y$~be
the category whose objects are oriented cycles $y\in C_d(Y)$ which represent
the fundamental class $[Y]\in H_d(Y)$, and with a unique morphism~$y\to y'$
for all~$y,y'\in \cat Y$.  Define a functor $\fun Y\:\cat Y\to\tcat1$ by
$\fun Y(y)=\TT$ for each~$y$ and $\fun Y(y\to y')$ acts as multiplication
by~$\etpi{\alpha (x)}$, where $x$~is any $(d+1)$-chain with $y'=y + \partial
x$.  An easy argument shows that $\alpha (x) = \alpha (x')$ for any two
choices of such a chain.  Define~$\iline Y$ as the {\it inverse limit\/}
of~$\fun Y$.\footnote{See the beginning of~\S{2} for a discussion of inverse
limits.} That is, an element of~$\iline Y$ is a function $i(y)\in \fun
Y(y)=\TT$ on the objects in~$\cat Y$ such that $i(y') = \fun Y(y\to
y')\,i(y)$ for all morphisms $y\to y'$.  It is easy to check that $\iline
Y$~exists.

Next, suppose $\alpha $~is a $(d+1)$-cocycle on a closed oriented
$(d-1)$-manifold~$S$.  Then we claim that the integral
  $$ I_{S,\alpha }=\exp(\tpi\int_{S}\alpha )\in \tcat2  $$
now makes sense as a \T-gerbe.  The construction is entirely analogous to the
previous one except there is one more layer of argument.  So consider the
category~$\cat S$ whose objects are oriented cycles $s\in C_{d-1}(S)$ which
represent the fundamental class $[S]\in H_{d-1}(S)$, and with a unique
morphism between any two objects.  Now if $s,s'\in\cat S$, construct a
category~$\cat{s,s'}$ whose objects are $d$-chains~$y$ which satisfy
$s'=s+\partial y$, and with a unique morphism between any two objects.
Define a functor $\fun{s,s'}\:\cat{s,s'}\to\tcat1$ by $\fun{s,s'}(y)=\TT$ for
each~$y$ and $\fun{s,s'}(y\to y')$ acts as multiplication by~$\etpi{\alpha
(x)}$, where $x$~is any $(d+1)$-chain with $y'=y+\partial x$.  An easy
argument shows that $\alpha (x)= \alpha (x')$ for any two choices of such a
chain.  Define the \T-torsor~$\ilines{s,s'}$ to be the inverse limit
of~$\fun{s,s'}$.  Now define a functor $\fun S\:\cat S\to \tcat2$ by $\fun
S(s)=\tcat1$ for each~$s$ and $\fun S(s\to s')$ acts as multiplication
by~$\ilines{s,s'}$.  The \T-gerbe~$\iline S$ is defined to be the inverse
limit of~$\fun S$.

It is clear how to continue to higher codimensions.  Now we turn to manifolds
with boundary.

If $\alpha $~is a $(d+1)$-cocycle on a compact oriented $(d+1)$-manifold~$X$,
then in~\cite{FQ,Proposition~B.1} we describe the integral
  $$ \exp(\tpi\int_{X}\alpha )\in I_{\partial X,i^*\alpha }, $$
where $i\:\partial X\hookrightarrow X$ is the inclusion of the boundary, and
$I_{\bX,i^*\alpha }$ is the \T-torsor described previously.  We will not
review that here, but rather go on to the next case.  Namely, suppose that
$\alpha $~is a $(d+1)$-cocycle on a compact oriented $d$-manifold~$Y$.  The
we claim that the integral
  $$ \exp(\tpi\int_{Y}\alpha )\in I_{\partial Y,i^*\alpha } $$
makes sense, where now $I_{\bY,i^*\alpha }$~is the \T-gerbe described
previously.  Call $S=\bY$ and let $s\in C_{d-1}(S)$ represent the fundamental
class, i.e., $s\in \cat S$.  By the definition of~$I_{\bY,i^*\alpha }$ above
we must construct a torsor $\ilines{Y,s}\in \tcat1$ and for any $s,s'\in \cat
S$ an isomorphism
  $$ \ilines{Y,s}\otimes \ilines{s,s'}\longrightarrow \ilines{Y,s'}.
     \tag{A.3}$$
To construct~$\ilines{Y,s}$ let $\cat{Y,s}$~be the category whose objects are
$d$-chains $y\in C_d(Y)$ such that $y$~represents the fundamental class
$[Y,\bY]\in H_d(Y,\bY)$ and $\partial y=i_*s$.  We postulate a unique
morphism $y\to y'$ between any two objects of~$\cat{Y,s}$.  Define a functor
$\fun{Y,s}\:\cat{Y,s}\to\tcat1$ by $\fun{Y,s}(y)=\TT$ for each ~$y$ and
$\fun{Y,s}(y\to y') $ is multiplication by~$\etpi{\alpha (x)}$, where $x$~is
any $(d+1)$-chain with $y'=y+\partial x$.  As before, this is independent of
the choice of~$x$.  Set $\ilines{Y,s}$~to be the inverse limit
of~$\fun{Y,s}$.  To construct the isomorphism~\thetag{A.3}, suppose that
$y\in \cat{Y,s}$ and~$a\in \cat{s,s'}$, i.e., $y\in C_d(Y)$
represents~$[Y,\bY]$ with~$\partial y=s$, and $a\in C_d(S)$ with $\partial
a=s'-s$.  Then $y+a\in \cat{Y,s'}$.  The isomorphism~\thetag{A.3} is defined
to be the identity relative to the trivializations of the torsors determined
by~$y$, $a$, and~$y+a$.

This discussion indicates the constructions contained in the following
assertion, which we boldly state for arbitrary codimension.

     \proclaim{\protag{A.4} {Assertion}}
 Let $Y$~be a closed oriented $(d+1-n)$-manifold ($n>0$) and $\alpha \in
C^{d+1}(Y;\RZ)$ a singular cocycle.  Then there is an element~$I_{Y,\alpha
}\in \tcat n$ defined.  If $X$~is a compact oriented $(d+2-n)$-manifold,
$i\:\partial X\hookrightarrow X$ the inclusion of the boundary, and $\alpha
\in C^{d+1}(X;\RZ)$ a cocycle, then
  $$ \eint X{\alpha }\in I_{\partial X,i^*\alpha } $$
is defined.  These ``higher $\TT$-torsors'' and integrals satisfy:\newline
 \rom(a\rom)\ \rom({\it Functoriality\/}\rom)\ If $f\:Y'\to Y$ is an
orientation preserving diffeomorphism, then there is an induced isomorphism
  $$ f_*\:\intline {Y'}{f^*\alpha } \longrightarrow \iline{Y}  $$
and these compose properly.  If $F \:X'\to X$ is an orientation preserving
diffeomorphism, then there is an induced isomorphism\footnote{If $n=1$ then
\thetag{A.5}~is an {\it equality\/} of elements in a $\ZZ$-torsor.  For~$n>1$
it is an {\it isomorphism\/} between elements in a ``higher $\ZZ$-torsor''.
A similar remark holds for~\thetag{A.7}, \thetag{A.9}, and~\thetag{A.10}.}
  $$ (\partial F)_*\left[ \eint {X'}{F^*\alpha }\right] \longrightarrow \eint
     {X}{\alpha } . \tag{A.5} $$
 \rom(b\rom)\ \rom({\it Orientation\/}\rom)\ There are natural isomorphisms
  $$ I_{-Y,\alpha }\mstrut  \cong (\iline{Y})\inv , \tag{A.6}$$
and
  $$ \eint{-X}{\alpha }   \cong \[{\eint X{\alpha }}\]\inv . \tag{A.7} $$
 \rom(c\rom)\ \rom({\it Additivity\/}\rom)\ If $Y=Y_1\sqcup Y_2$ is a disjoint
union, then there is a natural isomorphism
  $$ \intline{Y_1\sqcup Y_2}{ \alpha _1\sqcup  \alpha _2} \cong
     \intline{Y_1}{ \alpha _1}\cdot \intline{Y_2}{ \alpha _2}. \tag{A.8}$$
If $X=X_1\sqcup X_2$ is a disjoint union, then there is a natural isomorphism
  $$ \eint{X_1\sqcup X_2}{\alpha _1\sqcup \alpha _2} \cong
     \eint{X_1}{\alpha _1} \cdot \eint{X_2}{\alpha _2}. \tag{A.9}$$
 \rom(d\rom)\ \rom({\it Gluing\/}\rom)\ Suppose $j\:Y\hookrightarrow X$ is a
closed oriented codimension one submanifold and $X\cut$~is the manifold
obtained by cutting~$X$ along~$Y$.  Then $\partial X\cut = \partial X\sqcup Y
\sqcup -Y$.  Suppose $\alpha \in C^{d+1}(X;\RZ)$~is a singular
$(d+1)$-cocycle on~$S$, and $\alpha \cut \in C^{d+1}(X\cut;\RZ)$~the induced
cocycle on~$X\cut$.  Then there is a natural isomorphism
  $$ \Tr_{Y,j^* \alpha }\left[\eint{X\cut}{\alpha \cut}\right]\longrightarrow
     \eint X{\alpha } , \tag{A.10} $$
where $\Tr_{Y,j^* \alpha  }$ is the contraction
  $$ \Tr_{Y,j^* \alpha } \:\intline{\partial X\cut}{ \alpha \cut} \cong
     \intline{\partial X}{i^* \alpha }\otimes \intline{Y}{j^* \alpha }
     \otimes {\intline{Y}{j^* \alpha }}\inv \longrightarrow \intline{\partial
     X}{i^* \alpha }. $$
 \rom(e\rom)\ \rom({\it Stokes' Theorem \rom I\/}\rom)\ Let $\alpha \in
C^{d+1}(W;\RZ)$~be a singular cocycle on a compact oriented
$(d+3-n)$-manifold~$W$.  Then there is a natural isomorphism\footnote{Note
that $\tcat{n-2}$~is the identity element in~$\tcat{n-1}$.  If~$n=1$, then
\thetag{A.11}~should be interpreted as
  $$ \eint{\partial W}{\alpha } = 1. $$
A similar remark applies to~\thetag{A.12} below.}
  $$ \eint{\partial W}{\alpha } \cong \tcat{{n-2}}.  \tag{A.11} $$
 \rom(f\rom)\ \rom({\it Stokes' Theorem \rom{II}\/}\rom)\ A singular
$d$-cochain~$\beta \in C^{d}(Y;\RZ)$ on a closed oriented
$(d+1-n)$-manifold~$Y$ determines a trivialization
  $$ \intline{Y}{\delta \beta }\cong \tcat{n-1}.   $$
A singular $d$-cochain~$\beta \in C^{d}(X;\RZ)$ on a compact oriented
$(d+2-n)$-manifold~$X$ satisfies
  $$ \exp\(\tpi \int_{X}\delta \beta \) \cong \tcat{n-2} \tag{A.12} $$
under this isomorphism.
     \endproclaim

\flushpar
 The assertion in~(e) only has real content for~$n=1$.  If~$n>1$, then
$\iline{\partial W}$~is trivialized by $\eint{W}{\alpha }$.

We leave the reader to contemplate higher order gluing laws analogous
to~\cite{FQ,Proposition~B.10} and those discussed in~\S{9}.

\enddocument